\documentclass[article]{aastex631}
\usepackage{mathrsfs, mathtools}
\usepackage{captcont,xcolor}
\usepackage{blindtext}
\usepackage{subfigure}
\usepackage{graphicx}
\usepackage{CJKutf8}
\usepackage{appendix}
\usepackage{multirow}
\usepackage{nth}
\usepackage{makecell}
\shortauthors{Zhou et al.}

\newcommand{\R}{\textup{\textrm{R}}}
\def \um {$\rm \mu m$}
\def \HCNto {HCN\,$J=2\rightarrow1$}
\def \HCOto {HCO$^+$\,$J=2\rightarrow1$}
\def \HCOp  {HCO$^+$}

\def \Joz {$J=1\rightarrow0$}
\def \Jto {$J=2\rightarrow1$}
\def \kms  {$\rm km\,s^{-1}$}
\def \Lsun {$\rm L_\odot$}

\def\purple#1 {{\textcolor{purple}{#1}}\ }
\def\red#1 {\textcolor{red}{#1}}
\def\new#1 {{\bf #1 }}
\def\blue#1 {{\textcolor{blue}{#1}}\ }

\begin{document}
\begin{CJK*}{UTF8}{gbsn}

\title{Dense Gas and Star Formation in Nearby Infrared Bright Galaxies:  APEX survey of HCN and HCO$^+$ $J$=2$\rightarrow$1 }
\correspondingauthor{Zhi-Yu Zhang}
\email{zzhang@nju.edu.cn}

\author[0000-0002-0818-1745]{Jing Zhou}
\affil{School of Astronomy and Space Science, Nanjing University, Nanjing 210093, China.}
\affil{Key Laboratory of Modern Astronomy and Astrophysics (Nanjing University), Ministry of Education, Nanjing 210093, China.}

\author[0000-0002-7299-2876]{Zhi-Yu Zhang}
\affil{School of Astronomy and Space Science, Nanjing University, Nanjing 210093, China.}
\affil{Key Laboratory of Modern Astronomy and Astrophysics (Nanjing University), Ministry of Education, Nanjing 210093, China.}

\author[0000-0003-0007-2197]{Yu Gao}
\affil{Department of Astronomy, Xiamen University, Xiamen, Fujian 361005, China.}
\affil{Purple Mountain Observation $\&$ Key Laboratory for Radio Astronomy, Chinese Academy of Science, 10 Yuanhua Road, Nanjing 210033, China.}

\author[0000-0001-6106-1171]{Junzhi Wang}
\affil{Shanghai Astronomical Observatory, Chinese Academy of Science, 80 Nandan Road, Shanghai 200030, China.}
\affil{Department of Physics, Guangxi University, Nanning 530004, China.}

\author[0000-0002-8614-6275]{Yong Shi}
\affil{School of Astronomy and Space Science, Nanjing University, Nanjing 210093, China.}
\affil{Key Laboratory of Modern Astronomy and Astrophysics (Nanjing University), Ministry of Education, Nanjing 210093, China.}

\author[0000-0002-3890-3729]{Qiusheng Gu}
\affil{School of Astronomy and Space Science, Nanjing University, Nanjing 210093, China.}
\affil{Key Laboratory of Modern Astronomy and Astrophysics (Nanjing University), Ministry of Education, Nanjing 210093, China.}

\author[0000-0002-8117-9991]{Chentao Yang}
\affil{Department of Space, Earth and Environment, Chalmers University of Technology, Onsala Space Observatory, 439 92 Onsala, Sweden}
\affil{European Southern Observatory, Alonso de C\'ordova 3107, Vitacura, Casilla 19001, Santiago de Chile, Chile.}

\author[0000-0002-2504-2421]{Tao Wang}
\affil{School of Astronomy and Space Science, Nanjing University, Nanjing 210093, China.}
\affil{Key Laboratory of Modern Astronomy and Astrophysics (Nanjing University), Ministry of Education, Nanjing 210093, China.}

\author[0000-0003-3032-0948]{Qing-Hua Tan}
\affil{Purple Mountain Observation $\&$ Key Laboratory for Radio Astronomy, Chinese Academy of Science, 10 Yuanhua Road, Nanjing 210033, China.}

\begin{abstract}
Both Galactic and extragalactic studies on star formation suggest that stars
form directly from dense molecular gas. To trace such high volume density gas,
HCN and HCO$^+$ \Joz\ have been widely used for their high dipole moments,
relatively high abundances, and often being the strongest lines after CO.
However, HCN and HCO$^+$ \Joz\ emission could be arguably dominated by the gas
components at low volume densities. \HCNto\ and \HCOto, with more suitable
critical densities ($1.6 \times 10^{6}$ and $2.8 \times 10 ^5~ \rm cm^{-3}$)
and excitation requirements, would trace typical dense
gas closely related to star formation. Here we report new
observations of \HCNto\ and \HCOto\ towards 17 nearby infrared-bright galaxies
with the APEX 12-m telescope. The correlation slopes between luminosities of
\HCNto, and \HCOto\ and total infrared emission are 1.03$\pm 0.05$ and 1.00$\pm
0.05$, respectively. The correlations of their surface densities, normalised
with the area of radio/sub-millimeter continuum, show even tighter relations
(Slopes: $0.99 \pm 0.03$ and $1.02 \pm 0.03$). The eight AGN-dominated galaxies
show no significant difference from the eleven star-formation dominated
galaxies in above relations. The average HCN/HCO$^+$ ratios are 1.15$\pm$0.26
and 0.98$\pm$0.42 for AGN-dominated and star-formation dominated galaxies,
respectively, without obvious dependencies on infrared luminosity, dust
temperature, or infrared pumping. The Magellanic Clouds roughly follow the same
correlations, expanding to eight orders of magnitude. On the other hand,
ultra-luminous infrared galaxies with active galactic nucleus (AGN)
systematically lay above the correlations, indicating potential biases
introduced by AGNs.

\end{abstract}  

\section{Introduction}\label{sect:introduction}

Stars, as building blocks of galaxies, contribute most radiation of galaxies,
dominate metal enrichment of galaxies, drive galactic outflows, and essentially
construct galaxy structures. Therefore, star formation activity is one of the
most important evolutionary processes in galaxies \citep{kennicutt12}. It has
been found that molecular gas supplies the raw material of star formation,
while the majority of molecular gas is not directly observable from the H$_2$
emission \citep{Bolatto13}. Molecular gas is often traced with the rotational
transitions of other molecules excited by collisions with H$_2$ molecules.
Among them, CO transitions mostly trace the bulk of molecular gas, due to its
weak permanent dipole moment \citep[$\mu_{10}^e = 0.11~\rm
Debye$;][]{Solomon05} and low upper energy levels \citep[5.5 K for
\Joz;][]{moleculardata2005}, while the dense gas tracers, e.g.,
rotational transitions of HCN, HCO$^+$, CS, N$_2$H$^+$, etc., can trace denser
molecular clouds, due to their much higher dipole moments \citep[$\mu_{10}^e$ =
2.98\,Debye and 3.92\,Debye for HCN, HCO$^+$\,\Joz,
respectively;][]{Papadopoulos2007}.

The star formation rates in galaxies, on the other hand, are usually traced
with ultraviolet continuum emission, optical line tracers such as H$\rm \alpha$
and [O {\sc II}], dust emission at infrared (IR) wavelengths, radio continuum
emission, or X-ray emission \citep{Kennicutt98a,kennicutt12}. Among them, the
IR emission from dust normally traces the bolometric energy heated up by the
young stars.  The IR emission is in general extinction-free and often covered
by multi-IR-wavelength space telescopes such as  \textrm{ Herschel}
\citep{Herschel}, \textrm{ Spitzer} \citep{Spitzer}, Infrared Astronomical
Satellite \citep[IRAS;][]{IRAS1984}, etc.  Therefore, the total IR luminosity
is widely adopted as a star-formation rate tracer for gas-rich galaxies
\citep[e.g.,][]{Galametz16,Galametz20}.

A long-standing question remains: which gas is forming stars?
\cite{Kennicutt98} found a super-linear correlation between the surface
densities of star formation rate (SFR) and total gas mass, where the area was
defined by CO \Joz\ or IR images. This correlation, however,
turns to a linear shape with a slope index of unity, when the sample is limited
to nearby normal spiral galaxies beyond 500-pc scales
\citep[e.g.,][]{Bigiel2008}. On the other hand, \cite{Gao2004b} found that the
dense molecular gas traced by HCN\,\Joz\ is the direct source of
star formation. This correlation is further connected to Galactic massive
star-forming regions on sub-pc scales \citep{Wu2005,Wu2010}. Furthermore,
\citet{Heiderman10} also found a similar result by counting young stellar
objects in Galactic dense molecular clouds.

However, the emission of HCN\,\Joz\ and
HCO$^+$\,\Joz\ could be largely contaminated by diffuse gas with
relatively high column densities \citep{Evans2020}.  Although the high-$J$
transitions (e.g., $J=3\rightarrow2$ and $J=4\rightarrow3$) of HCN and
HCO$^+$ \citep{zhang14apj,Tan2018,Lifei3-2} are found to have linear
correlations with star formation rate, these high-$J$ transitions are mainly
from the densest molecular cores heated by massive stars, and leave out most
emission from cold clumps.

\HCNto\ and \HCOto, which have moderate critical densities \citep[$n^{\rm HCN
2-1}_{\rm crit}$= $\sim 1.6\times 10^6~\rm cm^{-3}$, $n^{\rm HCO^+ 2-1}_{\rm
crit}$=  $\sim 2.8\times 10^5~\rm cm^{-3}$, for conditions of a kinetic
temperature $T_{\rm kin} \sim$ 50\,K, optically thin, and no
background][]{Shirley2015}  and suitable upper energy levels \citep[$E^{\rm HCN
2-1}_{\rm up}= 12.76 \rm K$ and $E^{\rm HCO^+ 2-1}_{\rm up}= 12.84 \rm K$,
respectively;][] {Shirley2015}, could avoid aforementioned disadvantages of
other transitions.  On the other hand, the rest frequency of \HCNto\ is close
to the H$_2$O 177.3-GHz line in the Earth atmosphere, so the observation needs
excellent weather conditions for nearby galaxies.



In this paper, we present Atacama Pathfinder Experiment (APEX) observations of
\HCNto\ and \HCOto\ in a sample of 17 IR bright galaxies. In Sect.\ref{sec:obs}
we describe observations with APEX 12-m telescope and ancillary data adopted in
this paper. In Sect.\ref{sec:methods} we describe methods to derive line
luminosities, photometry, and dust properties. In Sect.4 we present the
obtained spectra, correlations between star formation rate and dense gas
tracers, and line ratios. In Sect.5 we discuss the assumptions, caveats, and
physical implications of our results. In Sect.6 we summarize our work. We adopt
cosmological parameters of $H_0=71~\rm km\,s^{-1}\, Mpc^{-1}$, $\Omega_{\rm
M}=0.27$, $\Omega_\Lambda=0.73$ throughout this work \citep{Spergel2007}.

\section{Observation and data reduction}\label{sec:obs}
\begin{deluxetable*}{cccccccccccc}
\tablenum{1}
\tablecaption{Basic Information} \label{table:basicinfo}
\tablewidth{0pt}
\setlength\tabcolsep{3pt}
\tablehead{
\colhead{Source name} & \colhead{R.A.} & \colhead{Dec.} & \colhead{Redshift} &
\colhead{Distance $^a$} & \colhead{Ref.} & \colhead{HPBW$^{1.4\rm GHz}$} & \colhead{Ref.}& \colhead{$S^{\rm 1.4GHz}_{\rm VLA}$ $^d$} & \colhead{$S^{\rm 1.4GHz}_{\rm VLBI}$} & \colhead{$S_{\rm CO J=1-0}$} & \colhead{Type}  \\
\colhead{} & \colhead{(h,m,s)} & \colhead{(d,m,s)} & \colhead{($z$)} & \colhead{(Mpc)} & \colhead{} & \colhead{(arcsec)} &  \colhead{} & \colhead{(mJy)} &\colhead{(mJy)} & \colhead{($\rm Jy~km~s^{-1}$)} & \colhead{} }
\startdata
NGC\,4945        & 13:05:27.51 & $-$49:28:06.0 & 0.00188                   & 3.8  $\pm$   0.3                  &  1                & $7.9 \times 4.3$   & 9                          & 6450                & 33                     & 9887 $\pm$    28              & AGN \\
NGC\,1068        & 02:42:40.70 & $-$00:00:48.0 & 0.00379                   & 10.1 $\pm$   2.0                  &  1              & $9.2  \times 2.2$  & 9                          & 4848                 & 2.6                    & 1903 $\pm$    77              & AGN \\
NGC\,7552        & 23:16:10.70 & $-$42:35:05.0 & 0.00536                   & 14.8 $\pm$   1.3                  &  2               & $7.3  \times 7.1$  & 9                          & 280                  & ~~... $^e$             & 652    $\pm$  88                & SF \\
NGC\,4418        & 12:26:54.61 & $-$00:52:39.0 & 0.00727                   & 23.9 $\pm$   2.2                  &  3               & $0.5$              & 10                         & 41                   & ~~... $^e$             & 132    $\pm$  28                  & SF \\
NGC\,1365        & 03:33:36.40 & $-$36:08:25.0 & 0.00546                   & 17.5 $\pm$   3.5                  &  1                & $21.7 \times 10.5$ & 9                          & 376                  & ~~ 4.57 $^f$           & 2166 $\pm$    102             & AGN \\
NGC\,3256        & 10:27:51.30 & $-$43:54:13.0 & 0.00935                   & 37.4 $\pm$   6.0                  &  4                & $12.2 \times 9.3$  & 9                          & 668                  & ~~... $^e$             & 1223 $\pm$    8               & SF \\
NGC\,1808        & 05:07:42.30 & $-$37:30:47.0 & 0.00332                   & 12.3 $\pm$   2.5                  &  1                & $14   \times 8.8$  & 9                          & 528                  & ~~... $^e$             & 1898 $\pm$    137             & SF \\
IRAS\,13120-5453 & 13:15:06.30 & $-$55:09:23.0 & 0.031249                  & 129.3 $\pm$  9.1                  &  1               & $1.5  \times 0.5$  & 9                          & 118                    & ~~... $^e$             & 126    $\pm$  13                  & SF \\
IRAS\,13242-5713 & 13:27:23.80 & $-$57:29:22.0 & 0.009788                  & 37.6 $\pm$   2.6                  &  5                & $6    \times 3.7$  & 9                          & 97                     & ~~... $^e$             & ...                               & SF \\
MRK\,331         & 23:51:26.80 & $+$20:35:10.0 & 0.01848                   & 53 $\pm$   5                      &  3                & $2.53$             & 11                         & 71                   & $<7.5$                 & 49.5   $\pm$  3.7                 & SF \\
NGC\,6240A       & 16:52:58.90 & $+$02:24:03.5 & \multirow{2}{*}{0.02488 } & \multirow{2}{*}{103 $\pm$  7}     & \multirow{2}{*}{ 6 } & $0.66 \times 0.42$ & \multirow{2}{*}{~~12 $^b$} & \multirow{2}{*}{396} & \multirow{2}{*}{5.4}   & \multirow{2}{*}{333.2  $\pm$  33} & \multirow{2}{*}{AGN} \\
NGC\,6240B       & 16:52:58.91 & $+$02:24:04.2 &                           &                                   &                    & $1.03\times 0.41$  &                            &                        &                        &                                   & \\
NGC\,3628        & 11:20:17.00 & $+$13:35:23.0 & 0.00281                   & 10.3 $\pm$   0.4                  &  7                & $71$               & 13                         & 476                    & $<4.5$                 & 1309 $\pm$    25              & SF \\
NGC\,3627        & 11:20:14.90 & $+$12:59:30.0 & 0.00243                   & 10.7 $\pm$   0.5                  &  1               & $160$              & 13                         & 459                    & $<3.39$                & 4477 $\pm$    75              & AGN \\
IRAS\,18293-3413 & 18:32:41.10 & $-$34:11:27.0 & 0.017996                  & 74.8 $\pm$   5.3                  &  5                & $5.9  \times 4.9$  & 9                          & 226                  & ~~... $^e$             & 686.1  $\pm$  7.6                 & SF \\
NGC\,7469        & 23:03:15.60 & $-$08:52:26.0 & 0.01632                   & 59.7 $\pm$   1.6                  &  3               & $3.3$              & 11                         & 181                  & 32.5                   & 48.1   $\pm$  3.5                 & AGN \\
IRAS\,17578-0400 & 18:00:31.90 & $-$04:00:53.0 & 0.013325                  & 60   $\pm$   6                    &  3                & $3    \times 2.3$  & 9                          & 80                     & ~~... $^e$             & ...                               & SF \\
IC\,1623         & 01:07:47.20 & $-$17:30:25.0 & 0.02007                   & 80.9 $\pm$   5.7                  &  6                & $15$               & 11                         & 249                  & 4.7                    & 291  $\pm$  45                & AGN \\
Arp\,220A        & 15:34:57.29 & $+$23:30:11.3 & \multirow{2}{*}{0.01813}  & \multirow{2}{*}{84.1 $\pm$   5.9} & \multirow{2}{*}{ 6 } & $0.27 \times 0.24$ & \multirow{2}{*}{~~12 $^c$} & \multirow{2}{*}{515}   & \multirow{2}{*}{91.25} & \multirow{2}{*}{515    $\pm$  51} & \multirow{2}{*}{SF} \\
Arp\,220B        & 15:34:57.22 & $+$23:30:11.5 &                           &                                   &                    & $ 0.49\times 0.31$ &                            &                        &                        &                                   & \\
IRAS\,19254-7245 & 19:31:21.40 & $-$72:39:18.0 & 0.06171                   & 273  $\pm$   18                   &  8                & $0.5$              & 14                         & ...                    & ...                    & 58.8   $\pm$  5.9                 & AGN \\
\enddata
\tablecomments{ 
$^a$ We adopt redshift-independent distances measures with Tully-Fisher
relation, Tip of the Red-Giant Branch (TRGB) stars, supernova Ia (SN Ia), and
Cepheids from NED for most of the galaxies. NGC\,1068, NGC\,7552, NGC\,4418, NGC\,3265, NGC\,1808,
MRK\,331, NGC\,3628, and IRAS\,17578-0400 are measured using the Tully-Fisher
relation \citep{Nasonova11,Russell02,Theureau07,Tully88,Tully13}. NGC\,4945 and
NGC\,3627 are measured with TRGBs \citep{Tully15,Jang17}.  NGC\,1365 is
measured with Cepheid \citep{Willick01}. NGC\,7469 is measured with SN Ia
\citep{Koshida17}. We adopt Hubble Flow Distance of the rest galaxies in our sample
\citep{Karachentsev96,Vaucouleurs91,Mould00}.\\
$^b$ Size of NGC\,6240 is estimated from two-component Gaussian fitting of ALMA 480 GHz continuum observation (Project code: 2015.1.00717.S).\\
$^c$ Measurement of Arp\,220 uses data of combination of VLA A configuration and MERLIN from \cite{Varenius16}. \\
$^d$ Fluxes of 1.4-GHz continuum only have uniform error of the survey and the errors of individual galaxy is hard to determine.\\
$^e$ These galaxies are all star-formation
dominated, so their SMBHs would not affect 1.4\,GHz continuum
size. \citep{Varenius14NGC4418,Saikia1990NGC1808,Herrera2017IRAS17578}\\
$^f$ High resolution 1.4\,GHz flux of  NGC\,1365 comes from VLA
observation in \cite{Sandqvist1995NGC1365} }
\tablerefs{(1) \citet{Nasonova11}; (2) \citet{Russell02}; (3) \citet{Theureau07}; (4) \citet{Tully88}; (5) \citet{Vaucouleurs91}; (6) \citet{Mould00}; (7) \citet{Tully13}; (8) \citet{distref8}; (9) \citet{Condon2021}; (10) \citet{Costagliola13}; (11) \citet{liu2015}; (12) This work; (13) \citet{Condon87}; (14) \citet{Imanishi16}}

\end{deluxetable*}

\subsection{Sample selection}

Our sample was selected from the survey of CS $J=7\rightarrow6$, HCN
$J=4\rightarrow3$, and HCO$^+~J=4\rightarrow3$ \citep{zhang14apj}, which
contains nearby normal galaxies, luminous and ultra-luminous infrared galaxies
(ULIRGs).  These galaxies are originally selected from the \textrm{ Infrared
Astronomical Satellite (IRAS)} Revised Bright Galaxy Sample
\citep{sanders2003}.  All galaxies have $S_{\nu}(100\,\rm \mu m)>100\,\rm Jy$,
and declination $<20^{\circ}$ to be accessible from APEX. We exclude three
targets without any detection of HCN $J=4\rightarrow3$ and
HCO$^+~J=4\rightarrow3$.  The final sample consists of 19 galaxies, which
include 17 newly observed galaxies and two ULIRGs from the literature, Arp~220
\citep{Galametz16} and Superantennae \citep{Imanishi22}.

The total IR luminosities range from $1.8\times 10^{10}$ \Lsun\, to $1.8\times
10^{12}$ \Lsun, which implies a range of SFR from $3.6~\rm M_{\odot}\,yr^{-1}$
to $360~\rm M_{\odot}\,yr^{-1}$. The distance range is $\sim$ 3.72 -- 273
Mpc. The molecular gas masses (estimated from CO \Joz) range
from $4\rm \times 10^8~M_{\odot}$ to $2.1 \rm \times 10^{10}~M_{\odot}$. We
further divided the sample into AGN-dominated and Star formation (SF)-dominated
galaxies according to the classifications on NASA/IPAC Extragalactic Database
(NED)\footnote{\url{http://ned.ipac.caltech.edu/}}. The final sample consists
of eight AGN-dominated and eleven SF-dominated galaxies. The basic information
of the sample is shown in Table \ref{table:basicinfo}.

\subsection{\HCNto\ and \HCOto\ observations}

\begin{figure*}[ht]
\includegraphics[height=1.45in]{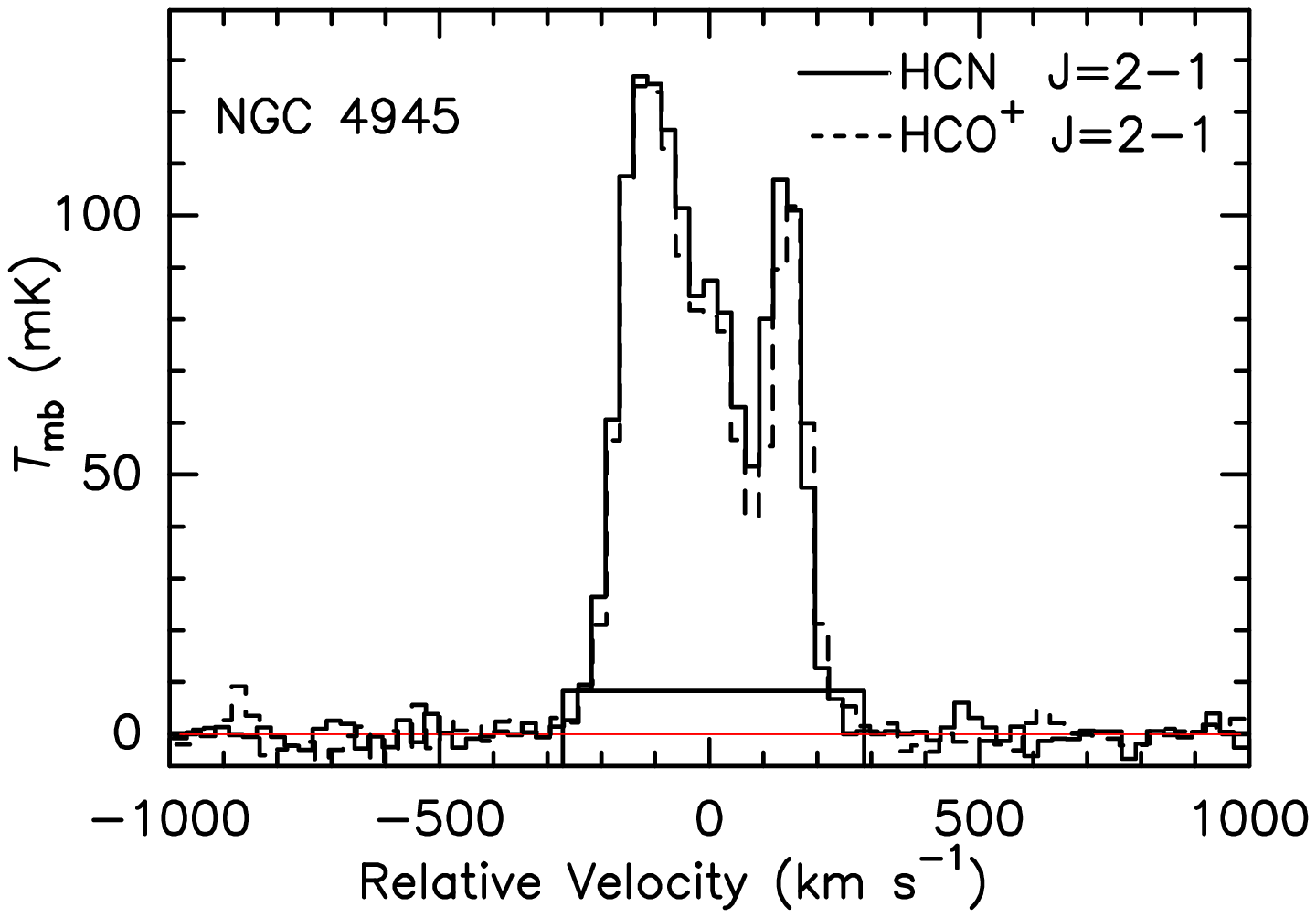}
\includegraphics[height=1.45in]{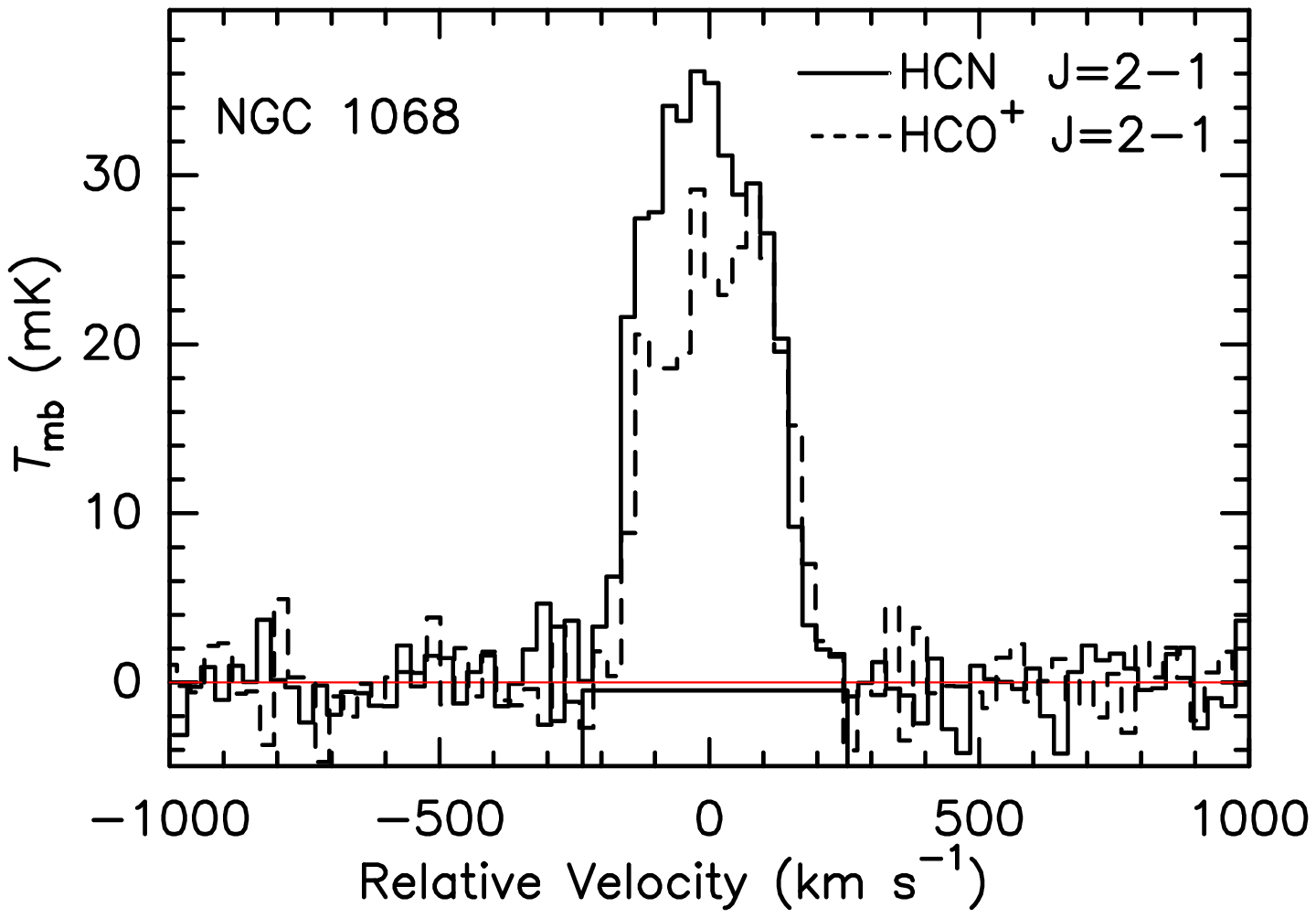}
\includegraphics[height=1.45in]{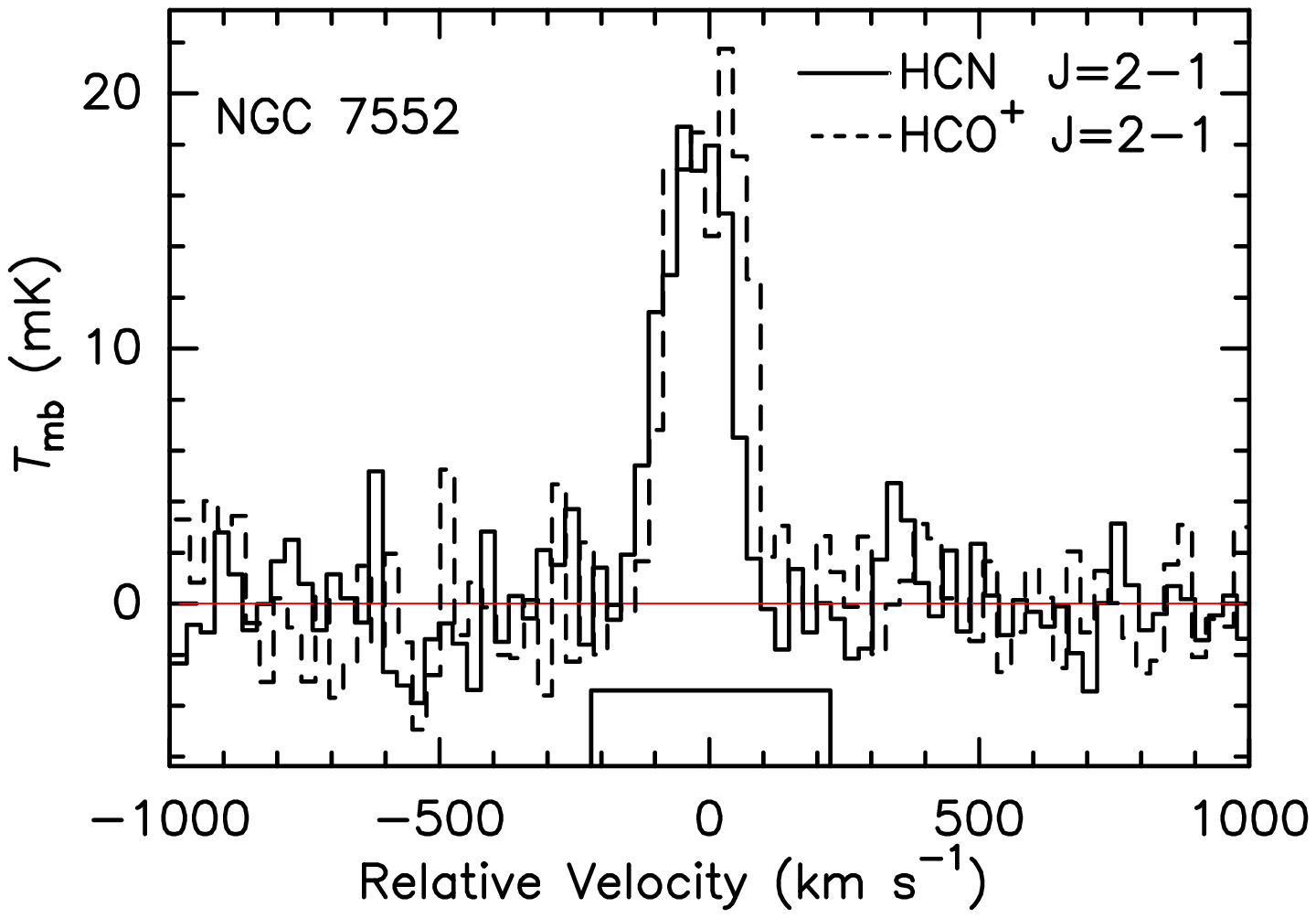}
\includegraphics[height=1.45in]{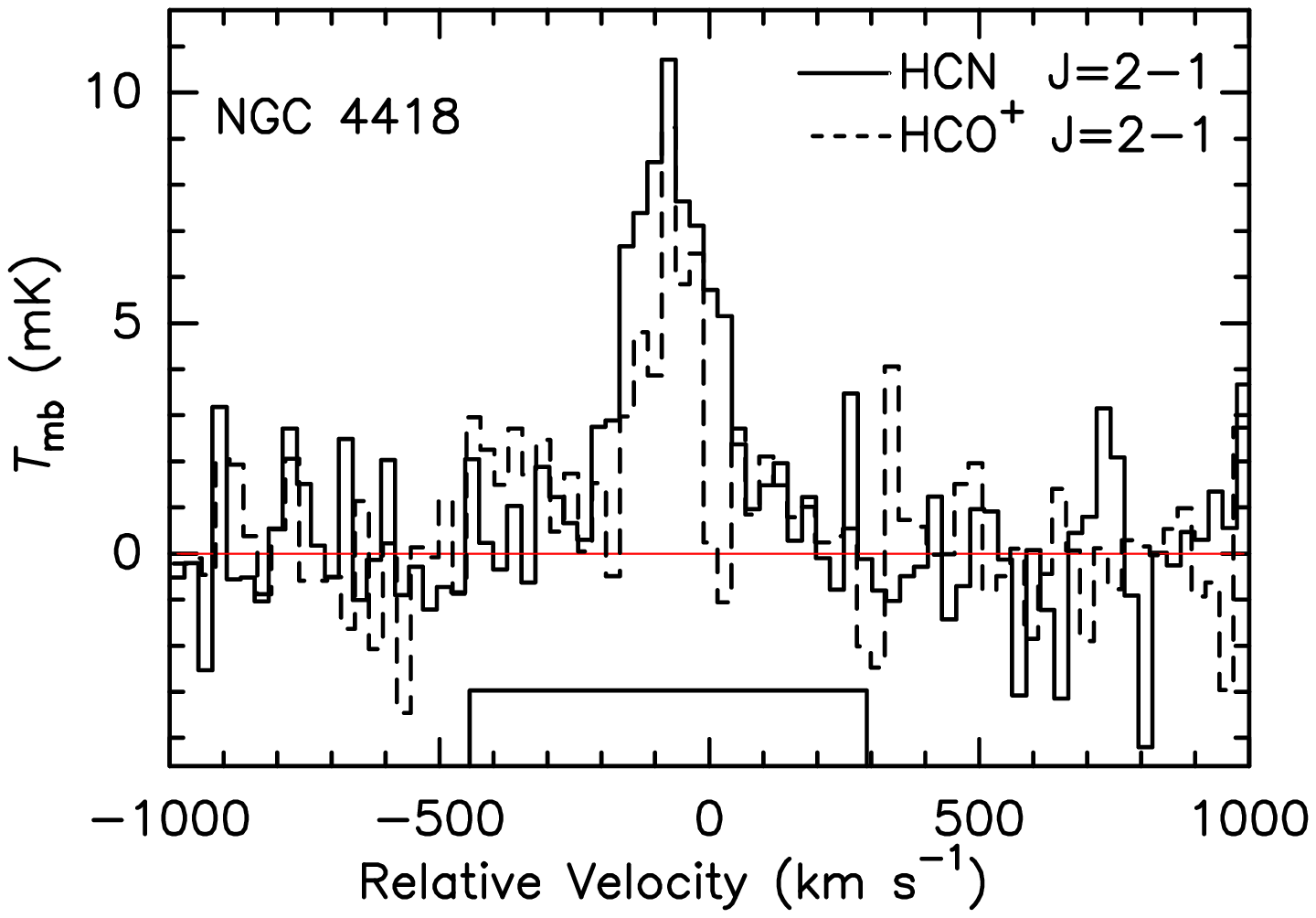}
\includegraphics[height=1.45in]{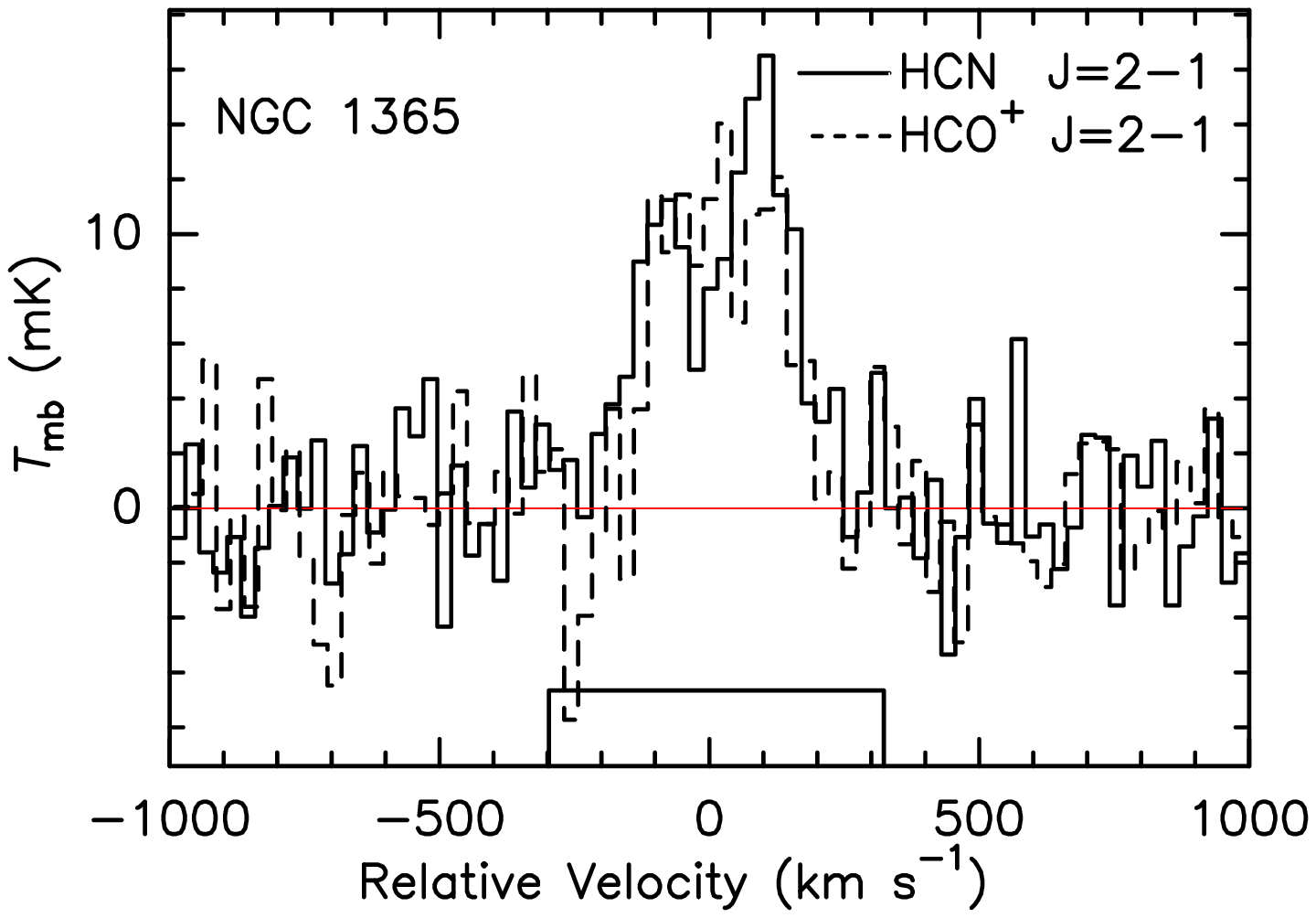}
\includegraphics[height=1.45in]{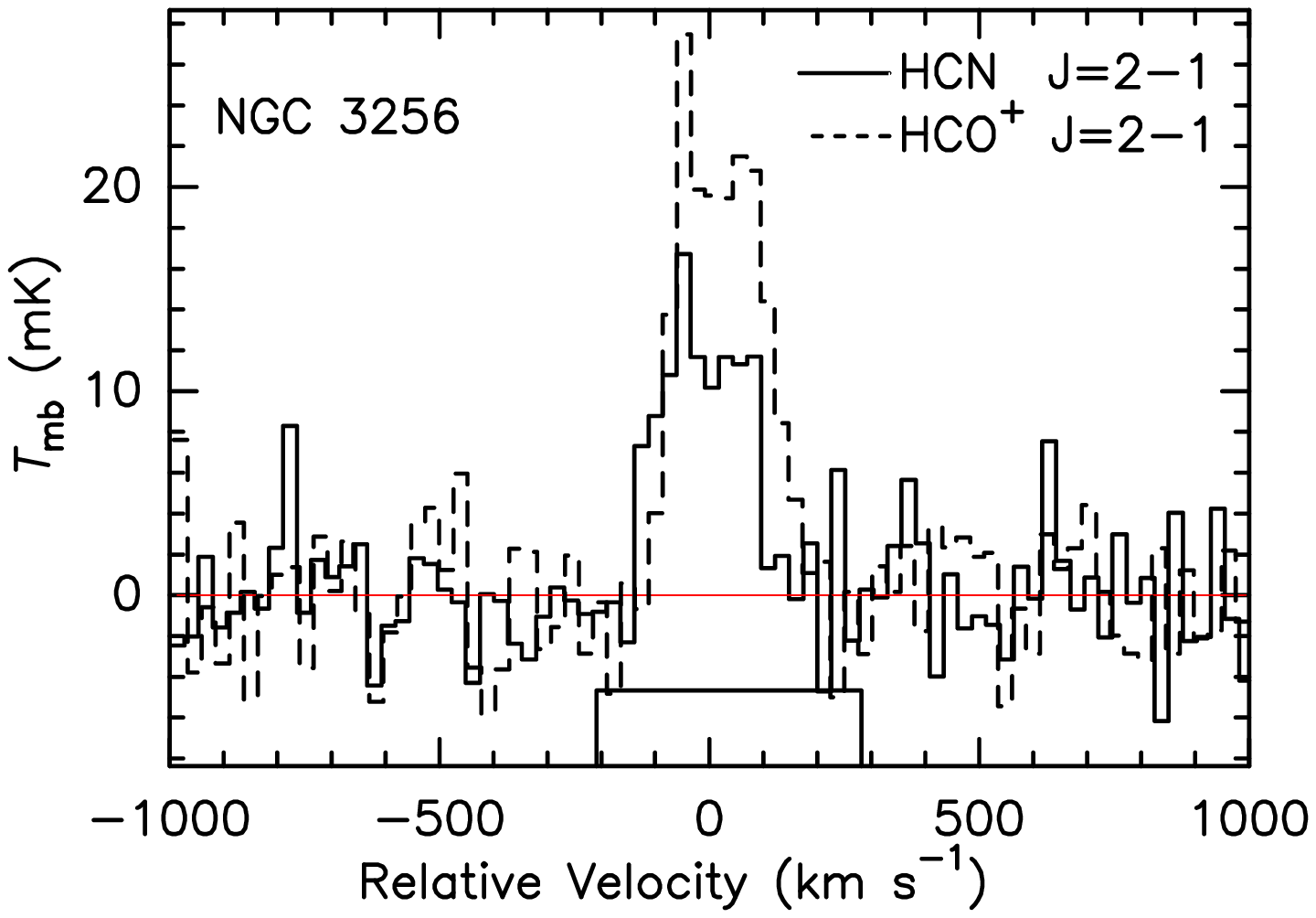}
\includegraphics[height=1.45in]{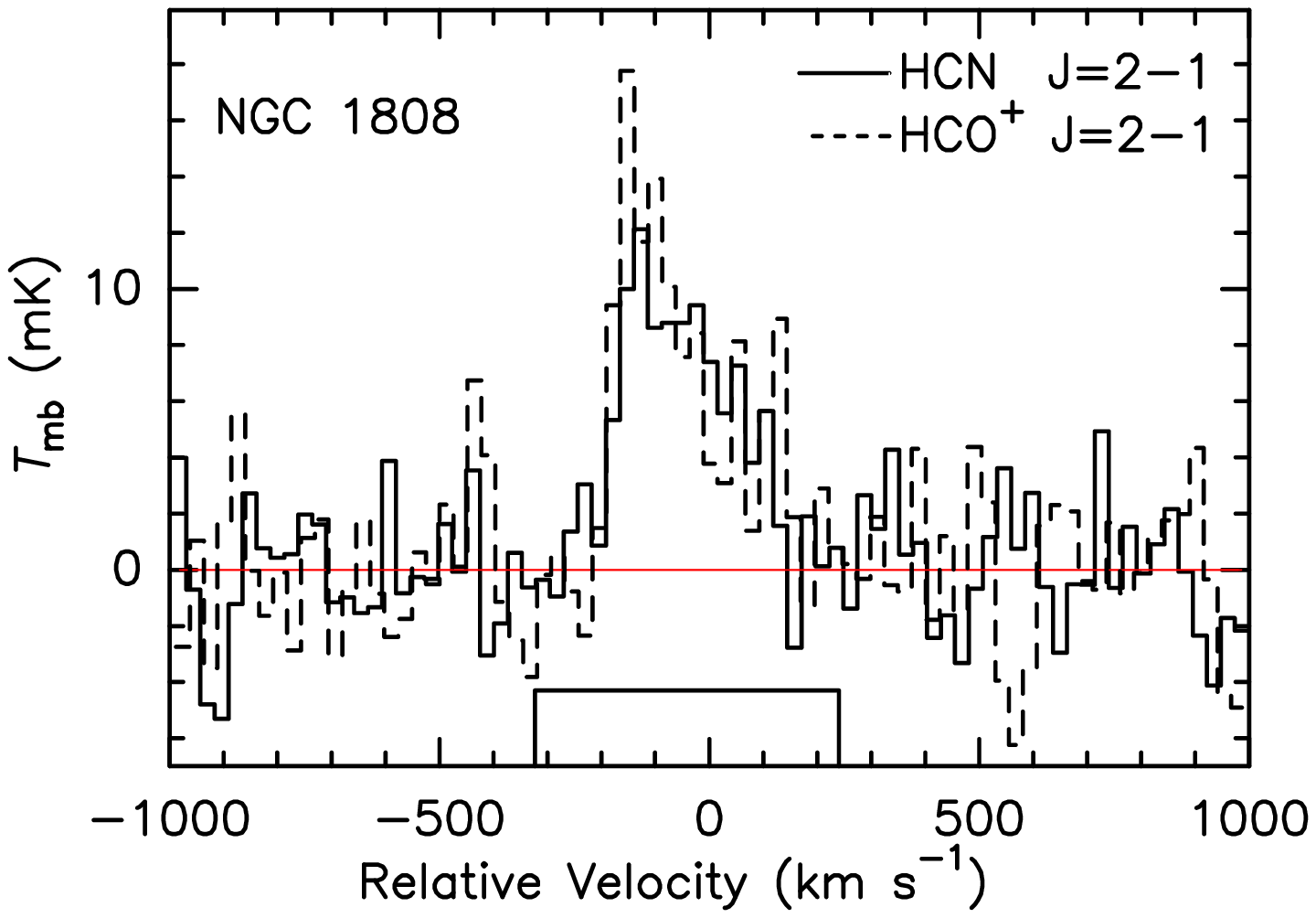}
\includegraphics[height=1.45in]{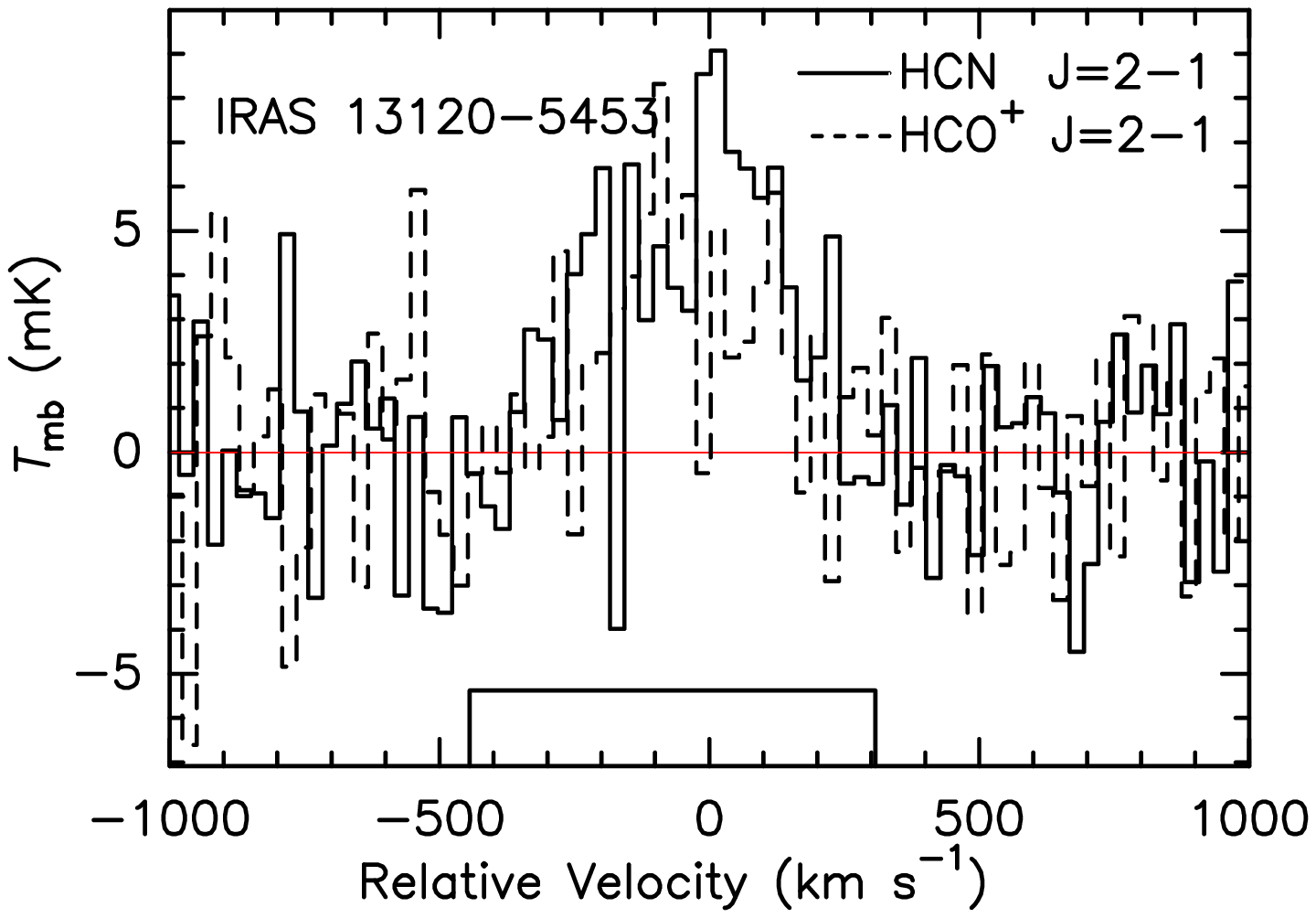}
\includegraphics[height=1.45in]{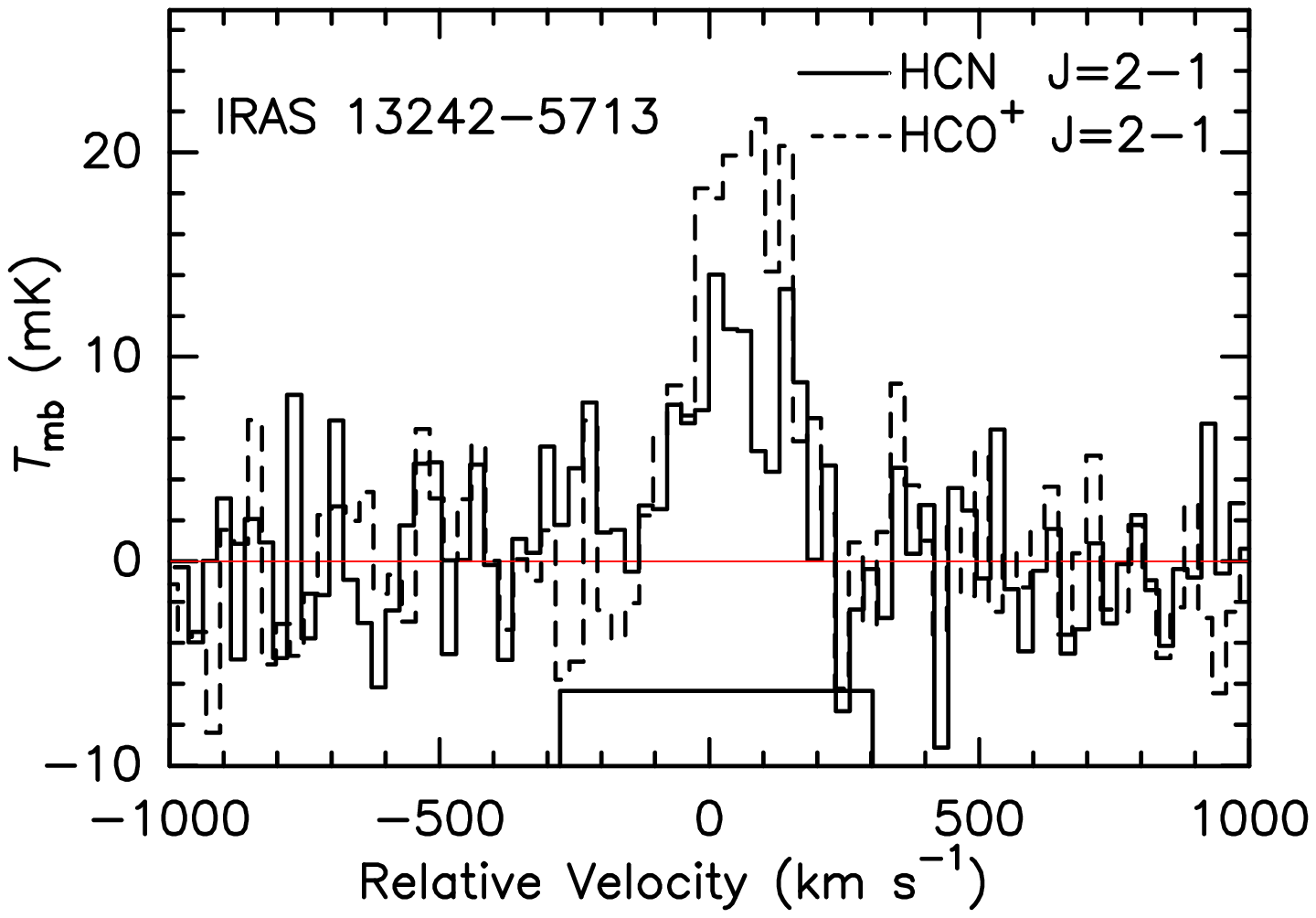}
\includegraphics[height=1.45in]{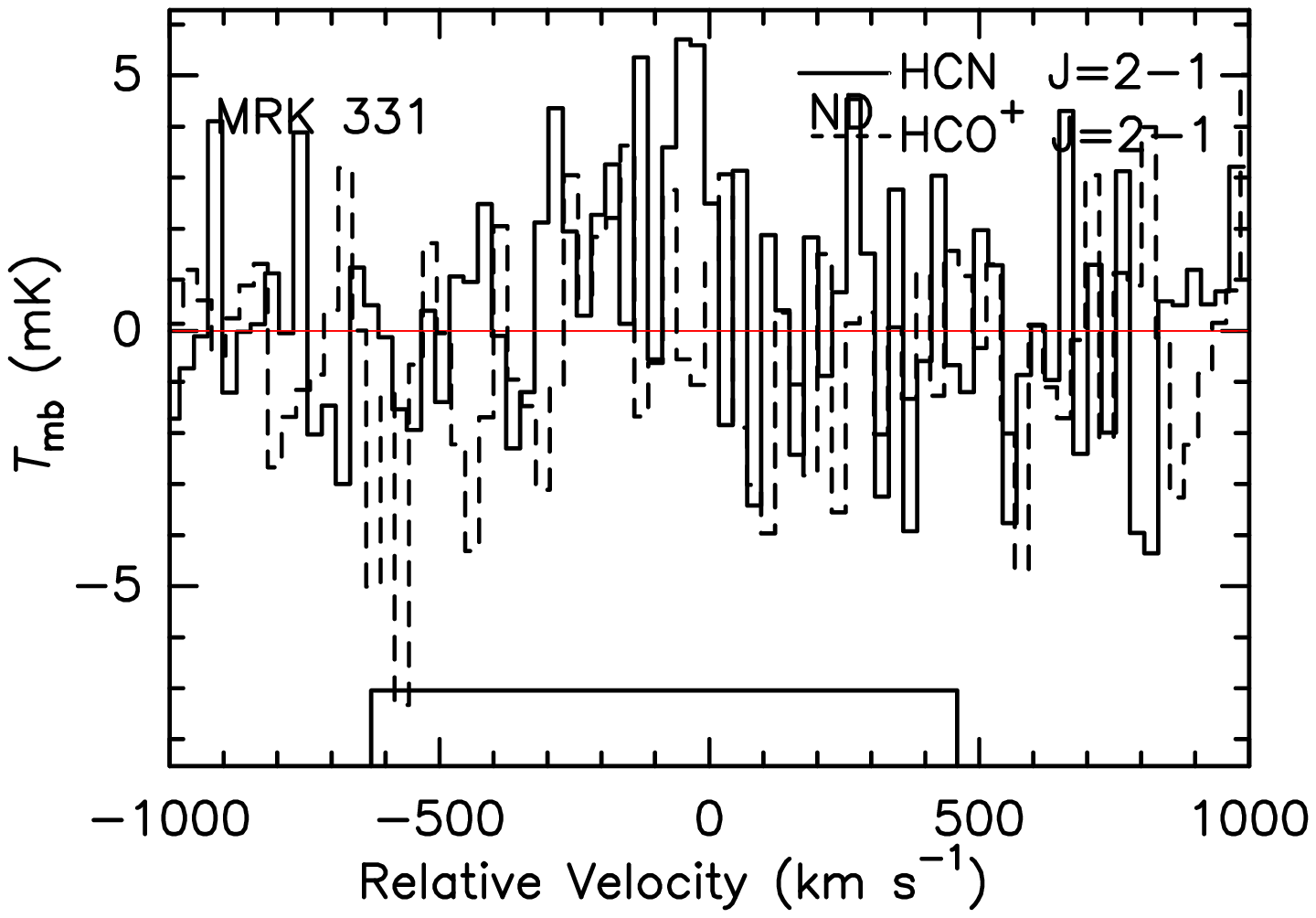}
\includegraphics[height=1.45in]{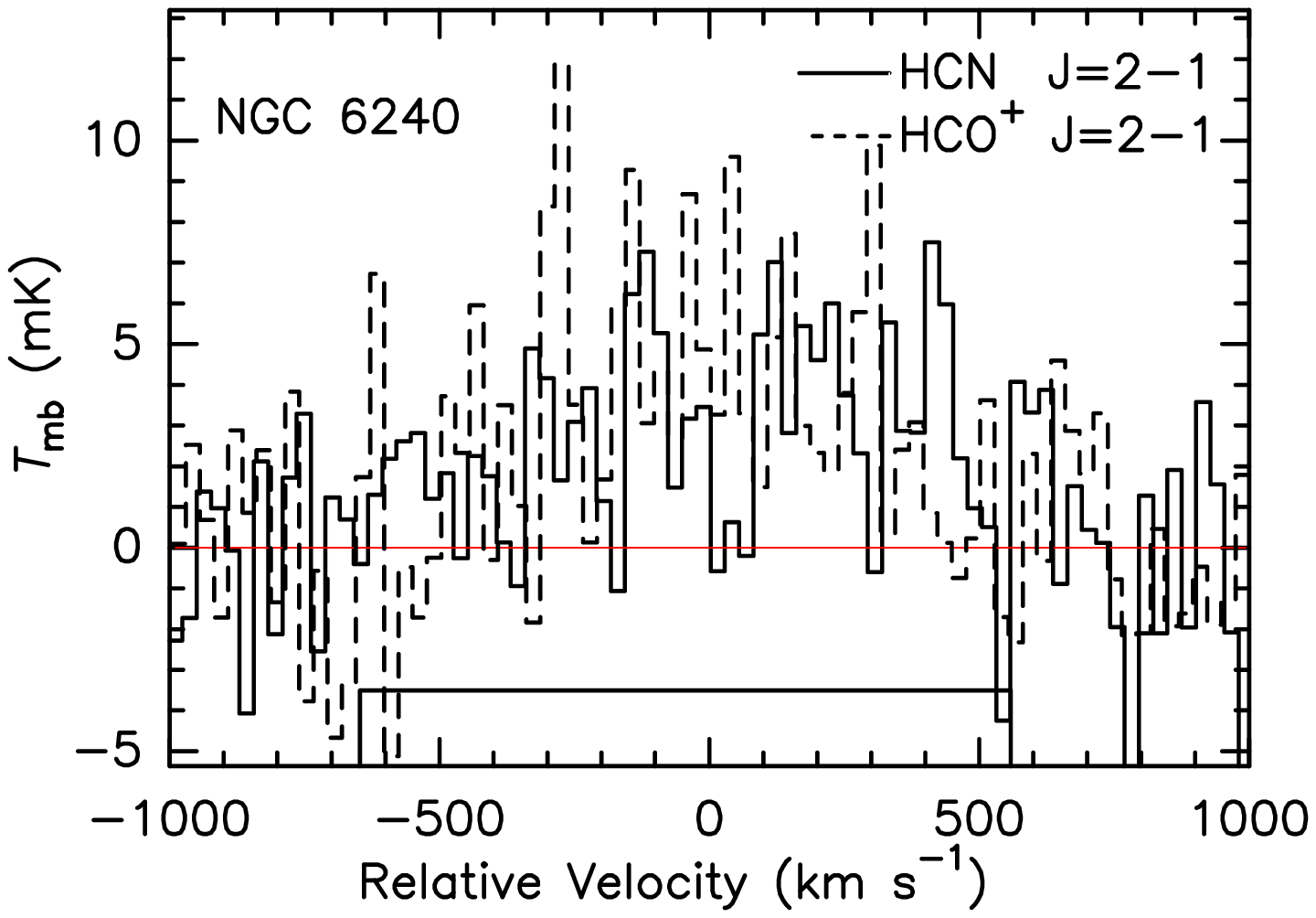}
\includegraphics[height=1.45in]{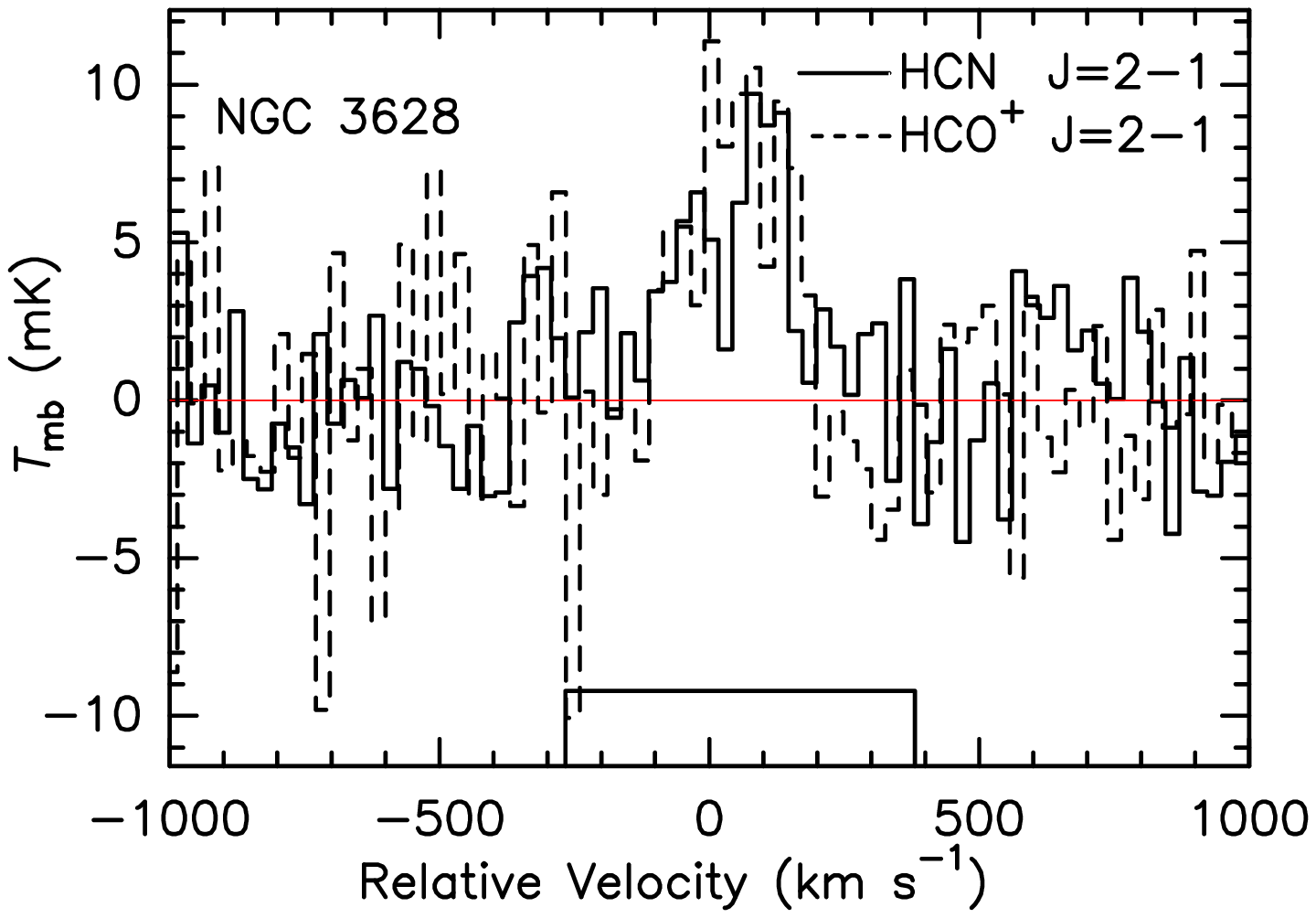}
\includegraphics[height=1.45in]{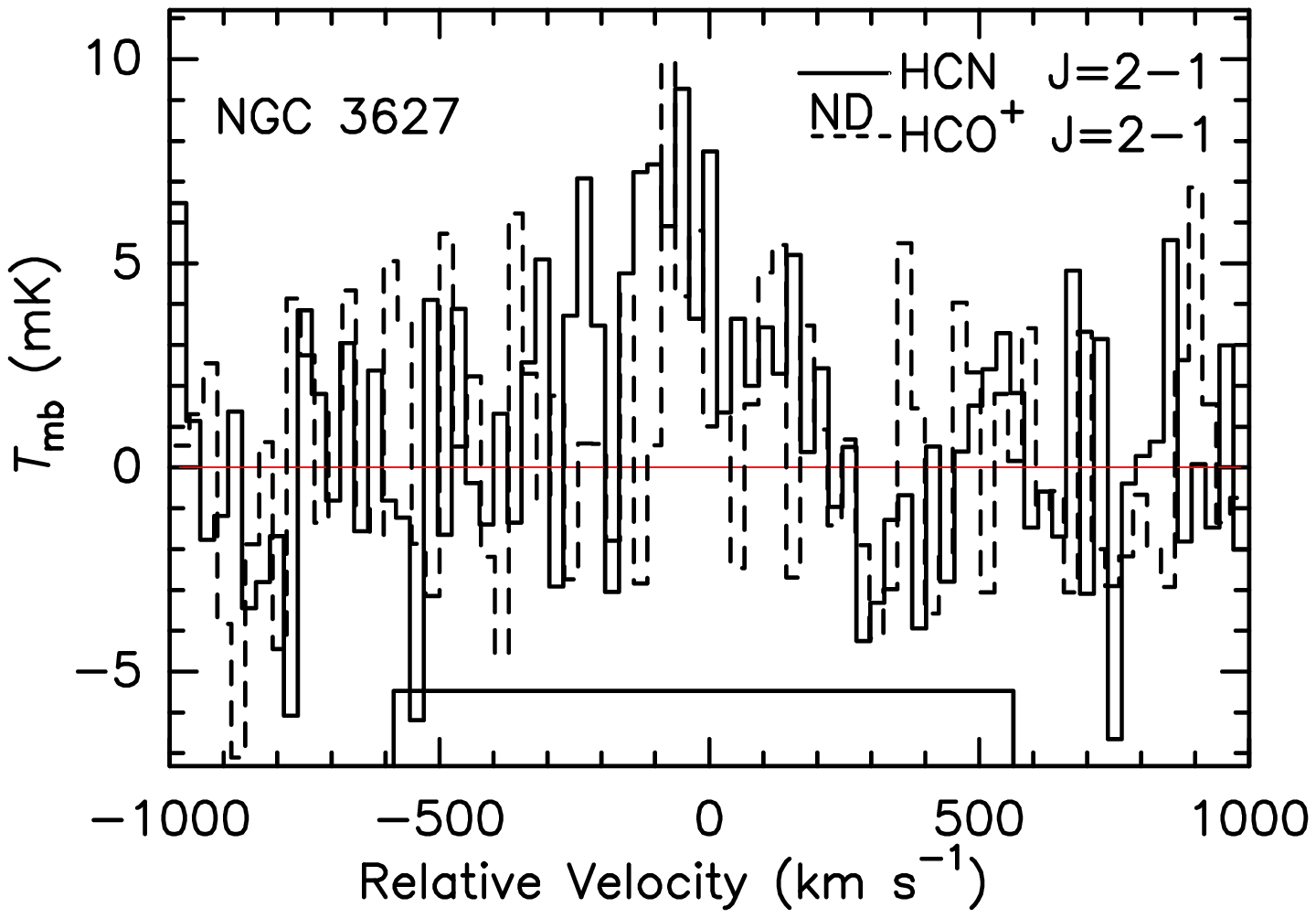}
\includegraphics[height=1.45in]{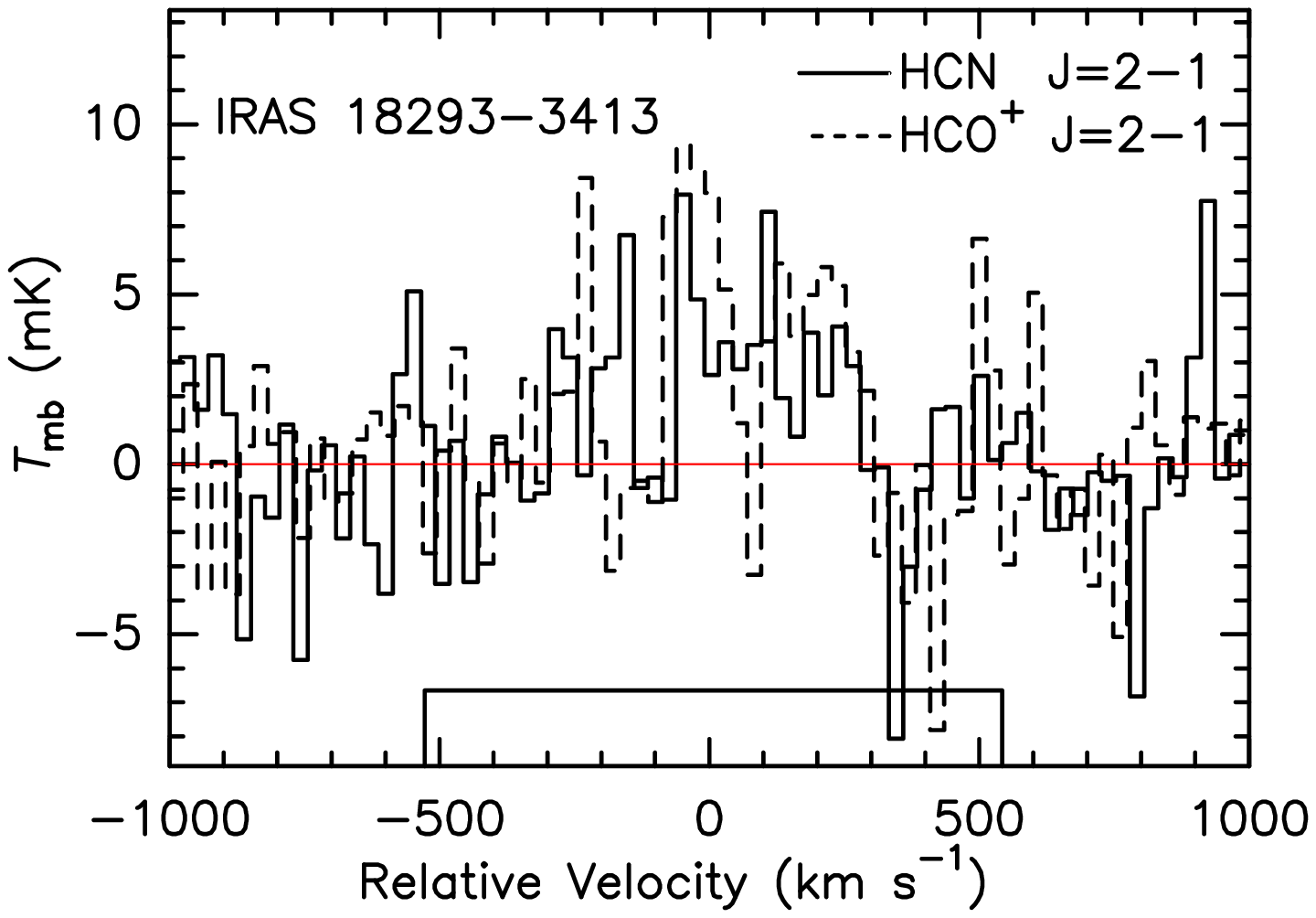}
\includegraphics[height=1.45in]{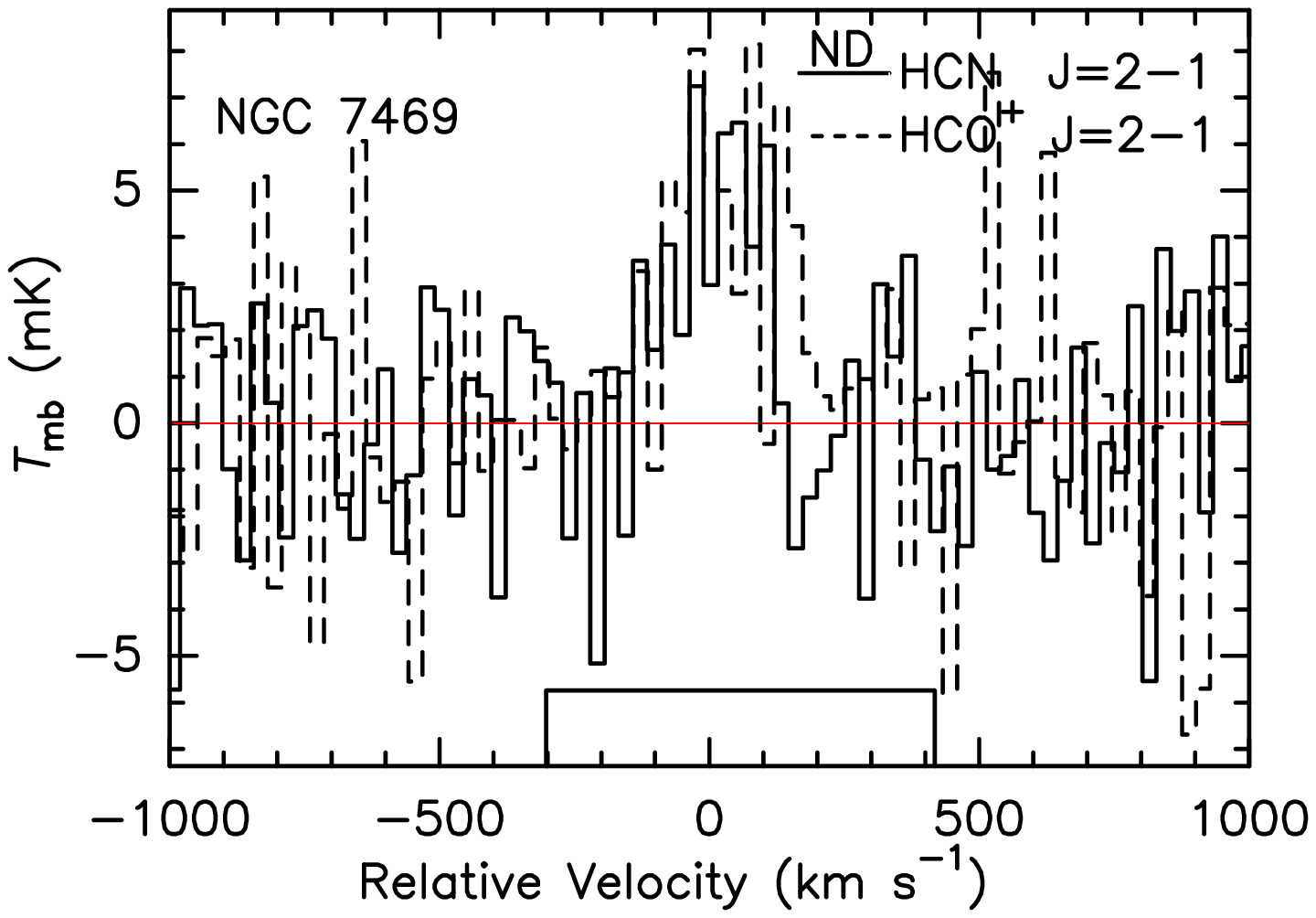}
\includegraphics[height=1.45in]{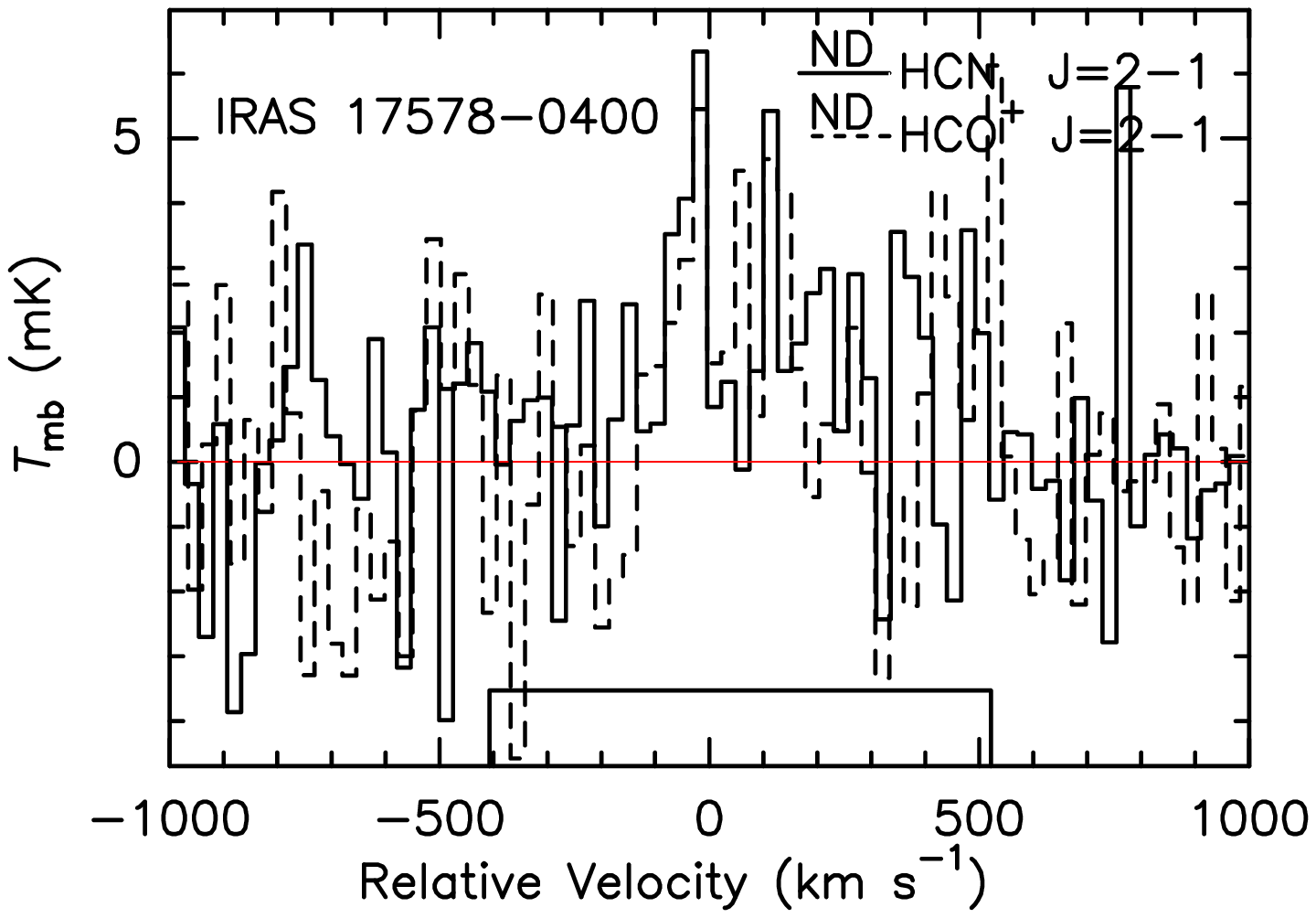}
\includegraphics[height=1.45in]{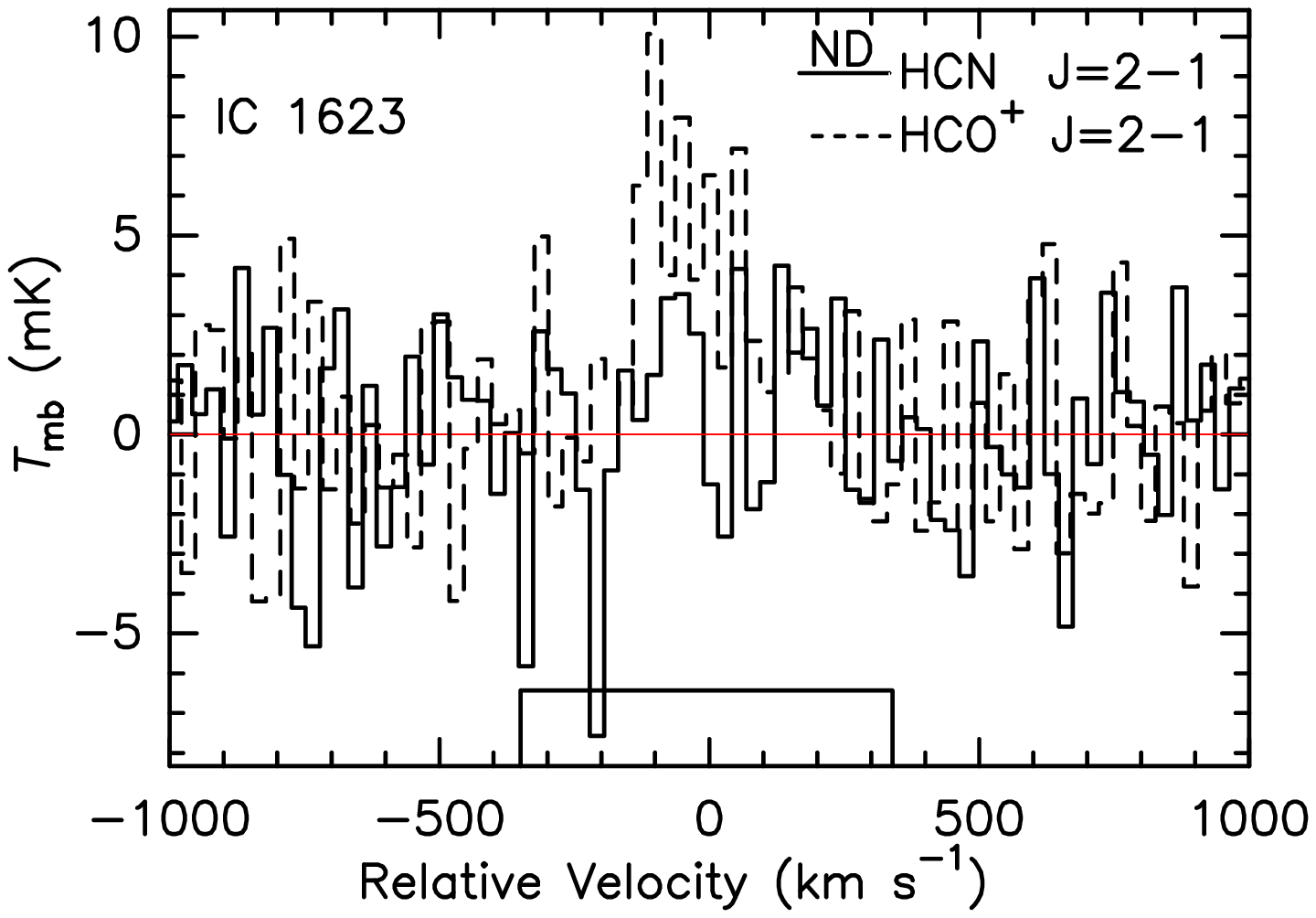}
\caption{\HCNto\ and \HCOto\ spectra of 17 galaxies observed with the APEX 12-m
        telescope. Solid and dashed lines show line profiles of \HCNto\ and
\HCOto, respectively. We detect both lines in 12 galaxies. Five sources only
have one detected line with a velocity-integrated line intensity $>$
3$\sigma$. The non-detection are tagged as ``ND'' above the line labels.} 
\label{spectrum}
\end{figure*}

The observation was conducted with the APEX 12-m telescope during 2016 July and
August (Project ID: E-097.B-0986A-2016). The weather was in good (precipitable
water vapour; PWV $<$ 0.9 mm) conditions for ten sources, and in normal
conditions (PWV $\sim$ 1--1.4 mm) for the rest. The wobbler switching mode was
adopted for all observations, with a switching frequency of 2\,Hz and a beam
throw of 120$''$ at each side of the target. The beam size is $\sim$34.7$''$
(from 34.2$''$ to 35.2$''$) on average, with a slight variation between
targets, depending on the specific line and redshift. 

We employed the Swedish-ESO PI Instrument for APEX (SEPIA) receivers
\citep{Belitsky18} to observe \HCNto\ and \HCOto\ simultaneously, which have rest
frequencies of 177.261 GHz and 178.375 GHz, respectively. The SEPIA-180
receiver offers two sidebands, double polarisations, and a 4-GHz bandwidth for
each sideband. The targeted two lines are configured in the lower sideband.
Focusing calibrations were conducted on Mars, Jupiter, IRAS~15194-5115, IK-Tau,
and R-Dor, every 3--4\,hrs. Pointing calibrations were conducted on Mars,
Jupiter, or nearby ($<10^\circ$) carbon stars every hour.



We used the {\sc class} package in {\sc gildas}
\footnote{\url{https://www.iram.fr/IRAMFR/GILDAS/}} to reduce the spectral line
data. We checked the quality of all spectra by eye, fitted the baseline with a
first-order polynomial profile, and averaged them together.  Each sideband has
104,851 channels with an initial velocity resolution of 0.0644 $\rm
km\,s^{-1}$. We smoothed each spectrum to a velocity resolution of 26\,$\rm
km\,s^{-1}$, at which the final r.m.s. noise ranges from 1.5\,mK to 2\,mK. The
main beam temperature, $T_{\rm mb}$, is converted from the antenna temperature,
$T^*_{\rm A}$, using $T_{\rm mb}=T^*_{\rm A}\eta_{\rm f}/\eta_{\rm mb}$, where
$\eta_{\rm f}=0.95$ is the forward efficiency, and $\eta_{\rm mb}$=0.73 is the
main beam efficiency. We adopt Kelvin to Jansky conversion factor 39$\,\rm
Jy/K$ to convert $T^*_{\rm A}$ to flux density from the beam-covered regions.
The observational results are shown in Table
\ref{table:obsresult}, including integrated flux, Gaussian fitting Full Width
Half Maximum (FWHM) velocity width, and Gaussian fitting Peak flux density.


\subsection{Infrared data}

We obtained multi-wavelength photometric data of the Photodetector Array Camera
and Spectrometer \citep[PACS; for 70\,\um, 100\,\um, 160\,\um;][]{PACS2010} and
the Spectral and Photometric Imaging REceiver \citep[SPIRE; for 250\,\um,
350\,\um, 500\,\um;][] {SPIRE2010}, on board the Herschel Space Observatory.
These data were processed to level 2.5 and downloaded from European Space
Agency (ESA)\footnote{\url{http://archives.esac.esa.int/hsa/whsa/}}. We also
downloaded available 24\,\um~data from the archival Spitzer Space Telescope
\citep[MIPS;][]{MIPS2004}, which were processed to level 2. The beam sizes, in
Half Power Beam Width (HPBW), are 13.9$''$, 6.4$''$, 5.7$''$, 7$''$, 11.2$''$,
18.2$''$, 24.9$''$, and 36.1$''$ for WISE 22\,\um, MIPS 24\,\um, PACS 70\,\um,
100\,\um, 160\,\um, SPIRE 250\,\um, 350\,\um, and 500\,\um, respectively.

Because some nearby galaxies can not be fully enclosed by the APEX beam, we further
measured the emission size of the  PACS 70\,\um\ images with the diameter which
encloses 90\% flux of the entire galaxy and listed them in Table
\ref{table:obsresult}. From the PACS 70 \um\ images, seven galaxies (NGC\,4945,
NGC\,1068, NGC\,7552, NGC\,1365, NGC\,1808, NGC\,3627, and NGC\,3628) have
sizes (in diameter) larger than the APEX beam or can not be fully covered by
the APEX beam, while the other 12 galaxies are fully enclosed by the APEX beam
of 34.7$''$ (FWHM). For these 12 point-like galaxies, we adopt the 25\,\um,
60\,\um, and 100\,\um\ fluxes from the RBGS survey \citep{sanders2003}, which
was obtained with angular resolutions of 0.7$'$, 1.7$'$, $\sim$ 3$'$,
respectively. For the seven extended galaxies, we adopt IRAS fluxes scaled from
other wavelengths, as described in Section \ref{section:photometry}.

\subsection{Radio continuum}
\label{radiocontinuum}

We assume that the 1.4\,GHz radio continuum emission is dominated by
synchrotron radiation from supernova remnants  \citep{White1985}  and free-free
radiation from {H\/{\sc ii}} regions \citep{essential2016}, both would trace
recent star-forming activities.  We use the 1.4\,GHz radio continuum to measure
the size of star-forming region, assuming that star formation contributes
majority of the radio flux. To verify this assumption and to check possible AGN
contamination, very high-resolution radio data is needed. Therefore, we
collected very long baseline interferometric (VLBI) data in the literature,
which are only available for a few targets.

For galaxies with multiple measurements of the radio size, we adopt the ones
observed with highest angular resolutions (Table \ref{table:basicinfo}).  Ten
galaxies have sizes measured from MeerKAT data, which has an angular resolution
of $\sim$7.5$''$ \citep{Condon2021}.  Sizes of NGC\,3627 and NGC\,3628 were
measured using NRAO VLA Sky Survey \citep[NVSS;][]{Condon87}, which has an
angular resolution of 45$''$.  The size of NGC\,4418 is estimated from MERLIN
observation at a resolution of $0''.35\times 0''.16$ \citep{Costagliola13}. We
adopt the 250-GHz ALMA continuum size for IRAS\,19254-7245, which was measured
by \cite{Imanishi16}.

For other galaxies, we measured their sizes using data downloaded from ALMA
data archive and from the literature.  We fit the maps with a 2-D Gaussian
profile, using task {\sc imfit} in {casa} \citep{CASAdoc}.  The 1.4\,GHz
continuum data of Arp\,220 was combined with MERLIN data and VLA
A-configuration data in the literature \citep{Varenius16}.  NGC\,6240 and
IRAS\,19254-7245 have strong AGN contribution to the 1.4 GHz continuum, we
estimate their sizes using ALMA dust continuum data.  We download ALMA 480-GHz
data of NGC~6240 (Project code: 2015.1.00717.S) and fit its sizes with a
two-component Gaussian model (NGC~6140A and NGC~6240B in Table
\ref{table:basicinfo}). 

The sizes and fluxes of the radio continuum are from the whole galaxies.  When
computing luminosity surface densities in Section \ref{sec:surfacedensity}, we
adopt the radio area for galaxies with radio sizes smaller than the APEX beam.
For galaxies with radio sizes larger than the APEX beam, we adopt the
APEX beam area to compute the surface densities.

\subsection{Ancillary CO data}

Most of the velocity-integrated CO \Joz\ fluxes come from
\cite{Baan2008}, observed with the Institut de Radioastronomie Millimetrique
(IRAM) 30-m telescope (HPBW $\sim$21$''$) and the Swedish-ESO Submillimeter
Telescope (SEST) 15-m telescope (HPBW $\sim$45$''$). CO fluxes of
IRAS\,13120-5453 and NGC\,4418 are from \cite{sliwa17} and
\cite{Papadopoulos12} observed with ALMA and IRAM 30 m, respectively. However,
there is no low-$J$ CO data available for IRAS\,13242-5713 and
IRAS\,17578-0400. 

\begin{deluxetable*}{cccccccc}
\tablenum{2}
\tablecaption{Observational Result}\label{table:obsresult}
\tablewidth{0pt}
\tablehead{
\colhead{Source name} & \multicolumn{3}{c}{\HCNto} & \multicolumn{3}{c}{\HCOto} & \colhead{Diameter$_{70}$ $^a$}\\
\colhead{} & \colhead{Integrated flux}      & \colhead{FWHM}              & \colhead{Peak} & \colhead{Integrated flux}      & \colhead{FWHM}              & \colhead{Peak} & \\
\colhead{} & \colhead{($\rm Jy~km~s^{-1}$)} & \colhead{($\rm km~s^{-1}$)} & \colhead{(mJy)}& \colhead{($\rm Jy~km~s^{-1}$)} & \colhead{($\rm km~s^{-1}$)} & \colhead{(mJy)} & \colhead{(arcsec)}
}
\startdata
NGC\,4945        & 885 $\pm$ 11     & 328.2 $\pm$  1.9 & 3120 $\pm$ 27 & 855 $\pm$ 11     & 328.1 $\pm$  2.7 & 2917 $\pm$ 35 & ~~ 85.2 $^b$    \\
NGC\,1068        & 243.5 $\pm$  6.4 & 257.9 $\pm$  6.3 & 1076 $\pm$ 35 & 180 $\pm$  6     & 253.4 $\pm$  8.7 & 807  $\pm$ 39 & ~~ 38.6    $^b$  \\
NGC\,7552        & 63.7 $\pm$  6.0  & 138.5 $\pm$  8.6 & 546  $\pm$ 47 & 85.8 $\pm$  5.9  & 166.7 $\pm$ 11.1 & 593  $\pm$ 55 & ~~ 21.9 $^c$     \\
NGC\,4418        & 54.4 $\pm$  5.9  & 193 $\pm$ 20     & 261  $\pm$ 35 & 38.7 $\pm$  5.9  & 107 $\pm$ 17     & 215  $\pm$ 43 &  11.3     \\
NGC\,1365        & 95.7 $\pm$  8.7  & 328 $\pm$ 29     & 343  $\pm$ 39 & 71.1 $\pm$  8.5  & 241 $\pm$ 24     & 351  $\pm$ 47 & ~~ 19.2 $^c$    \\
NGC\,3256        & 57.9 $\pm$  7.6  & 180 $\pm$ 18     & 398  $\pm$ 55 & 106.1 $\pm$  7.6 & 181 $\pm$ 11     & 683  $\pm$ 59 &  19.5     \\
NGC\,1808        & 55.7 $\pm$  7.5  & 236 $\pm$ 25     & 289  $\pm$ 43 & 58.0 $\pm$  7.4  & 156 $\pm$ 29     & 394  $\pm$ 90 & ~~ 22.9 $^c$    \\
IRAS\,13120-5453 & 48 $\pm$ 11      & 333 $\pm$ 52     & 187  $\pm$ 39 & 43 $\pm$ 11      & 362 $\pm$ 85     & 152  $\pm$ 47 &  11.5     \\
IRAS\,13242-5713 & 57 $\pm$ 11      & 223 $\pm$ 42     & 300  $\pm$ 78 & 108 $\pm$ 11     & 194 $\pm$ 18     & 636  $\pm$ 82 &  16.7     \\
MRK\,331         & 30.4 $\pm$  8.3  & 214 $\pm$ 83     & 101  $\pm$ 27 & $<$ 32           & ...              & ...           &  11.0     \\
NGC\,6240        & 61 $\pm$ 15      & 809 $\pm$ 142    & 101  $\pm$ 23 & 83 $\pm$ 15      & 640 $\pm$ 84     & 152  $\pm$ 27 &  11.4     \\
NGC\,3628        & 41 $\pm$ 11      & 235 $\pm$ 58     & 187  $\pm$ 59 & 55 $\pm$ 11      & 210 $\pm$ 30     & 289  $\pm$ 59 & ~~ 42.5    $^b$ \\
NGC\,3627        & 46 $\pm$ 12      & 332 $\pm$ 62     & 179  $\pm$ 43 & $<$ 54           & ...              & ...           & ~~ 63.8    $^b$ \\
IRAS\,18293-3413 & 35 $\pm$ 10      & 332 $\pm$ 105    & 117  $\pm$ 51 & 60 $\pm$ 10      & 418 $\pm$ 96     & 164  $\pm$ 47 &  12.3     \\
NGC\,7469        & $<$ 50.2         & ...              & ...           & 39.8 $\pm$  9.9  & 287 $\pm$ 61     & 164  $\pm$ 47 &  11.7     \\
IRAS\,17578-0400 & $<$ 38.2         & ...              & ...           & $<$ 33           & ...              & ...           &  11.7     \\
IC\,1623         & $<$ 32.1         & ...              & ...           & 30.6 $\pm$  8.8  & 208 $\pm$ 47     & 187  $\pm$ 55 &  14.0     \\
\enddata
\tablecomments{ 
$^a$ Diameter of PACS\,70\um\ image which encloses 90\% flux of the entire galaxy, which is convolved with PACS\,70\um\ 5.7$''$ beam. \\
$^b$ NGC\,4945, NGC\,1068, NGC\,3628, and NGC\,3627 have larger diameters than APEX beam.\\
$^c$ NGC\,7552, NGC\,1365, and NGC\,1808 have long structure disks which cannot be enclosed by APEX beam.\\
}
\end{deluxetable*}

\section{Method and Analysis}\label{sec:methods}

\subsection{HCN and HCO$^+$ \Jto\ line luminosity}

We calculate the velocity-integrated main beam temperature and the associated
thermal noise following \cite{trgreve2009}:

\begin{equation}
\centering
\sigma\left(\int _{\Delta_ {\rm v}}T_{\rm mb}dv\right)=\sqrt{N_{\Delta _{\rm v}}}\left(1+\frac {N_{\Delta _{\rm v}}}{N_{\rm bas}}\right)^{1/2} \sigma (T_{\rm mb,ch})\Delta v_{\rm ch} + 10\,\% \int _{\Delta_ {\rm v}}T_{\rm mb}dv,
\end{equation} 

where $v_{\rm ch}~$= 26 \kms\, is the final velocity resolution,
$N_{\Delta_{\rm v}}=(\Delta v/\Delta v_{\rm ch})$ is the number of channels
that cover the line, $N_{\rm bas}$ is the number of the line-free channels, and
$\sigma (T_{\rm mb,ch})$ is the channel-to-channel r.m.s. noise.  We adopt
10\,\% as the absolute flux calibration error during observation.

Then we computed the line luminosities of the region observed in the whole
galaxy using the equation from \cite{Gao2004a}:

\begin{equation}
L^\prime=\pi/(4{\rm ln}2)\theta_{\rm mb}^2\int_{\Delta _{\rm v}}T_{\rm mb}dv~D_{\rm L}^2(1+z)^{-3},
\end{equation}

where luminosity distance, $D_{\rm L}$, and redshift, $z$, are taken from NED
(Table \ref{table:basicinfo}). We further propagate uncertainties from distance
and flux to the final error of line luminosities, using the following formula:

\begin{equation}
\centering
\sigma_{\rm flux}(L^\prime)=\pi/(4{\rm ln}2)\theta_{\rm mb}^2(1+z)^{-3}D_{\rm L}^2~\sigma\left(\int _{\Delta_ {\rm v}}T_{\rm mb}dv\right),
\end{equation}
\begin{equation}
\centering
\sigma_{\rm dist}(L^\prime)=\pi/(4{\rm ln}2)\theta_{\rm mb}^2(1+z)^{-3}\times 2D_{\rm L}\int _{\Delta_ {\rm v}}T_{\rm mb}dv~\sigma\left(D_{\rm L}\right),
\end{equation}
\begin{equation}
\centering
\sigma(L^\prime)=\sqrt{\sigma^2_{\rm flux}(L^\prime)+\sigma^2_{\rm dist}(L^\prime)},
\end{equation}

where $\sigma_{\rm flux}(L^\prime)$ is the luminosity error propagated from
flux error, $\sigma_{\rm dist}(L^\prime)$ is the luminosity error propagated
from distance estimation, and $\sigma (D_{\rm L})$ is the distance error.
Because all luminosities adopt the same distance, their ratios do not include
errors of distances.

\begin{deluxetable*}{cccccccccc}
\tablenum{3}
\tablecaption{Infrared Fluxes inside the APEX beam} \label{infrared}
\tablewidth{0pt}
\setlength\tabcolsep{3pt}
\tablehead{
\colhead{Source name} & \colhead{MIPS 24}        & \colhead{IRAS 25}        & \colhead{IRAS 60}         & \colhead{PACS 70}         & \colhead{PACS 100}         & \colhead{PACS 160}        & \colhead{SPIRE 250}       & \colhead{SPIRE 350}       & \colhead{SPIRE 500} \\
\colhead{} & \colhead{(Jy beam$^{-1}$)} & \colhead{(Jy)}             & \colhead{(Jy)}              & \colhead{(Jy beam$^{-1}$)}  & \colhead{(Jy beam$^{-1}$)}   & \colhead{(Jy beam$^{-1}$)}  & \colhead{(Jy beam$^{-1}$)}  & \colhead{(Jy beam$^{-1}$)}  & \colhead{(Jy beam$^{-1}$)}
}
\startdata
NGC\,4945        & 9.7 $\pm$  1.0  & 42.3$\pm$  4.3  & 625.5 $\pm$ 62.7 & 734.5 $\pm$ 73.5 & 1060 $\pm$ 106     & 883.8 $\pm$ 88.4 & 333  $\pm$ 33    & 124  $\pm$ 12    & 33.1$\pm$ 3.3 \\
NGC\,1068        & 31.4 $\pm$  3.2 & 87.6$\pm$  8.9  & 196.4 $\pm$ 19.7 & 180.7 $\pm$ 18.1 & ...                & 146.4 $\pm$ 14.7 & 51.1 $\pm$ 5.1   & 18.3 $\pm$ 1.8   & 5.0 $\pm$ 0.51 \\
NGC\,7552        & ...             & 11.9$\pm$  1.2  & 77.4 $\pm$ 7.8   & 82.5 $\pm$  8.3  & 95.6 $\pm$  9.6    & 67.1 $\pm$ 6.9   & 22.5 $\pm$ 2.3   & 7.6 $\pm$ 0.8    & 2.0 $\pm$ 0.21 \\
NGC\,4418        & 5.9 $\pm$  0.7  & 9.7 $\pm$  1.1  & 43.9 $\pm$ 4.5   & 41.5 $\pm$  4.2  & 33.0 $\pm$  3.3    & 18.1 $\pm$ 1.8   & 6.2 $\pm$ 0.6    & 2.3 $\pm$ 0.2    & 0.7 $\pm$ 0.08 \\
NGC\,1365        & 8.2 $\pm$  0.8  & 14.3$\pm$  1.5  & 94.3 $\pm$ 9.5   & 95.7 $\pm$  9.6  & 132.4 $\pm$ 13.3   & 111.4 $\pm$ 11.2 & 46.5 $\pm$ 4.7   & 17.8 $\pm$ 1.8   & 5.1 $\pm$ 0.52 \\
NGC\,3256        & 11.1 $\pm$  1.1 & 15.7$\pm$  1.6  & 102.6 $\pm$ 10.3 & 111.3 $\pm$ 11.1 & 120.7 $\pm$ 12.1   & 82.5 $\pm$ 8.3   & 26.9 $\pm$ 2.7   & 8.7 $\pm$ 0.9    & 2.5 $\pm$ 0.26 \\
NGC\,1808        & 9.7 $\pm$  1.0  & 17.0$\pm$  1.7  & 105.5 $\pm$ 10.6 & 113.1 $\pm$ 11.3 & 130.5 $\pm$ 13.1   & 94.7 $\pm$ 9.5   & 33.6 $\pm$ 3.4   & 11.6 $\pm$ 1.2   & 3.1 $\pm$ 0.32 \\
IRAS\,13120-5453 & 2.4 $\pm$  0.2  & 3.0 $\pm$  0.3  & 41.1 $\pm$ 4.2   & 47.7 $\pm$  4.8  & 52.3 $\pm$  5.2    & 33.7 $\pm$ 3.4   & 11.7 $\pm$ 1.2   & 4.1 $\pm$ 0.4    & 1.0 $\pm$ 0.12 \\
IRAS\,13242-5713 & 5.5 $\pm$  0.6  & 7.6 $\pm$  0.8  & 81.4 $\pm$ 8.2   & 91.5 $\pm$  9.2  & 98.1 $\pm$  9.8    & 66.0 $\pm$ 6.7   & 23.0 $\pm$ 2.4   & 7.9 $\pm$ 0.9    & 2.2 $\pm$ 0.25 \\
MRK\,331         & 1.9 $\pm$  0.2  & 2.5 $\pm$  0.3  & 18.0 $\pm$ 1.8   & 20.0 $\pm$  2.0  & 22.5 $\pm$  2.3    & 16.3 $\pm$ 1.7   & 5.8 $\pm$ 0.6    & 2.1 $\pm$ 0.2    & 0.5 $\pm$ 0.06 \\
NGC\,6240        & 2.9 $\pm$  0.3  & 3.5 $\pm$  0.4  & 22.9 $\pm$ 2.3   & 25.0 $\pm$  2.5  & 26.2 $\pm$  2.6    & 17.2 $\pm$ 1.9   & 5.8 $\pm$ 0.6    & 2.0 $\pm$ 0.2    & 0.6 $\pm$ 0.07 \\
NGC\,3628        & 2.1 $\pm$  0.4  & 4.8 $\pm$  0.5  & 54.8 $\pm$ 5.6   & 50.6 $\pm$  5.1  & 52.0 $\pm$  5.2    & 67.8 $\pm$ 6.8   & 29.1 $\pm$ 2.9   & 10.8 $\pm$ 1.1   & 3.1 $\pm$ 0.32 \\
NGC\,3627        & ...             & 8.6 $\pm$  0.9  & 66.3 $\pm$ 6.7   & 15.2 $\pm$  1.5  & 22.8 $\pm$  2.3    & 19.0 $\pm$ 1.9   & 7.0 $\pm$ 0.7    & 2.5 $\pm$ 0.3    & 0.8 $\pm$ 0.08 \\
IRAS\,18293-3413 & 3.1 $\pm$  0.3  & 4.0 $\pm$  0.4  & 35.7 $\pm$ 3.6   & 42.8 $\pm$  4.3  & 55.1 $\pm$  5.5    & 42.6 $\pm$ 4.3   & 15.0 $\pm$ 1.5   & 5.6 $\pm$ 0.6    & 1.5 $\pm$ 0.16 \\
NGC\,7469        & 4.6 $\pm$  0.5  & 6.0 $\pm$  0.6  & 27.3 $\pm$ 2.8   & 28.9 $\pm$  3.1  & 32.9 $\pm$  3.7    & 23.3 $\pm$ 2.8   & 8.2 $\pm$ 1.0    & 2.9 $\pm$ 0.4    & 0.7 $\pm$ 0.10 \\
IRAS\,17578-0400 & 0.7 $\pm$  0.1  & 1.1 $\pm$  0.2  & 27.7 $\pm$ 3.0   & 31.4 $\pm$  3.3  & 34.5 $\pm$  3.7    & 23.3 $\pm$ 2.6   & 8.5 $\pm$ 1.0    & 3.0 $\pm$ 0.4    & 0.8 $\pm$ 0.10 \\
IC\,1623         & 2.7 $\pm$  0.3  & 3.6 $\pm$  0.4  & 22.9 $\pm$ 2.4   & 24.6 $\pm$  2.5  & 27.6 $\pm$  2.8    & 20.3 $\pm$ 2.4   & 7.3 $\pm$ 0.7    & 2.7 $\pm$ 0.3    & 0.7 $\pm$ 0.08 \\
Arp\,220         & 5.3 $\pm$  0.5  & 8.0 $\pm$  0.8  & 104.1 $\pm$ 10.5 & 132.8 $\pm$ 13.4 & ...                & 83.3 $\pm$ 8.4   & 30.6 $\pm$ 3.1   & 10.7 $\pm$ 1.1   & 3.7 $\pm$ 0.39 \\
IRAS\,19254-7245 & 1.1 $\pm$  0.1  & 1.4 $\pm$  0.2  & 5.2 $\pm$ 0.6    & 5.5 $\pm$  0.6   & 5.8 $\pm$  0.6     & 4.1 $\pm$ 0.4    & 1.5 $\pm$ 0.2    & 0.5 $\pm$ 0.1    & 0.1 $\pm$ 0.02 \\
\enddata
\tablecomments{Flux densities of MIPS, PACS, and SPIRE maps are extracted at
        the same positions as the APEX pointings. We first convolve their
        native angular resolutions to the same beamsize of our APEX
        observations and scale the map unit to Jy beam$^{-1}$, by a factor of $\rm
        \pi/4ln\,2 \times (HPBW_{beam}/pixelsize)^2$. The IRAS data are taken
        from \cite{sanders2003} as flux densities of the entire galaxies.}
\end{deluxetable*}

\subsection{Photometry, Infrared Luminosity, and Dust Temperature}
\label{section:photometry}

Because our APEX observations have different beam sizes compared to that of the
IR photometric data, it is not possible to directly compare dense gas tracers
with IR data. Therefore, we convolve the Spitzer and Herschel maps to match
the beam size of the APEX 12-m data, a 34.7$''$ (diameter) Gaussian beam, using
the convolution kernels provided by \cite{Aniano2011}. Compared to aperture
photometry, this method is more robust and is much less affected by the spatial
distribution of the target \citep[e.g.,][]{Tan2018}. Details of the method were
described in \cite{Tan2018}.

To measure the photometric flux at each wavelength, and to estimate their
associated noise levels, our photometry procedures are listed as follows:
First, we subtract the background of an annulus region from 1.5 to 2 $\times$
the maximum source size, which is estimated using the curve of growth (see
Appendix \ref{app:bkg} for details). Second, we scale the image units to
Jy\,beam$^{-1}$ by a factor of $\rm \pi/4ln\,2 \times
(HPBW_{beam}/pixelsize)^2$. Third, we measure the value of the central pixel to
obtain the flux density at each IR band. Last, the error is estimated from the
background regions. The photometric results are listed in Table \ref{infrared}.
The flux error consists of photometric error, flux calibration error, and
systematic error. We adopt an absolute flux calibration error of 7\,\% and a
systematic error of 3\,\%, following \cite{Balog14}. 

Since {\it IRAS} and {\it Herschel} SPIRE 500\,\um~data have lower resolutions
than our APEX data, we scale the IRAS\,25\,\um, IRAS\,60\,\um, and
SPIRE\,500\,\um\ data to obtain the beam-matched fluxes, using aperture
correction factors obtained from MIPS\,24\,\um, PACS\,70\,\um, and
SPIRE\,350\,\um\ maps, respectively. For example, we measure the flux of entire galaxy ($S^{\rm total}_{70}$) and flux in the 34.7$''$ gaussian beam of the PACS\,70\,\um\ ($S^{\rm beam}_{70}$). Then we adopt $S^{\rm total}_{60}*S^{\rm beam}_{70}/S^{\rm total}_{70}$ as the IRAS\,60\,\um\ flux in the beam.

We then build dust spectral energy distribution (SED) for each galaxy and fit
total IR luminosity and dust temperature. We fit the data with a two-component
modified blackbody (MBB) dust model, following \cite{Galametz2013}. Since the
warm dust component contributes a non-negligible fraction to the 70\,\um~and
100\,\um\ emission, a single MBB model would not match the Wien side of the SED
and would overestimate the cold dust component. The two-component MBB model is
described as follows:

\begin{equation}
S_\nu = A_{\rm w}~\lambda^{-2}B_\nu(T_{\rm w}) + A_{\rm c}~\lambda^{-\beta_{\rm c}}B_\nu(T_{\rm c}),
\end{equation}

where $S_\nu$ is the flux density obtained from photometry; $B_\nu$ is the
Planck function; $\beta_{\rm c}$ is the emissivity index of the cold component;
$T_{\rm w}$ and $T_{\rm c}$ are the temperature of warm and cold components,
respectively; $A_{\rm c}$ and $A_{\rm w}$ describe the peaks of the two
components, respectively. To limit the number of free parameters, we fix
$\beta_{\rm w}= 2.0$, as a good approximation of dust model of
\cite{LiDraine2001}. So, the estimate of $T_{\rm w}$ has an additional
systematic error compared to $T_{\rm c}$.

We fit the SEDs and estimate the uncertainties using the Markov Chain Monte
Carlo ({\sf MCMC}) package, {\sc emcee} \citep{emcee2013}. Details of the dust
SED fitting are shown in Appendix \ref{app:SED_fitting}. Parameters obtained
from the SED fitting are listed in Table \ref{table:fittingresult}.  Then we
compute the total IR luminosity by integrating from 3\,\um~to 1000\,\um, as
$L_{\rm IR}$, which is adopted as the SFR tracer. The beam-matched $L_{\rm IR}$
accounts for $\sim 10\%-100\%$ of $L^{\rm whole}_{\rm IR}$ from the whole
galaxy provided by \cite{sanders2003}. In Appendix \ref{app:compare_lum}, we
present $L^\prime_{\rm HCN}$-$L_{\rm IR}$ correlations between Far-IR
luminosity (from 100\,\um~to 1000\,\um), near-to-mid-IR luminosity (from
3\,\um~to 100\,\um) , and IR luminosity of the warm dust component.

$L_{\rm IR}$ within the 34.7$''$ beam area ranges from 2.7$\,\times 10^9 \, \rm
L_{\odot}$ to 1.7$\,\times 10^{12} \, \rm L_{\odot}$. Cold dust temperature
$T_{\rm c}$ ranges from 16.9 K to 32.3 K, with an average value of 24 K.
Compared with the result of the overlapped galaxies in \cite{Uvivian12}, our
dust temperature is lower, possibly because of the inclusion of the
Herschel/SPIRE data, which would better constrain the cold dust component. Our
fitting result is consistent with \cite{Galametz2013}.

\subsection{Molecular gas mass and SFR estimation}

In this work, we derive dense gas mass from $L'_{\rm HCN {\it
J}=2\rightarrow1}$ and $L'_{\rm HCO^+ {\it J}=2\rightarrow1}$ following
\cite{Gao2004a}: 

\begin{equation}
M_{\rm dense}({\rm H_2}) \approx 2.1 \frac{\langle \, n({\rm H_2}) \rangle ^{1/2}}{T_{\rm b}}L^\prime_{\rm HCN\,1-0}\sim 10~L^\prime_{\rm HCN\,1-0}~\rm M_{\odot}\,(K\,km\,s^{-1}\,pc^2)^{-1},
\end{equation}

where $\langle \, n({\rm H_2}) \rangle$ is the average density of the dense gas
and $T_{\rm b}$ is the brightness temperature of the dense gas tracer. The
estimation of dense gas mass relies on both brightness temperature and density,
where we adopt $T_{\rm b}\sim 35~\rm K$ and $\langle \, n({\rm H_2}) \rangle =
3 \times 10^4~\rm cm^{-3}$ for typical conditions of Galactic virialized cloud
cores \citep{Radford1991a}.  There are few \HCNto\ and \HCOto\ observations in
the literature \citep[e.g.,][]{Immer2016}.  Therefore, we can derive an average
HCN\,\Jto/HCN\,\Joz\  $T_{\rm b}$ from
HCN\,$J=3\rightarrow2$/HCN\,$J=1\rightarrow0$ ratios with RADEX, by assuming
the above physical conditions.  We adopt the average
HCN\,$J=3\rightarrow2$/HCN\,$J=1\rightarrow0$ $T_{\rm b}$ ratio of $\sim$
0.26$\pm$0.10 found in Galactic dense cores \citep{Wu2010}, and the derived
HCN\,\Jto/HCN\,\Joz\  $T_{\rm b}$ ratio is $\sim 0.67$, which is further
adopted to convert from luminosity to the dense gas mass.  We derive the dense
gas mass traced by \HCOp\ with the same method, by assuming that the two lines
have similar excitations. The derived dense gas mass ranges from
$4\times10^7~\rm M_\odot$ to $10^{10}~\rm M_\odot$.

Following \cite{Tan2018}, we adopt the SFR conversion calibrated by
\cite{Kennicutt98a} and \cite{Murphy2011}:

\begin{equation}
\left(\frac{\rm SFR}{\rm M_{\odot}yr^{-1}}\right)=1.50\times 10^{-10}\left(\frac{L_{\rm IR}}{\rm L_{\odot}}\right)
\end{equation}

The SFR is calculated based on \cite{Kroupa2001} IMF. The derived gas mass and SFR are shown in Table \ref{table:fittingresult}. 

\begin{figure*}[ht]
\includegraphics[height=3.5in,width=7in]{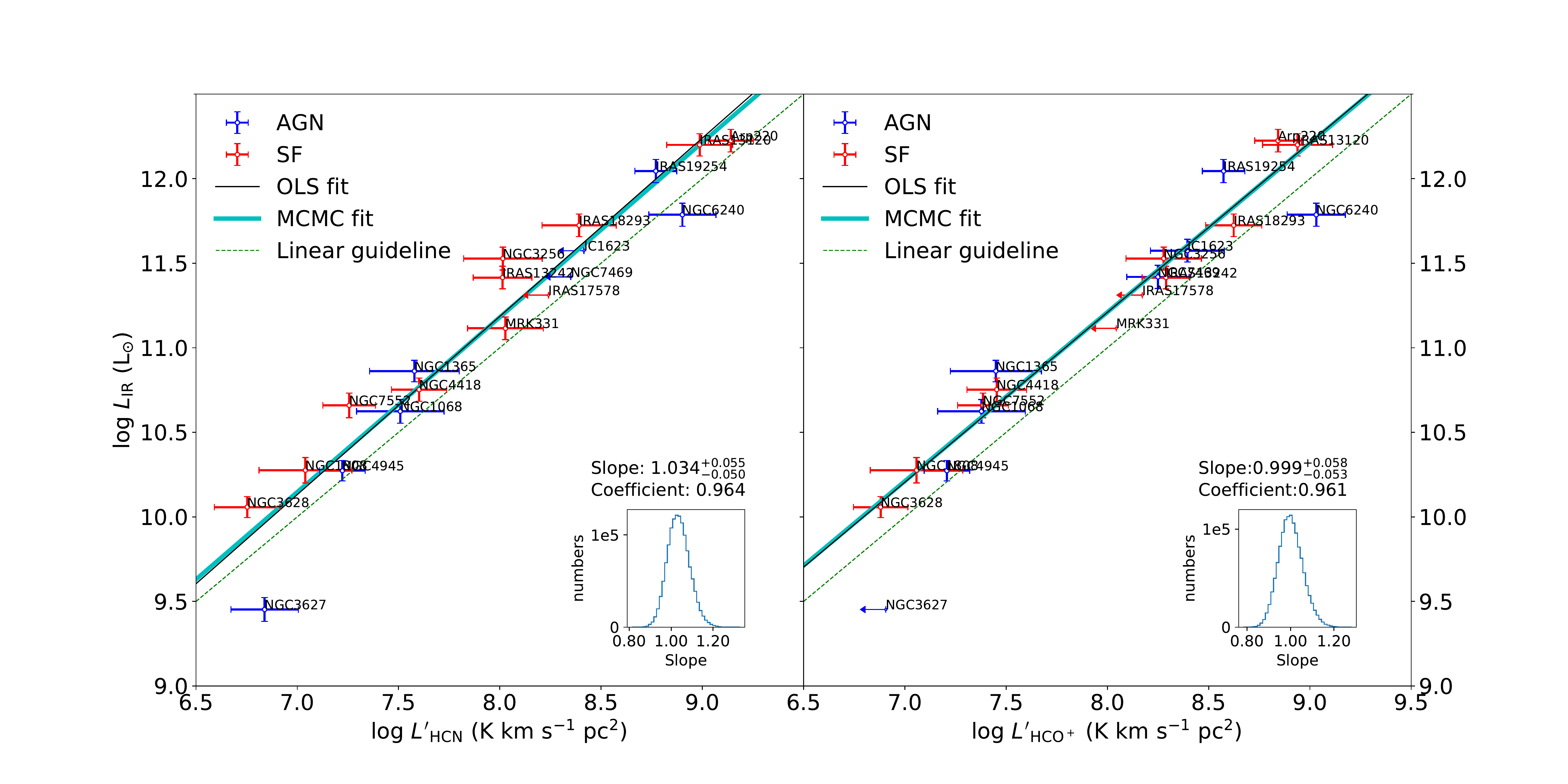}
\caption{{\it Left:} Correlation between $L^\prime_{\rm HCN\,{\it
J}=2\rightarrow1}$ and $L_{\rm IR}$. {\it Right:} Correlation between
$L^\prime_{\rm HCO^+\,{\it J}=2\rightarrow1}$ and $L_{\rm IR}$. AGN-dominated
and star-formation dominated galaxies are shown in blue and red points,
respectively.  The fitting results of Orthogonal Least Squares ({\sf OLS}) and
Markov chain Monte Carlo ({\sf MCMC})  are shown in black and cyan lines,
respectively.  The green-dashed line shows a linear relation for reference.
The insets present the probability density distributions of the slope
sampling.}
\label{SFLrelation}
\end{figure*}

\begin{figure*}[ht]
\includegraphics[height=3.5in,width=7in]{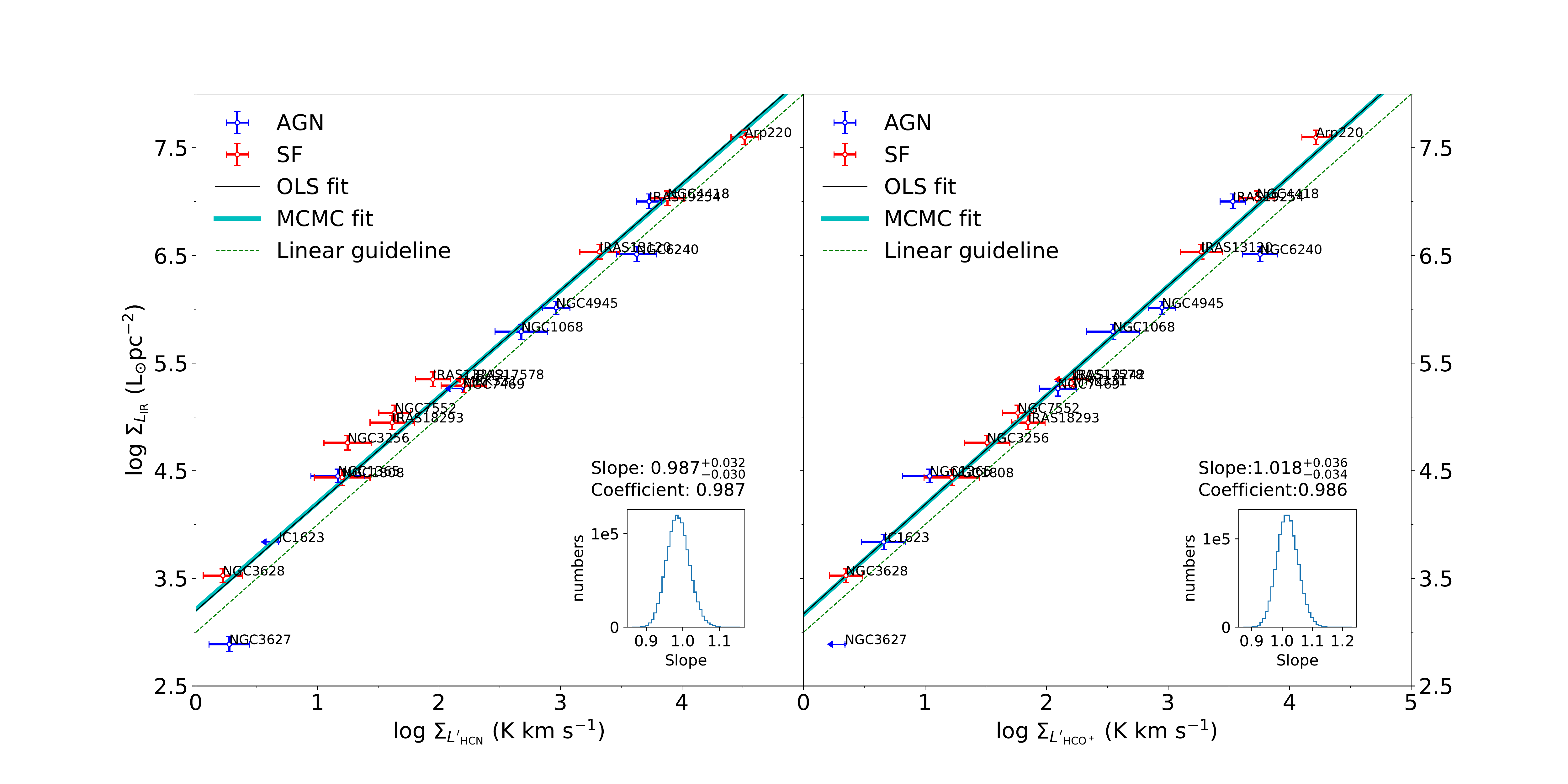}
\caption{{\it Left:} Correlation between $\Sigma_{L^\prime_{\rm HCN\,{\it
J}=2\rightarrow1}}$ and $\Sigma_{L_{\rm IR}}$. {\it Right:} Correlation between
$\Sigma_{L^\prime_{\rm HCO^+\,{\it J}=2\rightarrow1}}$ and $\Sigma_{L_{\rm
IR}}$. AGN-dominated and star-formation dominated galaxies are shown in blue
and red points, respectively.  The fitting results of Orthogonal Least Squares
({\sf OLS}) and Markov chain Monte Carlo ({\sf MCMC}) are shown in black and
cyan lines, respectively.  The green-dashed line shows a linear relation for
reference.  The insets present the probability density distributions of the
fitted slopes. } \label{densityrelation}
\end{figure*}

\begin{deluxetable*}{cccccccccc}
\tablenum{4}
\tablecaption{Derived molecular Properties and Fitting Results} \label{table:fittingresult}
\tablewidth{0pt}
\tablehead{
\colhead{Source name} & \colhead{$L'_{\rm HCN\,2-1}$} & \colhead{$L'_{\rm HCO^+\,2-1}$} &\colhead{$L_{\rm IR}$}&\colhead{$T_{\rm c}$} & \colhead{$\beta_{c}$} & \colhead{$T_{\rm w}$} & \colhead{$M^{\rm HCN}_{\rm dense}$} & \colhead{$M^{\rm HCO^+}_{\rm dense}$} &\colhead{$\rm SFR_{IR}$} \\
\colhead{}& \multicolumn{2}{c}{($\rm 10^7\,K\,km\,s^{-1}\,pc^{2}$)}& \colhead{($10^9\,\rm L_{\rm \odot}$)} & \colhead{(K)}&\colhead{}&\colhead{(K)}&\colhead{($\rm 10^8~M_\odot$)}&\colhead{($\rm 10^8~M_\odot$)}&\colhead{($\rm M_\odot\,yr^{-1}$)}\\
\colhead{(1)}&\colhead{(2)}&\colhead{(3)}&\colhead{(4)}&\colhead{(5)}&\colhead{(6)}&\colhead{(7)}&\colhead{(8)}&\colhead{(9)}&\colhead{(10)}
}
\startdata
NGC\,4945        & 1.7        $\pm$ 0.4 & 1.6          $\pm$ 0.4 & 19.0    $\pm$ 0.8 & 23 $\pm$ 1 & 2.4    $\pm$ 0.2 & 49   $\pm$ 2  & 1.8     $\pm$ 0.4  & 1.8    $\pm$ 0.4  & 3.5     $\pm$ 0.4  \\
NGC\,1068        & 3.2        $\pm$ 1.6 & 2.4          $\pm$ 1.2 & 42.3    $\pm$ 3.1 & 28 $\pm$ 1 & 2.1    $\pm$ 0.1 & 74   $\pm$ 10 & 3.4     $\pm$ 1.8  & 2.5    $\pm$ 1.3  & 7.3     $\pm$ 1.1  \\
NGC\,7552        & 1.8        $\pm$ 0.5 & 2.4          $\pm$ 0.7 & 40.0    $\pm$ 2.3 & 23 $\pm$ 2 & 2.5    $\pm$ 0.2 & 62   $\pm$ 2  & 1.9     $\pm$ 0.6  & 2.7    $\pm$ 0.7  & 7.9     $\pm$ 1.2  \\
NGC\,4418        & 4.0        $\pm$ 1.3 & 2.8          $\pm$ 1.0 & 55.2    $\pm$ 2.7 & 30 $\pm$ 4 & 1.8    $\pm$ 0.2 & 62   $\pm$ 2  & 4.3     $\pm$ 1.3  & 3.1    $\pm$ 1.0  & 9.6     $\pm$ 1.4  \\
NGC\,1365        & 3.8        $\pm$ 1.9 & 2.8          $\pm$ 1.5 & 69.7    $\pm$ 3.3 & 22 $\pm$ 1 & 2.3    $\pm$ 0.2 & 58   $\pm$ 2  & 4.2     $\pm$ 2.1  & 3.0    $\pm$ 1.6  & 12.2    $\pm$ 1.7  \\
NGC\,3256        & 10.4       $\pm$ 4.6 & 19.0         $\pm$ 8.1 & 351     $\pm$ 18  & 25 $\pm$ 2 & 2.4    $\pm$ 0.2 & 62   $\pm$ 3  & 11.2    $\pm$ 5.1  & 20.6   $\pm$ 8.8  & 53.3    $\pm$ 8.1  \\
NGC\,1808        & 1.1        $\pm$ 0.6 & 1.1          $\pm$ 0.6 & 20.7    $\pm$ 0.8 & 23 $\pm$ 2 & 2.5    $\pm$ 0.2 & 61   $\pm$ 2  & 1.2     $\pm$ 0.6  & 1.2    $\pm$ 0.6  & 3.5     $\pm$ 0.5  \\
IRAS\,13120-5453 & 97         $\pm$ 36  & 87           $\pm$ 35  & 1517    $\pm$ 79  & 25 $\pm$ 2 & 2.4    $\pm$ 0.2 & 54   $\pm$ 2  & 106     $\pm$ 39   & 94     $\pm$ 37   & 243.5   $\pm$ 36.7 \\
IRAS\,13242-5713 & 10.3       $\pm$ 3.4 & 19.5         $\pm$ 5.3 & 258     $\pm$ 14  & 24 $\pm$ 2 & 2.4    $\pm$ 0.2 & 55   $\pm$ 2  & 11.2    $\pm$ 3.7  & 21.2   $\pm$ 5.7  & 41.4    $\pm$ 6.2  \\
MRK\,331         & 10.7       $\pm$ 4.6 & $<$ 11.1               & 125.0 $\pm$ 6.5   & 23 $\pm$ 2 & 2.5    $\pm$ 0.2 & 59   $\pm$ 2  & 11.6    $\pm$ 4.9  & $<$ 11.9          & 21.3    $\pm$ 3.1  \\
NGC\,6240        & 80         $\pm$ 30  & 107          $\pm$ 36  & 586     $\pm$ 35  & 25 $\pm$ 2 & 2.3    $\pm$ 0.2 & 60   $\pm$ 2  & 87      $\pm$ 32   & 117    $\pm$ 39   & 95      $\pm$ 15   \\
NGC\,3628        & 0.6        $\pm$ 0.2 & 0.8          $\pm$ 0.2 & 11.7    $\pm$ 0.5 & 17 $\pm$ 1 & 2.8    $\pm$ 0.2 & 51   $\pm$ 1  & 0.6     $\pm$ 0.3  & 0.9    $\pm$ 0.3  & 2.2     $\pm$ 0.3  \\
NGC\,3627        & 0.7        $\pm$ 0.3 & $<$ 0.8                & 2.7   $\pm$ 0.1   & 23 $\pm$ 1 & 2.4    $\pm$ 0.1 & 57   $\pm$ 3  & 0.7     $\pm$ 0.3  & $<$ 0.9           & 0.7     $\pm$ 0.1  \\
IRAS\,18293-3413 & 25         $\pm$ 10  & 42           $\pm$ 13  & 521     $\pm$ 25  & 23 $\pm$ 2 & 2.4    $\pm$ 0.2 & 58   $\pm$ 2  & 26.7    $\pm$ 11.3 & 46     $\pm$ 15   & 83      $\pm$ 12   \\
NGC\,7469        & $<$  22.4            & 17.8       $\pm$ 6.3   & 250    $\pm$ 16   & 23 $\pm$ 2 & 2.4    $\pm$ 0.2 & 63   $\pm$ 3  & $<$ 24.3           & 19.3   $\pm$ 6.9  & 41.8    $\pm$ 6.4  \\
IRAS\,17578-0400 & $<$  17.4            & $<$ 14.9               & 205 $\pm$ 10      & 23 $\pm$ 2 & 2.4    $\pm$ 0.2 & 48   $\pm$ 1  & $<$ 18.8           & $<$ 16.1          & 32.8    $\pm$ 4.8  \\
IC\,1623         & $<$  26.0            & 25         $\pm$ 10    & 350     $\pm$ 19  & 23 $\pm$ 2 & 2.4    $\pm$ 0.2 & 60   $\pm$ 2  & $<$ 28.2           & 27.0   $\pm$ 11.3 & 59.1    $\pm$ 8.9  \\
Arp\,220         & 138        $\pm$ 36  & 70           $\pm$ 19  & 1650    $\pm$ 65  & 31 $\pm$ 3 & 1.7    $\pm$ 0.1 & 56   $\pm$ 4  & 151       $\pm$ 39 & 76     $\pm$ 21   & 258     $\pm$ 39   \\
IRAS\,19254-7245 & 58.9       $\pm$ 6.3 & 37.5         $\pm$ 4.3 & 1153    $\pm$ 97  & 22 $\pm$ 2 & 2.6    $\pm$ 0.2 & 64   $\pm$ 2  & 64        $\pm$ 15 & 40.7   $\pm$ 9.9  & 171     $\pm$ 27   \\
\enddata
\tablecomments{ Column 1: galaxy name. Column 2: \HCNto\ line luminosity.
        Column 3: \HCOto\ line luminosity. Column 4: total infrared luminosity.
        Column 5: cold component dust temperature. Column 6: cold component
        dust emissivity index.  Column 7: warm component dust temperature.
        Column 8: dense gas mass derived from \HCNto. Column 9: dense gas mass
        derived from \HCOto. Column 10: star formation rate derived from
        infrared luminosity.}
\end{deluxetable*}

\section{Results}
\subsection{Spectra}

We present APEX spectra of \HCNto\, and \HCOto\, in Figure \ref{spectrum}. In
total, we detect 14 \HCNto\, and 14 \HCOto\, lines, with their
velocity-integrated line intensities, $I > 3\,\sigma$. Twelve galaxies have
detections of both \HCNto\, and HCO$^+$\,\Jto. IRAS\,17578-0400
was not detected in either of the two lines.  NGC\,7469 and IC\,1623 were not
detected only in \HCNto,  possibly because of their large distances. On the
other hand, \HCOto\ is only marginally detected on $\sim$ 3-$\sigma$ levels in
these two galaxies. MRK\,331 and NGC\,3627 have non-detections of \HCOto. The
velocity-integrated fluxes of all galaxies are shown in Table
\ref{table:obsresult}. We show 3-$\sigma$ upper limits for those non-detected
lines.

\subsection{Correlation between luminosities of dense gas tracers and infrared emission}

In Figure \ref{SFLrelation}, we present the $L_{\rm IR}-L'_{\rm dense}$
correlation using \HCNto\, and \HCOto\, line luminosities, which trace total
dense gas mass. We fit linear regressions with both methods of Orthogonal Least
Squares ({\sf OLS}) and {\sf MCMC}, to avoid possible bias from the fitting algorithm. The
OLS was fitted with an IDL Astrolibrary {\sc sixlin},
which adopts orthogonal distances from data points to the fit line.  The code
first assumes a linear slope to compute orthogonal distance, and then performs
linear regression steps to compare with previous slopes,  until it reaches
steady state.  We perform {\sf MCMC} method using a Python package {\sc emcee}:
we first employ 64 walkers and each of them samples 5,000 steps. This would
ensure that most sampling are convergent. Then we burn in them, and perform
another 10,000 steps for each walker. This will offer a final sampling number
of 640,000 for statistics. We adopt the range encompassing 68\% of the data
about the median of the posterior probability density distribution to represent
approximate upper and lower 1-$\sigma$ limits for Gaussian-like distributions.
These fitted results from the two methods are shown in Figure
\ref{SFLrelation}, without significant difference between each other.  The
slope obtained from OLS fitting is within the 1-$\sigma$ range of that
from the Bayesian fitting.

We find linear correlations between $L_{\rm IR}$ with $L'_{\rm HCN {\it
J}=2\rightarrow1}$ and $L'_{\rm HCO^+ {\it J}=2\rightarrow1}$. The Pearson
correlation efficiencies, which quantifies linear correlations, are 0.96 and
0.97 with $p$-values of $1.7 \times 10^{-9}$ and $2.0 \times 10^{-10}$,
respectively. The Spearman correlation efficiencies, which quantifies monotone
correlations, are 0.98 and 0.96, with $p$-values of $1.0 \times 10^{-10}$ and
$4.8 \times 10^{-9}$, respectively. AGN-dominated and SF-dominated galaxies are
shown in blue and red points, respectively. The {\sc mcmc} fitting results are
listed below:

\begin{equation}
{\rm log}L_{\rm IR}=1.034^{+0.055}_{-0.051}~{\rm log}L'_{\rm HCN\,2-1}+2.91^{+0.4}_{-0.4},
\end{equation}
\begin{equation}
{\rm log}L_{\rm IR}=1.000^{+0.058}_{-0.054}~{\rm log}L'_{\rm HCO^+\,2-1}+3.21^{+0.43}_{-0.47},
\end{equation}

where the upper and lower errors are from probability density distributions of
parameters. Non-detections are not included during fitting.  We find no
definite difference between AGN-dominated and SF-dominated galaxies, which
essentially follow the same trend of $L_{\rm IR}$ and  $L'_{\rm dense}$. 

\subsection{Correlations of Luminosity Surface Densities}
\label{sec:surfacedensity}


Here we derive luminosity surface densities by adopting the area measured from
the 1.4-GHz radio continuum images, which can eliminate the degeneracy
introduced by the distance in luminosity correlations.  The 1.4-GHz radio
emission, which is contributed both by the synchrotron emission from supernova
remnants and by the free-free emission from H{\sc ii} regions,  originates from
the same star-forming regions as the IR emission \citep{Bell2003}. There exists
rich archival radio data at high angular resolutions, which would help
determine the 1.4\,GHz continuum sizes (see details in section
\ref{radiocontinuum}).

The correlations between the surface densities of IR luminosity ($\Sigma_{
L_{\rm IR}}$) and dense gas tracer line luminosity ($\Sigma_{L'_{\rm dense}}$)
are shown in Figure \ref{densityrelation}. The {\sf MCMC} fitting results are
shown below:

\begin{equation}
{\rm log}\Sigma_{L_{\rm IR}}=0.987^{+0.032}_{-0.030}~{\rm log}\Sigma_{L'_{\rm HCN\,2-1}}+3.21^{+0.09}_{-0.09}
\end{equation}
\begin{equation}
{\rm log}\Sigma_{L_{\rm IR}}=1.017^{+0.035}_{-0.034}~{\rm log}\Sigma_{L'_{\rm HCO^+\,2-1}}+3.17^{+0.09}_{-0.10}
\end{equation}

where $\Sigma_{L}=L/(2\pi r^2_{\rm RC})$, is the luminosity surface density.
The fitted slope indices of \HCNto\ and \HCOto\ are 0.99 and 1.02,
respectively. Both OLS and MCMC methods give identical fitting results.   The
Pearson correlation efficiencies are 0.99 and 0.99 with $p$-values of $2.7
\times 10^{-12}$ and $1.8 \times 10^{-13}$, for \HCNto\ and \HCOto,
respectively. The Spearman correlation efficiencies of these two lines are 0.99
and 0.98, with $p$-values of $8.1 \times 10^{-13}$ and $1.0 \times 10^{-10}$,
respectively. Both correlation coefficients are higher than those obtained
from the luminosity relations, showing much tighter correlations in the surface
density relations. 

Scatters in the surface density correlations are 0.150 and 0.113 dex for $\rm
HCN$ and \HCOto, respectively. These values are very close to the scatters in
the luminosity correlations (0.147 and 0.113 dex). This is likely because the
surface density correlations have larger dynamic ranges, and both types of
correlations share the same physical origin.

\subsection{Star formation efficiency of dense molecular gas}

Using HCN and HCO$^+$ $J=4\rightarrow3$ transitions in nearby star-forming
galaxies, \cite{Tan2018} found that $L_{\rm IR}/L^\prime_{ \rm dense}$, which
is a proxy of dense gas star formation efficiency (SFE), increases with $L_{\rm
IR}$ within individual galaxies, but not for galaxy-integrated ratios.
Furthermore, there seems to be also a correlation between $L_{\rm IR}/L^\prime_{ \rm
dense}$ and  PACS 70/100\um~ratio (as a proxy of warm dust temperature). 

\begin{figure*}[ht]
\includegraphics[height=3.5in]{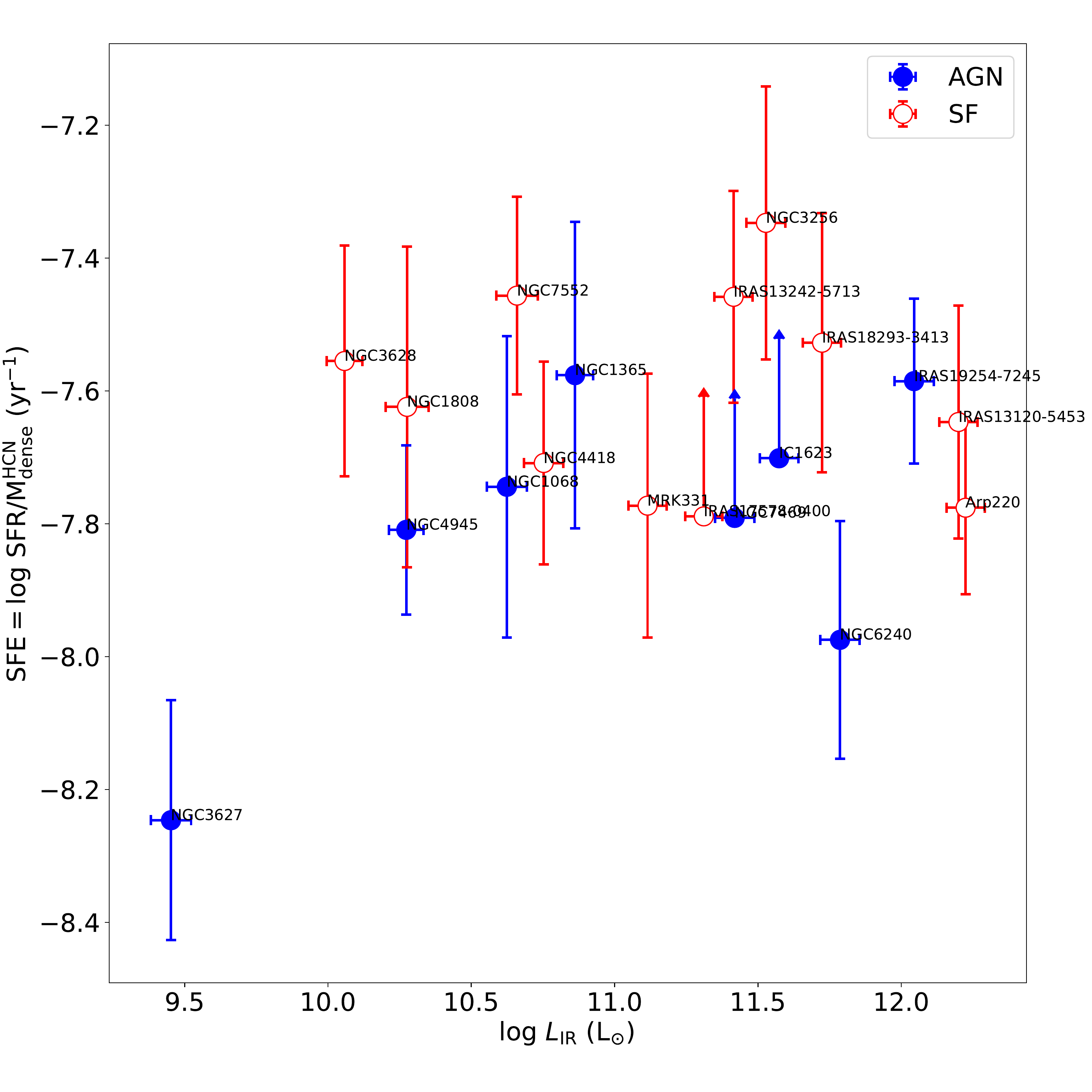}
\includegraphics[height=3.5in]{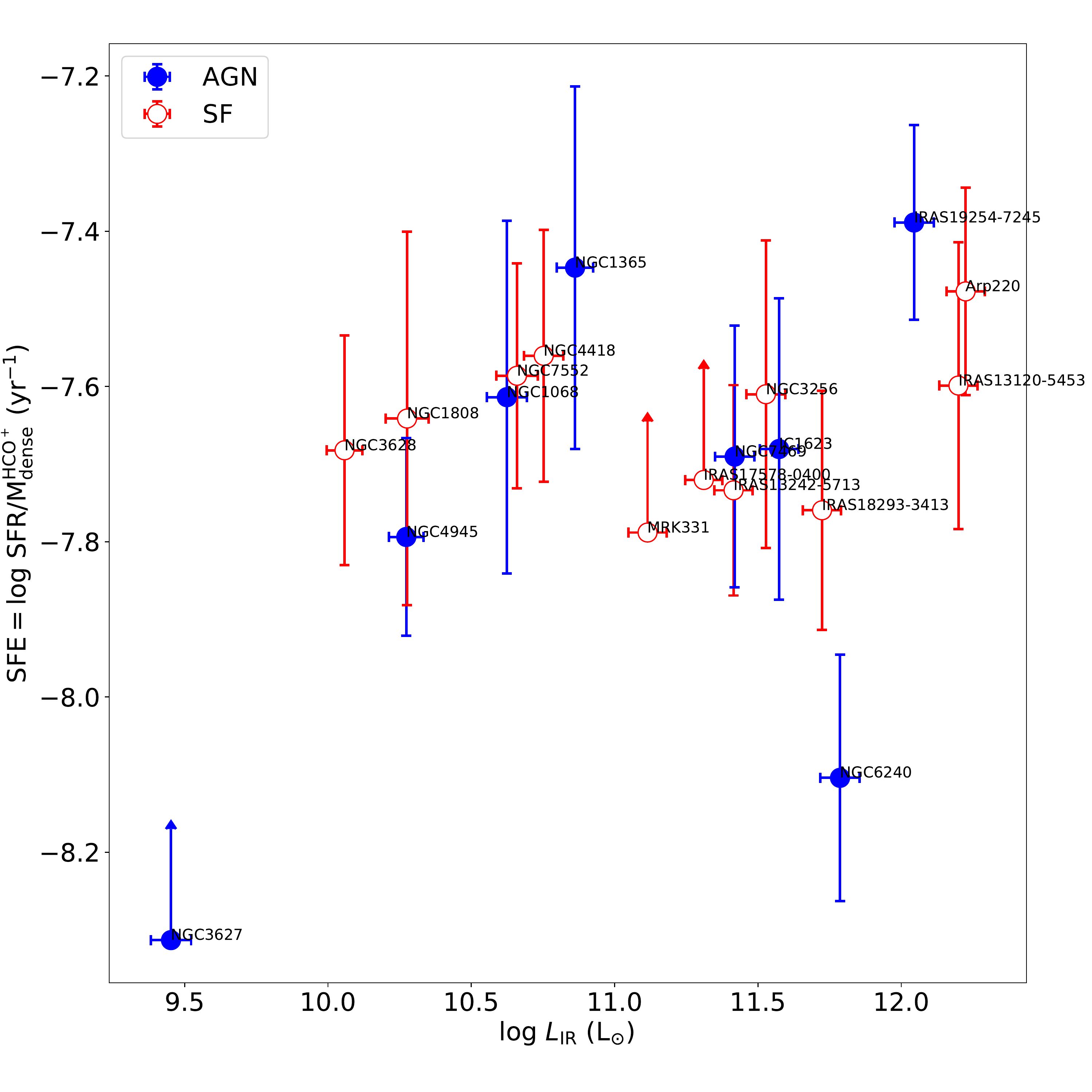}
\includegraphics[height=3.5in]{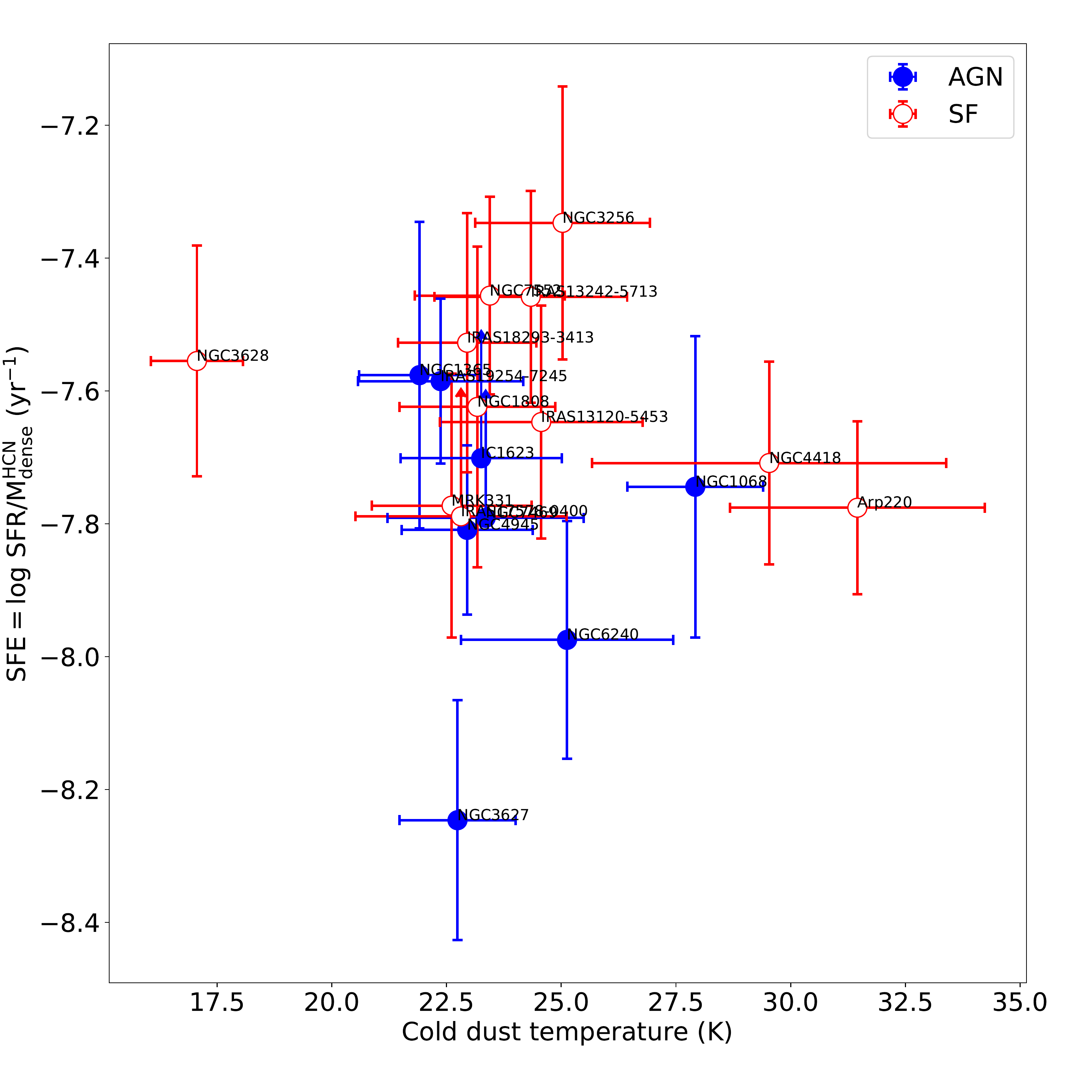}
\includegraphics[height=3.5in]{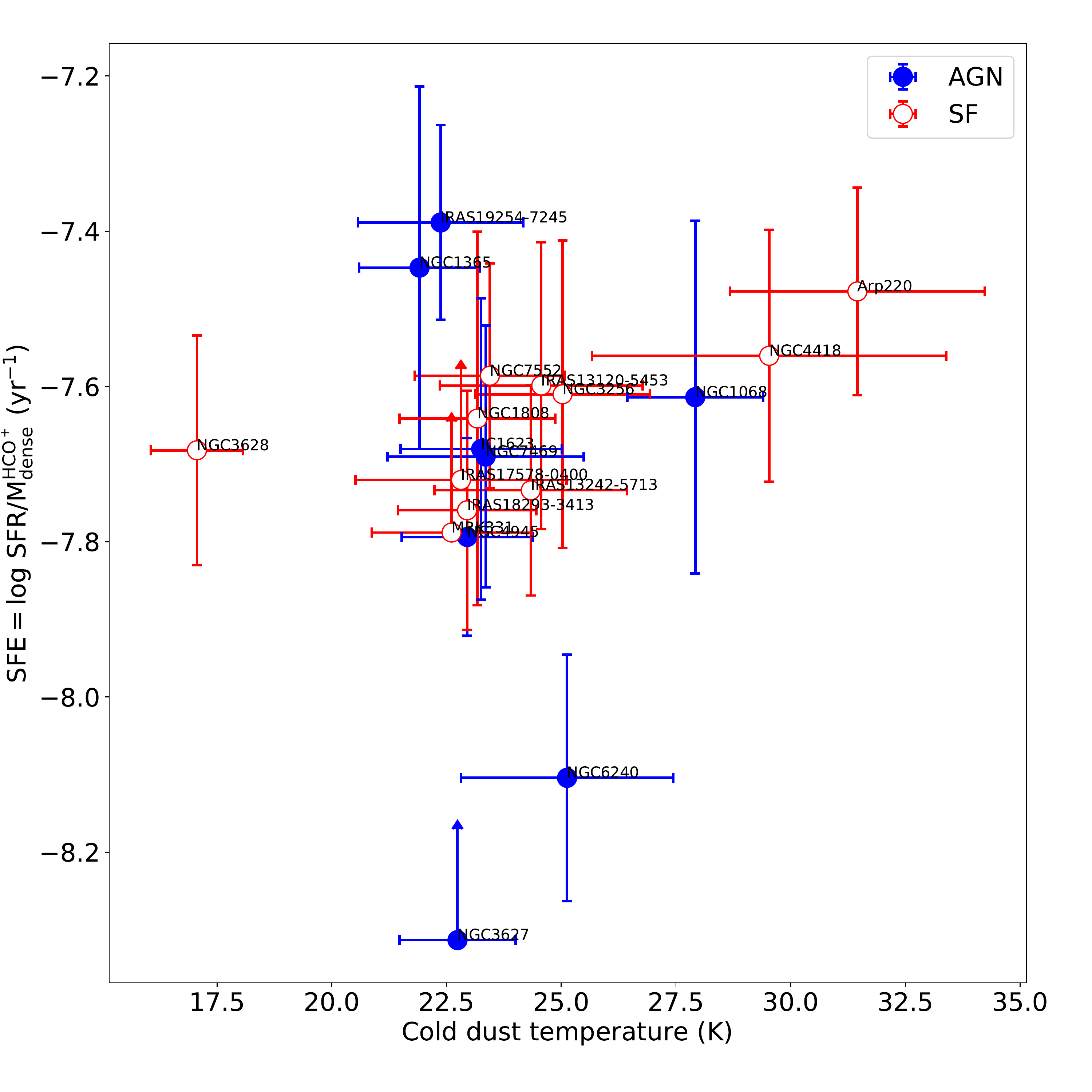}
\caption{{\it Top left:} SFR/$M_{\rm dense}$ traced by \HCNto\ as a function of
IR luminosity. {\it Top right} SFR/$M_{\rm dense}$ traced by \HCOto\ as a
function of IR luminosity. {\it Bottom left:} SFR/$M_{\rm dense}$ traced by
\HCNto\ as a function of cold dust temperature. {\it Bottom right:} SFR/$M_{\rm
dense}$ traced by \HCNto\ as a function of cold dust temperature. AGN-dominated
and star-formation dominated galaxies are shown in blue and red points,
respectively. } \label{SFE}
\end{figure*}

To further verify these trends, we plot SFE (derived from $L_{\rm IR}/L'_{\rm
HCN}$ and $L_{\rm IR}/L'_{\rm HCO^+}$) as a function of IR luminosity In Figure
\ref{SFE}.  Our data do not show any statistical correlation in the SFE-$L_{\rm
IR}$ diagram, with Spearman correlation $p$-values of 0.99 and 0.85 for \HCNto\
and \HCOto, respectively (see Table \ref{table:fittingresult}), meaning that
the data has a high chance to distribute randomly. These are consistent with
the integrated results of \cite{Tan2018}.    

In general, SFE obtained from HCN and \HCOp\ seem to be both confined within a
small range, with scatters of only 0.19 and 0.16. These small scatters are even
less than that of HCN \Joz\ \citep[$\sim 0.25$ found in $L_{\rm
IR}/L_{\rm HCN 1-0}$][]{Gao2004b}. This indicates that the \Jto\
transition of HCN and \HCOp\ are more robust in tracing the dense gas mass in
galaxies. However, the $L_{\rm IR}/L_{\rm HCN 4-3}$ ratios have a much more
diverged range of almost two orders of magnitude\citep{Tan2018}.

Unlike \cite{Tan2018}, on the other hand, the SFE-$T_{\rm c}$ diagram also
does not show any correlations, with Spearman correlation $p$-values of 0.46
and 0.91 for \HCNto\ and \HCOto, respectively. The $J=4\rightarrow3$
transitions are more sensitive to the dust temperature, because they need very
high density and relatively high temperature to be excited. These makes them to
trace the dense gas close to young massive stars, instead of the global
properties cold dense cores traced by the \Jto\ and
\Joz\ transitions.

\subsection{Ratios of $L'_{\rm HCN\, {\it J}=2\rightarrow1}$/$L'_{\rm HCO^+\, {\it J}=2\rightarrow1}$ }

\begin{figure*}[ht]
\includegraphics[width=3.5in]{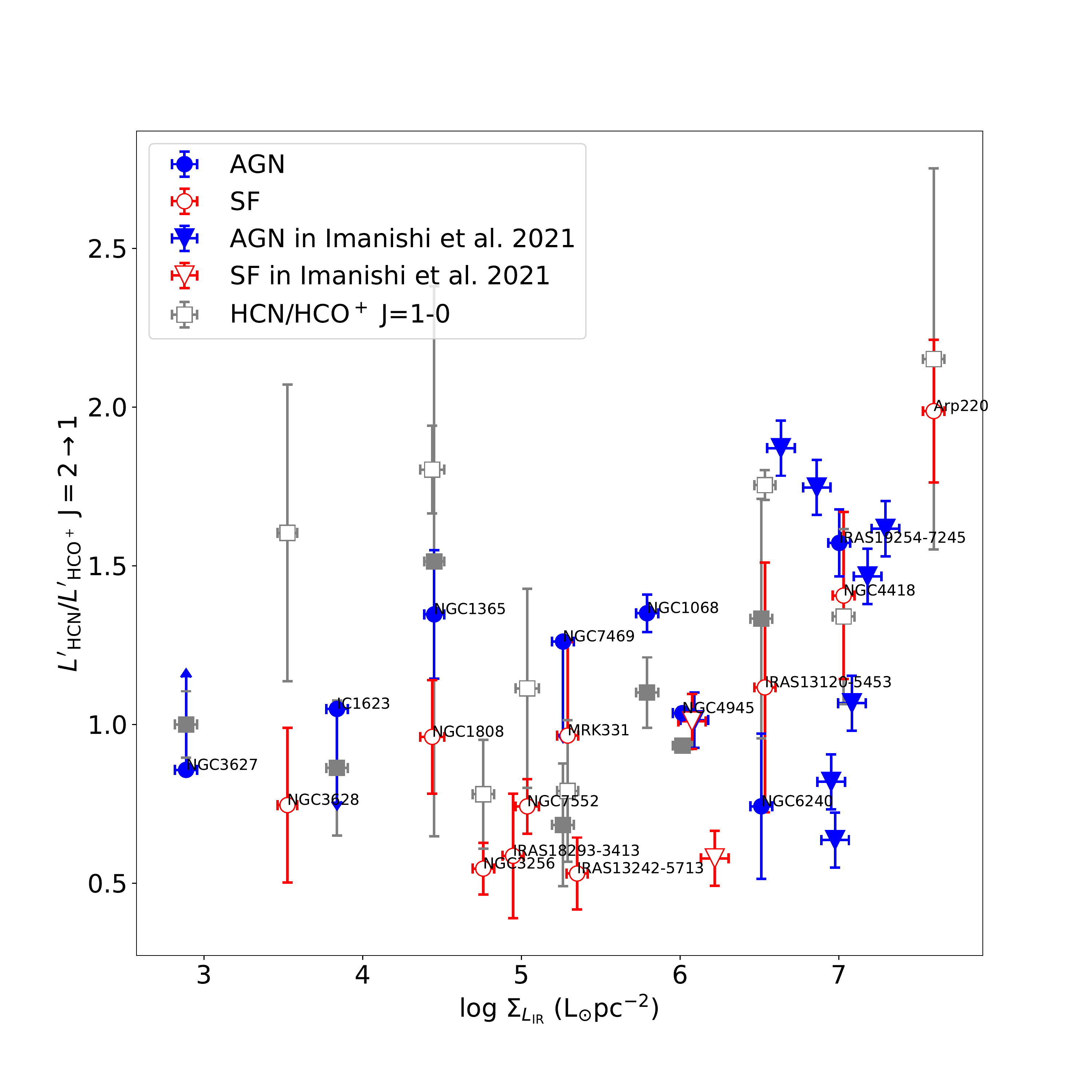}
\includegraphics[width=3.5in]{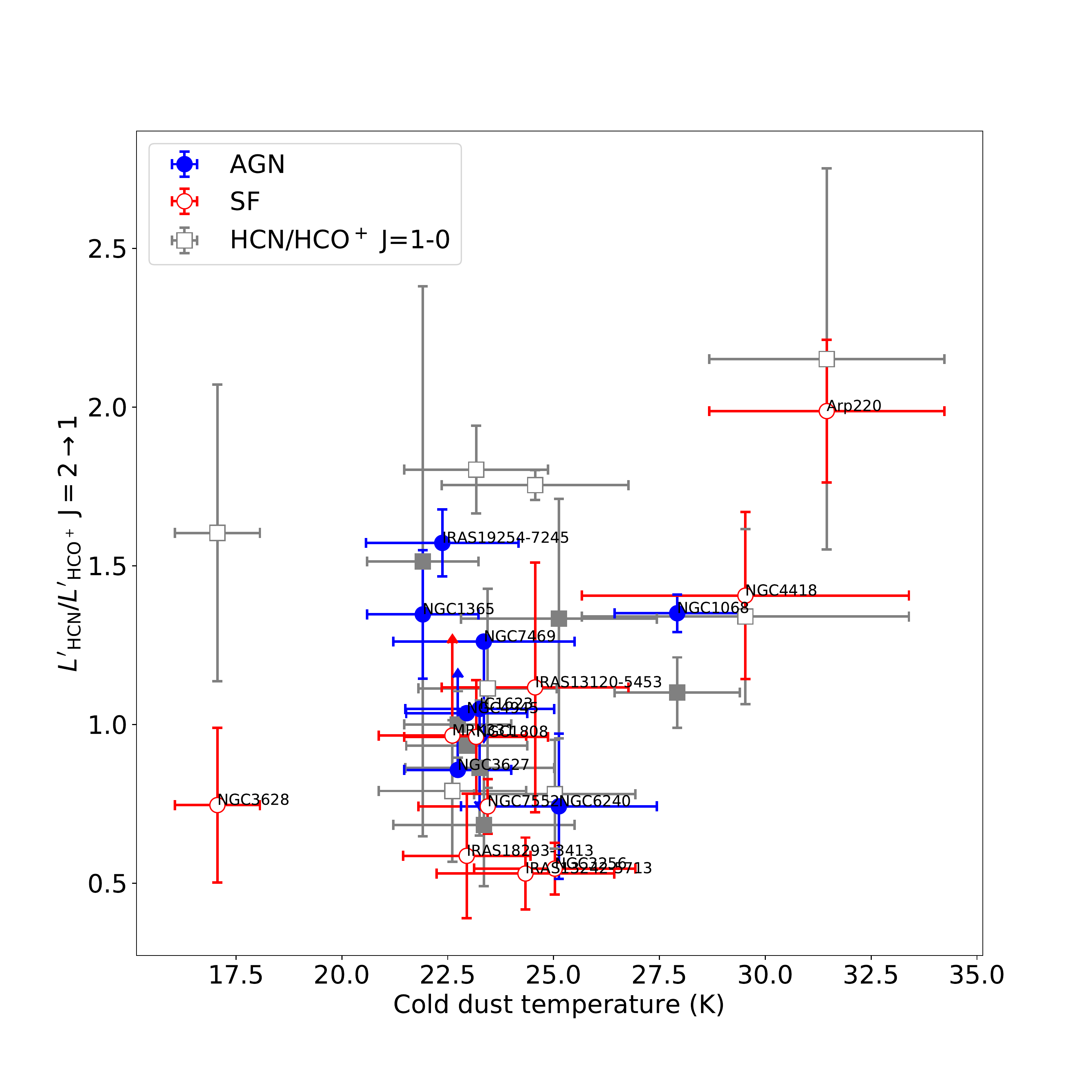}
\caption{{\it Left:} Ratio of $L'_{\rm HCN}/L'_{\rm HCO^+}$ \Jto\
as a function of infrared luminosity surface density. {\it Right:} Ratio of
$L'_{\rm HCN}/L'_{\rm HCO^+}$ \Jto\ as a function of cold
component dust temperature. The sample in this work is shown in circle, while
the ALMA-ULIRG sample from \citet{Imanishi22} is shown in triangle.
AGN-dominated galaxies are shown in blue filled points and SF-dominated
galaxies are shown in red empty points. Grey square points are HCN and HCO$^+$
\Joz\ from the literature for comparison
\citep{Wang2004,Baan2008,Gao2004a,Garcia06,Nguyen1992,Solomon1992,Krips08}.}
\label{lineratio}
\end{figure*}

Figure \ref{lineratio} presents the line ratio of $L'_{\rm HCN}/L'_{\rm HCO^+}$
as a function of IR luminosity surface density and cold dust temperature. We
also overlay HCN and HCO$^+$ \Joz\ ratios collected from the
literature \citep{Wang2004,Baan2008,Gao2004a,Garcia06,Nguyen1992,Solomon1992,Krips08}.
There seems no systematic difference between the ratios of both HCN/HCO$^+$
\Joz\ and HCN/HCO$^+$ \Jto. The average \Joz\ ratio is 1.25 $\pm$ 0.42
and \Jto\ ratio is 1.05 $\pm$ 0.42. \Joz\ ratio is higher, the difference is
smaller than the scatter.

The ratio of  $L'_{\rm HCN}/L'_{\rm HCO^+}$ may be changed by variation from
astrochemistry \citep{Imanishi07,Lintott06}, molecular excitation
\citep{Garcia06}, radiative transfer \citep{Knudsen07}, metallicities
\citep{Liang06}, and many other processes. Temperature and IR luminosity of
dust grains would reflect the energy from young stars and their surrounding
dense gas. Therefore, here we try to find possible dependence of this ratio
on star formation intensity (traced by the IR luminosity surface density) and on
of the cold dust temperature.  We choose the temperature of the
cold dust component, because 1). It comes from the bulk of the galactic dust
composition, taking the majority of the dust mass (>95\%; e.g., Appendix
\ref{app:SED_fitting}); 2). It is well sampled in our SED fitting (e.g., Fig.
\ref{SED_plot}), and 3). It contributes the majority of the far-IR luminosity,
which has an excellent correlation with the total IR luminosity
\citep[e.g.,][]{Zhu2008}. 

The average ratios of HCN/HCO$^+$ \Jto\ are 1.15$\,\pm\,$0.26 and
0.98$\,\pm\,$0.42 for  AGN-dominated galaxies and SF-dominated galaxies,
respectively.  Though it seems that AGN-dominated galaxies may systematically
have higher HCN/HCO$^+$ line ratios, the difference is still within
1\,$\sigma$. There seems to also exist a weak trend between line ratio and dust
temperature. But if we remove the data points of NGC\,3628 and Arp\,220, this
trend disappears.

\subsection{Dense gas fraction in galaxies}

\begin{figure*}[ht]
\includegraphics[height=3.5in,width=3.5in]{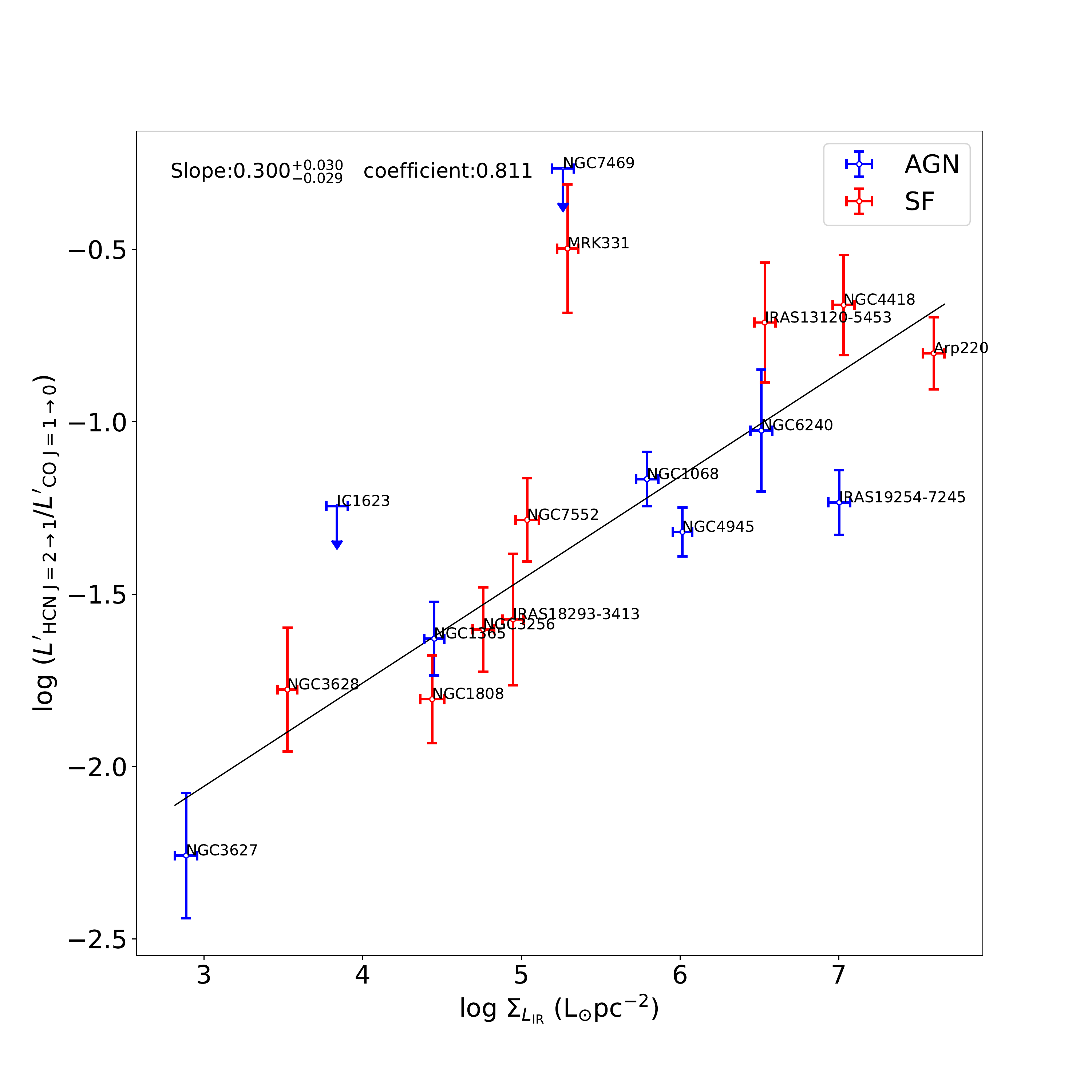}
\includegraphics[height=3.5in,width=3.5in]{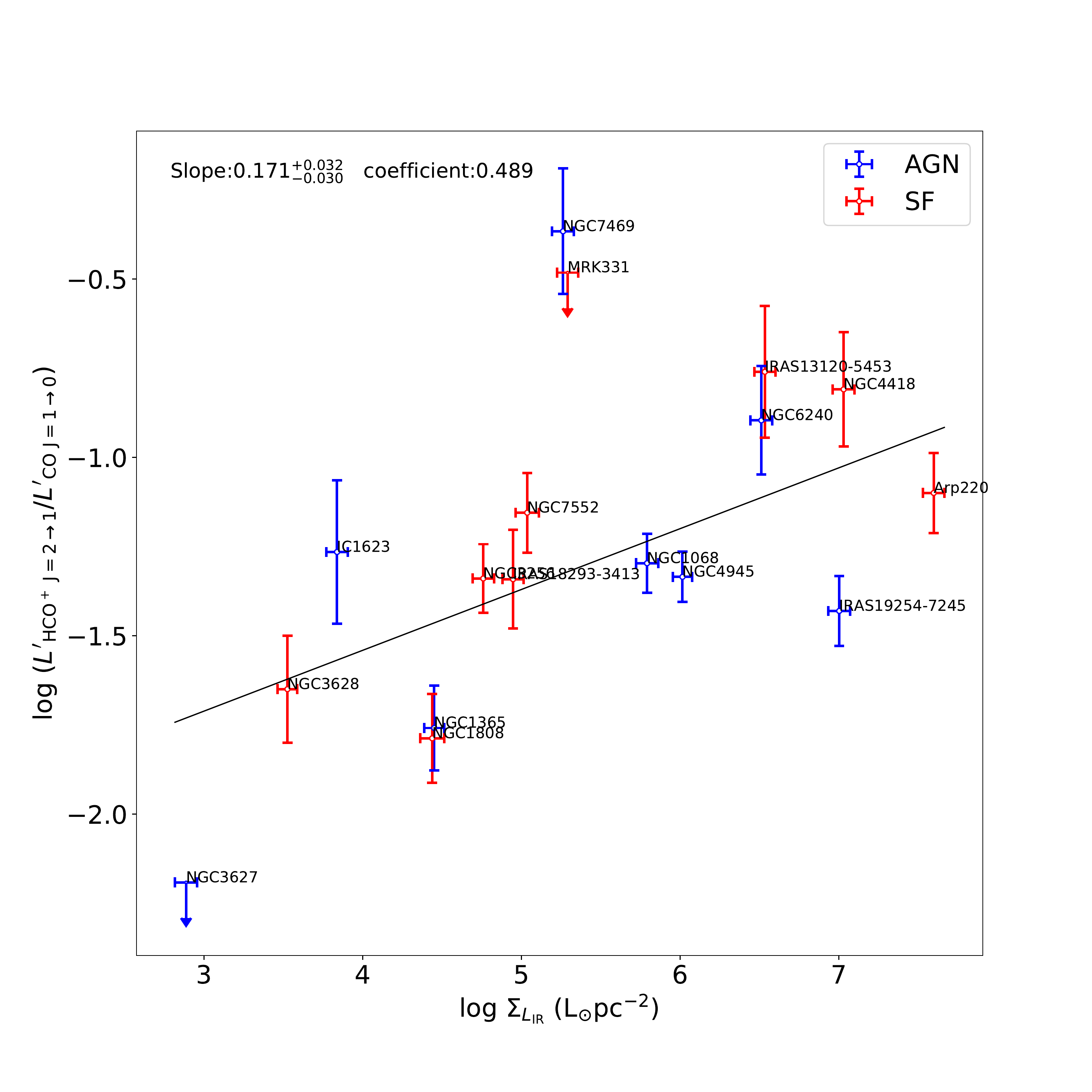}
\caption{ {\it Left:} $L'_{\rm HCN}/L'_{\rm CO}$ as a function of IR surface
density.  {\it Right:} $L'_{\rm HCO^+}/L'_{\rm CO}$ as a function of IR surface
density.  AGN-dominated and SF-dominated galaxies are shown in blue filled
points and  red empty points, respectively.  The best-fitted results are shown
with solid lines. Non-detections are not included in the fitting.}
\label{fraction}
\end{figure*}

The ratios of $L'_{\rm HCN}/L'_{\rm CO}$ and $L'_{\rm HCO^+}/L'_{\rm CO}$ would
roughly trace the dense gas fraction in galaxies, albeit the large
uncertainties of the conversion factors both in CO and dense gas tracers.
Similar to what was found in the $J$=1$\rightarrow$0 transition of HCN in
\citet{Gao2004b}, the ratios of $L'_{\rm HCN}/L'_{\rm CO}$ and $L'_{\rm
HCO^+}/L'_{\rm CO}$ both show increasing trends as a function of $L_{\rm IR}$
and $\Sigma_{L_{\rm IR}}$ (Figure \ref{fraction}).  The Pearson correlation
coefficient between $L'_{\rm HCN~J=2\rightarrow1}/L'_{\rm CO~J=1\rightarrow0}$
ratio and IR surface density is 0.81 ($p$-value=2.4$\times 10^{-4}$), which is higher than that of the $L'_{\rm
HCO^+~J=2\rightarrow1}/L'_{\rm CO~J=1\rightarrow0}$ ratio (0.49, $p$-value=0.064), indicating
that HCN/CO ratios might be more robust in tracing dense gas fractions.

These are also consistent with the positive correlations between $L'_{\rm
HCN\,1-0,3-2}/L'_{\rm CO\,1-0}$ ratio and $L_{\rm IR}$
\citep[e.g.,][]{Juneau2009}, who found a better correlation for the
$J=3\rightarrow2$ transition than that of \Joz\, indicating an
increased molecular gas density in more IR-bright galaxies.


We summarize the statistics parameters of correlations and list all of them in
Table \ref{table:fittingparam}, including {\sf MCMC} fitting results, {\sf OLS}
fitting results, Pearson (linear relation) ordered correlation coefficient and
$p$-value, Spearman (monotonic relation) ordered correlation coefficient and
$p$-value, and scatter of diversions from the {\sf OLS} fitting. 

Correlations of luminosities ($L'_{\rm dense}-L_{\rm IR}$) and luminosity
surface densities ($\Sigma_{L'_{\rm dense}}-\Sigma_{L_{\rm IR}}$) all show
significant correlation (Spearman $p$-value<10$^{-8}$). The correlations of
$L^\prime_{\rm HCN}/L^\prime_{\rm CO}-\Sigma_{L_{\rm IR}}$, $L^\prime_{\rm
HCN}/L^\prime_{\rm HCO^+}-\Sigma_{L_{\rm IR}}$, and $T_{\rm c}-\Sigma_{L_{\rm
IR}}$, are also statistically valid with $p$-values $<$ 0.05 for both Pearson
and Spearman rank. Those non-correlated relations show large error bars of
linear fitting and large $p$-values ($>$ 0.05), and sometimes also show large
difference between the results from {\sf MCMC} and {\sf OLS}.

\begin{figure*}
\includegraphics[width=1\textwidth]{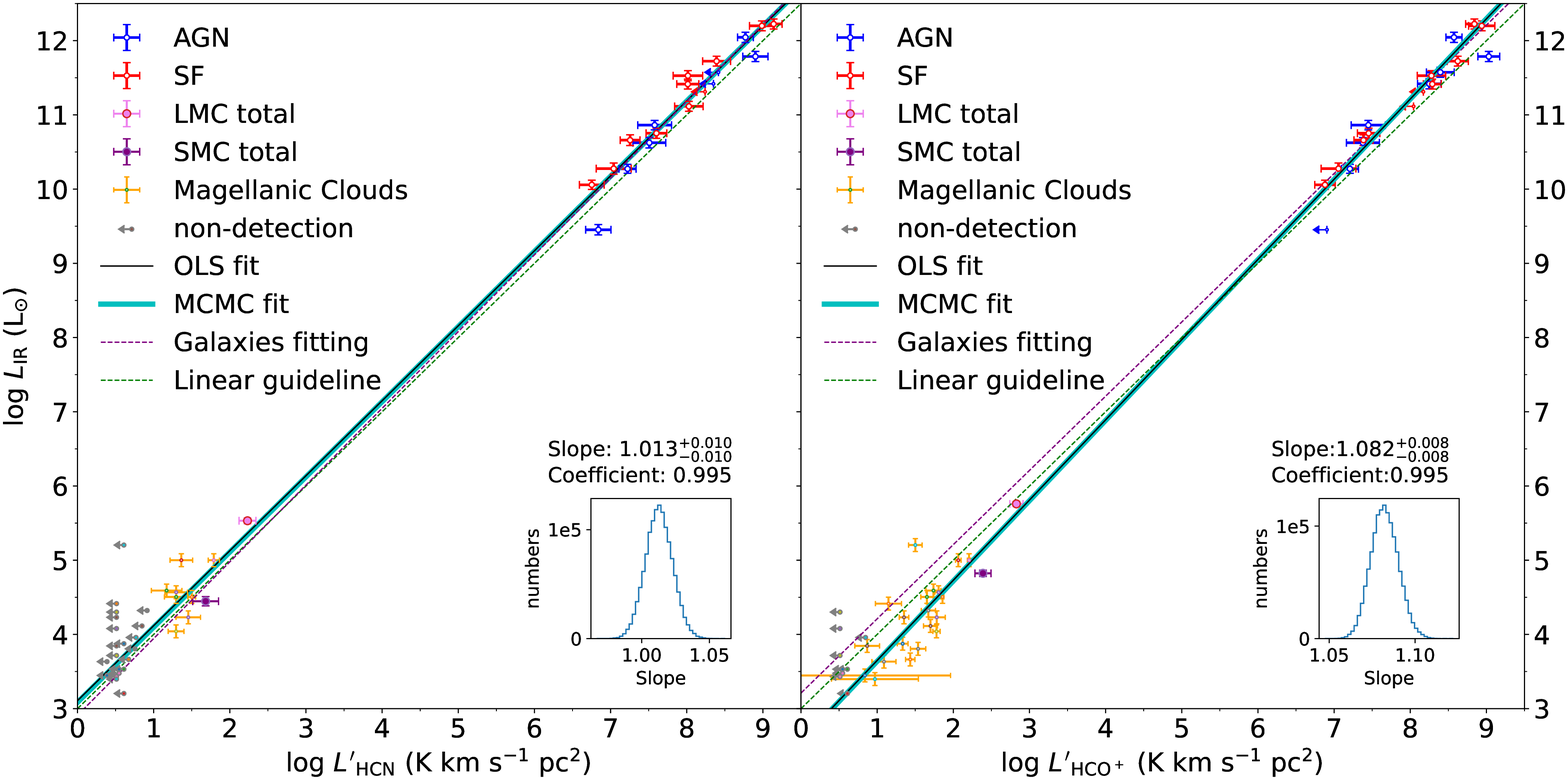}
\includegraphics[width=1\textwidth]{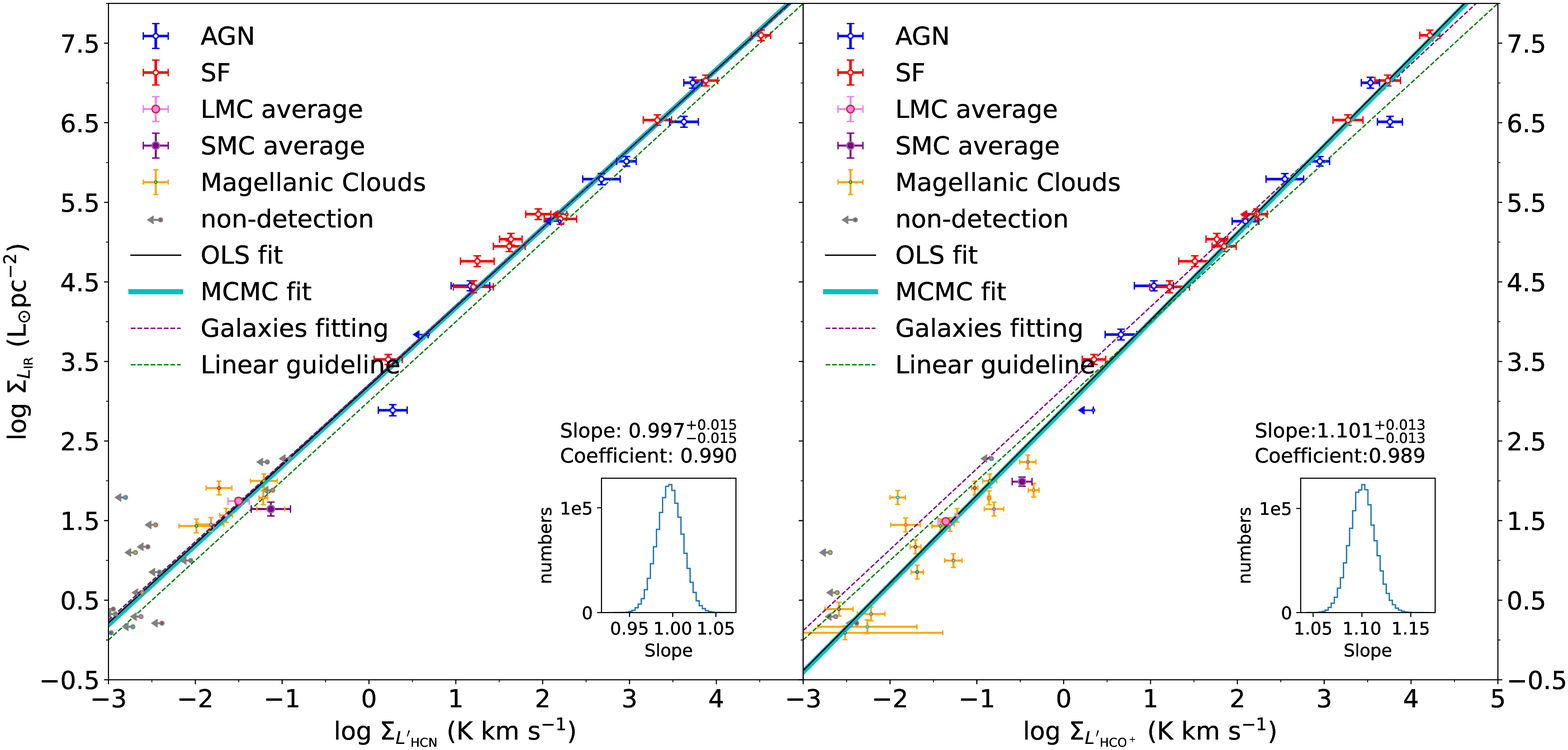}
\caption{ {\it Top:} Correlations of $L'_{\rm HCN}-L_{\rm IR}$ ({\it top left})
and $L'_{\rm HCO^+}-L_{\rm IR}$ ({\it top right}).  {\it Bottom:} Correlations
of $\Sigma_{L^\prime_{\rm HCN}}-\Sigma_{L_{\rm IR}}$ ({\it bottom left}) and
$\Sigma_{L^\prime_{\rm HCO^+}}-\Sigma_{L_{\rm IR}}$ ({\it bottom right}).  The
correlations include dense molecular clouds from Magellanic Clouds \citep{Galametz20},
which are shown in orange circles. The total luminosity and the averaged
luminosity surface density of detected targets in the Magellanic Clouds are
also shown for comparison. Luminosity upper limits of non-detection clouds are
shown in grey arrows. The purple dashed lines are taken from Figures
\ref{SFLrelation} and \ref{densityrelation} and present fitting results of
galaxies.}
\label{relation_append}
\end{figure*}

\begin{deluxetable*}{cccccccccc}
\tablenum{5}
\tablecaption{Statistic parameters of different fittings}\label{table:fittingparam}
\tablewidth{0pt}
\tablehead{
\colhead{} & \multicolumn{2}{c}{\sf MCMC} & \multicolumn{2}{c}{\sf OLS} & \multicolumn{2}{c}{Pearson} & \multicolumn{2}{c}{Spearman}  & \colhead{} \\
\colhead{Fitting name} & \colhead{slope} & \colhead{intercept} &\colhead{slope}&\colhead{intercept} & \colhead{$r_{\rm xy}$} & \colhead{$p$-value} &\colhead{$r_{\rm xy}$} & \colhead{$p$-value} & \colhead{scatter}\\
\colhead{(1)}&\colhead{(2)}&\colhead{(3)}&\colhead{(4)}&\colhead{(5)}&\colhead{(6)}&\colhead{(7)}&\colhead{(8)}&\colhead{(9)}&\colhead{(10)}
}
\startdata
$L_{\rm IR}-L^\prime_{\rm HCN}$                               & $1.03     _{-0.05 }    ^{+0.05 }$    & $2.9     _{-0.4 }    ^{+0.4 }$ & 1.05   $\pm$   0.04  & 2.8    $\pm$   0.3  & 0.96  & 1.7    $\times 10^{-9}$  & 0.98   & 1.0 $\times 10^{-10}$    & 0.15 \\
$L_{\rm IR}-L^\prime_{\rm HCO^+}$                             & $1.00     _{-0.05 }    ^{+0.06 }$    & $3.2     _{-0.5 }    ^{+0.4 }$ & 1.00   $\pm$   0.03  & 3.2    $\pm$   0.2  & 0.97  & 2.0    $\times 10^{-10}$ & 0.96   & 4.8    $\times 10^{-9}$  & 0.11 \\
$\Sigma_{L_{\rm IR}}-\Sigma_{L^\prime_{\rm HCN}}$             & $0.99     _{-0.03 }    ^{+0.03 }$    & $3.21    _{-0.09}    ^{+0.09}$ & 0.99   $\pm$   0.02  & 3.20   $\pm$   0.07 & 0.99  & 2.7    $\times 10^{-12}$ & 0.99   & 8.1    $\times 10^{-13}$ & 0.15 \\
$\Sigma_{L_{\rm IR}}-\Sigma_{L^\prime_{\rm HCO^+}}$           & $1.02     _{-0.03 }    ^{+0.04 }$    & $3.17    _{-0.10}    ^{+0.09}$ & 1.02   $\pm$   0.01  & 3.17   $\pm$   0.03 & 0.99  & 1.8    $\times 10^{-13}$ & 0.98   & 1.0    $\times 10^{-10}$ & 0.11 \\
$L^\prime_{\rm HCN}/L^\prime_{\rm HCO^+}-\Sigma_{L_{\rm IR}}$ & $0.33     _{-0.04 }    ^{+0.04 }$    & $-0.9    _{-0.2 }    ^{+0.2 }$ & 0.30   $\pm$   0.03  & -0.7   $\pm$   0.2  & 0.65  & 0.013                    & 0.59   & 0.027                    & 0.32 \\
$L^\prime_{\rm HCN}/L^\prime_{\rm HCO^+}-T_{\rm c}$           & $0.11     _{-0.26 }    ^{+0.03 }$    & $-1.6    _{-0.7 }    ^{+6.1 }$ & 0.06   $\pm$   0.01  & -0.4   $\pm$   0.3  & 0.53  & 0.053                    & 0.21   & 0.47                     & 0.36 \\
SFR/$M^{\rm HCN}_{\rm dense}-L_{\rm IR}$                      & $0.02     _{-0.05 }    ^{+0.05 }$    & $-7.7    _{-0.5 }    ^{+0.6 }$ & 0.03   $\pm$   0.03  & -7.8   $\pm$   0.3  & 0.14  & 0.62                     & -0.003 & 0.99                     & 0.19 \\
SFR/$M^{\rm HCO^+}_{\rm dense}-L_{\rm IR}$                    & $0.01     _{-0.05 }    ^{+0.05 }$    & $-7.5    _{-0.6 }    ^{+0.6 }$ & -0.01  $\pm$   0.02  & -7.3   $\pm$   0.3  & -0.13 & 0.62                     & -0.053 & 0.85                     & 0.16 \\
SFR/$M^{\rm HCN}_{\rm dense)}-T_{\rm c}$                      & $-0.02    _{-0.01 }    ^{+0.01 }$    & $-7.1    _{-0.3 }    ^{+0.3 }$ & -0.016 $\pm$   0.002 & -7.07  $\pm$   0.06 & -0.26 & 0.33                     & -0.20  & 0.46                     & 0.18 \\
SFR/$M^{\rm HCO^+}_{\rm dense}-T_{\rm c}$                     & $0.00     _{-0.01 }    ^{+0.01 }$    & $-7.5    _{-0.3 }    ^{+0.3 }$ & 0.003  $\pm$   0.003 & -7.52  $\pm$   0.08 & 0.052 & 0.85                     & -0.029 & 0.91                     & 0.16 \\
SFR/$M^{\rm HCN}_{\rm dense}-T_{\rm w}$                       & $-0.005   _{-0.08 }    ^{+0.008 }$   & $-7.8    _{-0.5 }    ^{+0.5 }$ & 0.003 $\pm$   0.002  & -7.63  $\pm$   0.14 & 0.050 & 0.85                     & 0.091  & 0.74                     & 0.19 \\
SFR/$M^{\rm HCO^+}_{\rm dense}-T_{\rm w}$                     & $0.007    _{-0.007 }   ^{+0.008 }$   & $-7.9    _{-0.5 }    ^{+0.4 }$ & 0.006  $\pm$   0.002 & -7.79  $\pm$   0.08 & 0.17  & 0.53                     & 0.35   & 0.18                     & 0.16 \\
$L^\prime_{\rm HCN}/L^\prime_{\rm CO}-\Sigma_{L_{\rm IR}}$    & $0.27     _{-0.03 }    ^{+0.03 }$    & $-2.8    _{-0.2 }    ^{+0.2 }$ & 0.29   $\pm$   0.02  & -2.89  $\pm$   0.09 & 0.81  & 2.4    $\times 10^{-4}$  & 0.84   & 8.0    $\times 10^{-5}$  & 0.27 \\
$L^\prime_{\rm HCO^+}/L^\prime_{\rm CO}-\Sigma_{L_{\rm IR}}$  & $0.10     _{-0.03 }    ^{+0.03 }$    & $-1.8    _{-0.2 }    ^{+0.2 }$ & 0.13   $\pm$   0.02  & -2.0   $\pm$   0.1  & 0.49  & 0.064                    & 0.58   & 0.023                    & 0.33 \\
$^*L_{\rm IR}-L^\prime_{\rm HCN}$                             & $1.01     _{-0.01 }    ^{+0.01 }$    & $3.08    _{-0.06}    ^{+0.06}$ & 1.010  $\pm$   0.006 & 3.11   $\pm$   0.05 & 0.997 & 4.2    $\times 10^{-26}$ & 0.97   & 2.3    $\times 10^{-14}$ & 0.17 \\
$^*L_{\rm IR}-L^\prime_{\rm HCO^+}$                           & $1.082    _{-0.008}    ^{+0.008}$    & $2.56    _{-0.04}    ^{+0.04}$ & 1.081  $\pm$   0.005 & 2.57   $\pm$   0.04 & 0.997 & 1.2    $\times 10^{-38}$ & 0.95   & 3.6    $\times 10^{-18}$ & 0.20 \\
$^*\Sigma_{L_{\rm IR}}-\Sigma_{L^\prime_{\rm HCN}}$           & $1.00     _{-0.01 }    ^{+0.02 }$    & $3.18    _{-0.04}    ^{+0.04}$ & 0.993  $\pm$   0.008 & 3.19   $\pm$   0.02 & 0.99  & 1.4    $\times 10^{-21}$ & 0.99   & 7.1    $\times 10^{-18}$ & 0.16 \\
$^*\Sigma_{L_{\rm IR}}-\Sigma_{L^\prime_{\rm HCO^+}}$         & $1.10     _{-0.01 }    ^{+0.01 }$    & $2.90    _{-0.02}    ^{+0.02}$ & 1.101  $\pm$   0.009 & 2.92   $\pm$   0.02 & 0.99  & 1.0    $\times 10^{-29}$ & 0.97   & 2.6    $\times 10^{-21}$ & 0.20 \\
$T_{\rm c}-\Sigma_{L_{\rm IR}}$                               & $1.4      _{-0.3  }    ^{+0.3  }$    & $16      _{-2   }    ^{+2   }$ & 3.5    $\pm$   0.6   & 6      $\pm$   3    & 0.67  & 0.002                    & 0.55   & 0.015                    & 0.87 \\
$T_{\rm w}-\Sigma_{L_{\rm IR}}$                               & $0.9      _{-0.4  }    ^{+0.4  }$    & $52      _{-2   }    ^{+2   }$ & 25     $\pm$   50    & -75    $\pm$   260  & 0.13  & 0.58                     & 0.091  & 0.71                     & 1.20 \\
\enddata
\tablecomments{ Column 1: fitting name. Column 2: slope fitted with Markov Chain Monte
Carlo ({\sf MCMC}). Column
3: intercept fitted with {\sf MCMC}. Column 4: slope fitted with Orthogonal
Least Squares ({\sf OLS}). Column 5: intercept fitted with {\sf OLS}. Column 6:
Pearson rank-order correlation coefficient. Column 7: Pearson rank-order
$p$-value. Column 8: Spearman rank-order correlation coefficient. Column 9:
Spearman rank-order $p$-value. Column 10: scatter of diversion from the
fitting.\\ $^*$ The fitting includes molecular clouds in LMC and SMC.
}
\end{deluxetable*}

\section{Discussion}


We adopt the total IR luminosity (integrated from 3\,\um~to 1000\,\um) as the
SFR tracer of galaxies. $L_{\rm IR}$ has been widely used as SFR tracer, while
it still has some limitations. AGN may contribute to mid-infrared emission by
their hot dusty tori \citep{Padovani2017} and thus $L_{\rm IR}$ may
overestimate SFR in strong AGN-dominated galaxies. On the other hand, $L_{\rm
IR}$ traces SFR on relatively long timescales up to 200 Myr \citep{kennicutt12}
and it may overestimate significantly the instantaneous SFR of recent-quenched
galaxies \citep{Hayward2014}. When converting from gas luminosity to dense gas
mass, we assume that all molecular lines are collisionally excited. However,
IR-pumping mechanism may also contribute to the line excitation and could
largely enhance HCN and HCO$^+$ emission. If  the observed HCN and HCO$^+$
fluxes are dominated by IR-pumping, the dense gas mass would be overestimated.
Furthermore, variation in the stellar initial mass function (IMF) between
different galaxies would also bias the SFR calculation from $L_{\rm IR}$
\citep[e.g.,][]{Jerabkova2017,Zhang2018}.

The sizes of star-forming regions may be biased by the compact radio emission
contributed by AGN. Therefore, we collect VLBI high-resolution data from the
literature
\citep{Lenc2009NGC4945,Roy1998NGC1068,Varenius14NGC4418,Smith1998MRK331IC1623arp220,Gallimore2004NGC6240,Cole1998NGC3628,Deller2014NGC3627,Lonsdale2003NGC7469}
and determine if our measured sizes are severely biased by AGN (see Table
\ref{table:basicinfo}).  For most galaxies, the central compact sources only
contribute $\lesssim$ 10\% of the total radio fluxes. Among all targets,
NGC~7469 has the highest AGN contribution, which is $\sim$ 18\%. For the rest
of the sample, most of them are classified as star-formation dominated galaxies
so AGN could not contribute much radio emission. For IRAS\,19254-7245 and
NGC\,6240, we adopt ALMA 250 GHz and 480 GHz continuum to estimate star-forming
region sizes, respectively.

\subsection{Star formation relation for dense gas tracers}

Our results show that SFR follows tight linear correlations with dense gas mass
traced by both \HCNto\ and \HCOto, with Slopes of 1.03 $\pm$ 0.05 and 1.00
$\pm$ 0.06, Pearson coefficients of 0.96 and 0.97, and dispersions of 0.15 and
0.11 dex, respectively (see Figure \ref{SFLrelation}). The star formation
efficiencies are roughly constant (Pearson coefficients: 0.14 and -0.13)
against SFR, with scatters of 0.19 dex and 0.16 dex (see Figure \ref{SFE}) for
\HCNto\ and \HCOto, respectively. These suggest that dense gas clouds are
direct sources of star formation, while \HCNto\ and \HCOto\ can be good tracers
of dense molecular gas. On the other hand, we find no systematic variation
between the slopes derived from the \Jto\ (1.03$\pm$0.05) and \Joz\
(1.00$\pm$0.05) transitions.

NGC\,6240 and NGC\,3627 are two outliers deviated from the relationship of IR
and dense gas tracer luminosity. This makes their dense-gas star formation
efficiencies, which are traced by infrared to dense gas luminosity ratio, about
three times lower than the average value of other galaxies. This may indicate
that star formation efficiency of these dense gas is suppressed, or the dense
gas tracer emission is enhanced by the extra heating mechanisms for the same
star-forming activity.  NGC\,6240 is a starburst galaxy in final merger stage
\citep{Tacconi99,Iono07,Papadopoulos14,Kollatschny2020}. The enormous shock
condition across NGC\,6240, shown both in the enhanced high-$J$ CO lines and in
large scale shocked gas distribution \citep{Meijerink2013,Cicone18, Lu15},
could contribute mechanical heating to  the dense gas, along with cosmic-ray
and far-ultraviolet (FUV) radiation from photon-dominated regions
\citep{Papadopoulos14}. These extreme conditions in NGC~6240 may also enhance
the \HCNto\ and \HCOto\ emission. Our observation of NGC\,3627 is consistent
with \cite{Murphy2015}, who found that star formation efficiency in the nuclear
region of NGC\,3627 is several times lower than that in off-nuclear star-forming region. This could explain why NGC~3627 also shows offsets from both
HCN and \HCOto\ correlations.

\subsection{Connecting to star-forming clouds on small scales}

The correlations obtained on large scales may have contamination from
non-star-forming activities such as shock-enhanced line emission, diffuse gas
component, etc. Therefore, it is necessary to check if the correlations can be
extended to pc scales that only contains star-forming dense molecular cores.
Unfortunately, it is difficult to find systematic observations of the
\Jto\ transition towards our Milky Way, while such observations
do exist towards the Magellanic Clouds. Therefore, we compare the galaxy survey
with the Magellanic Clouds data from \cite{Galametz20}, who presented \HCNto\
and \HCOto\ observations towards $\sim$30 LMC and SMC molecular clouds using
APEX.    

As shown in Figure \ref{relation_append}, after including molecular clouds of
the Magellanic Clouds into our galaxy sample,  the correlations between
$L'_{\rm HCN~J=2\rightarrow1}$--$L_{\rm IR}$  and $L'_{\rm
HCO^+~J=2\rightarrow1}$--$L_{\rm IR}$ still hold with semi-linear slopes of
1.021 and 1.088, respectively. The upper limits of non-detections seem off the
correlation, possibly due to different detection criteria in \cite{Galametz20},
who consider a 3-$\sigma$ peak Gaussian fitting (instead of integrated flux) as
the detection threshold.  We also include data of Magellanic Clouds into the
correlations of surface densities, by adopting the area from
\citep[e.g.,][]{Wong2011,Muller2010}.  The obtained slopes are 1.00  and 1.10
for $\Sigma_{L'_{\rm HCN}}$--$\Sigma_{L_{\rm IR}}$ and $\Sigma_{L'_{\rm
HCO^+}}$--$\Sigma_{L_{\rm IR}}$, respectively.  Both HCN and \HCOp\ data
averaged (summed) across the Magellanic Clouds are shown in Figure
\ref{relation_append} as individual data points, which seem to also follow the
correlations of $\Sigma_{L'_{\rm HCN}}$--$\Sigma_{L_{\rm IR}}$ ($L'_{\rm
HCN~J=2\rightarrow1}$--$L_{\rm IR}$) and $\Sigma_{L'_{\rm
HCO^+}}$--$\Sigma_{L_{\rm IR}}$ ($L'_{\rm HCO^+~J=2\rightarrow1}$--$L_{\rm
IR}$) found in galaxies. 

On the other hand, we also simply overlay the Magellanic Clouds data with our
fitted galaxy-only correlations. The HCN data from Magellanic Clouds well match
with the correlation fitted in the galaxy-only sample, while the \HCOp\ data
from Magellanic Clouds systematically lay below the fitted lines from galaxies,
indicating a systematic HCN/HCO$^+$ ratio variation possibly due to different
metallicities.

\subsection{$L'_{\rm HCN}/L'_{\rm HCO^+}J=2\rightarrow1$ ratio and its origins}

The average of HCN-to-HCO$^+$ \Jto\ flux ratio of our sample is
1.15$\pm$0.26 and 0.98$\pm$0.42 for AGN-hosting and SF-dominated galaxies,
respectively. Though it seems that the AGN-hosting galaxies may have relatively
higher ratios than SF-dominated galaxies, the difference is still within 1
$\sigma$. So, we could not use this ratio to separate populations with AGNs.
Our result of \Jto\ is consistent with that found in the \Joz\ lines
\citep{Privon15, Lifei20}.

\cite{Kohno2001} suggested that enhanced HCN emission originate from X-ray
dominated region and can be used to search for pure AGNs, which is also
supported by \cite{Imanishi2004,Imanishi07}.  \cite{Imanishi22} found
significantly higher HCN-to-HCO$^+$ \Jto\ flux ratios in a high fraction of,
but not all, AGN-important ULIRGs than that in starburst-classified sources.
Some starburst-dominated galaxies may have HCN enhancement, which seems not
driven by a single process. We include their flux ratios (Figure
\ref{lineratio}).  The ALMA-ULIRG sample\citep{Imanishi22} has extreme
contribution from AGNs, as discussed in last paragraph of Section
\ref{section:Imanishi}. In such conditions, the HCN/HCO$^+$  line ratios might
be enhanced by the chemistry from X-ray dominated regions
\citep{Kohno2001,Harada2015}.

The average HCN-to-HCO$^+$ \Jto\ flux ratio in Magellanic Clouds is
0.49$\pm$0.34, which seems to be systematically lower than the ratios found in
galaxies. This is consistent with \cite{Braine17}, who also found that
HCN/HCO$^+$\,\Joz\ is lower in low-metallicity environment. 

In both the Milky Way and external galaxies, the [N/O] abundance ratio has a
positive correlation with [O/H] and [Fe/H], because nitrogen mostly originates
from low-mass stars and its nuclear production is a long-time
process\citep{Pilyugin03,Liang06}.  Therefore, the HCN/HCO$^+$ abundance ratio
would also decrease with [N/O] in low metallicity environments such as the
Magellanic Clouds, compared to normal metal-rich star-forming galaxies.

\subsection{Influence of infrared pumping}

Vibrational transitions of HCN and HCO$^+$ are difficult to be excited by
collision, but they can be excited by absorbing infrared photons at $\sim$ 13
-- 14\,\um, through infrared pumping \citep{Imanishi17}. Then the vibrational
transitions can cascade to rotational transitions and enhance \HCNto\ and/or
\HCOto\ lines.  If infrared pumping becomes a dominant process, the dense gas
mass estimated from  \HCNto\ and/or \HCOto\ would be over-estimated.

\cite{Sakamoto2010,Sakamoto2021} found strong detections of HCN v$_2$=1
$J=3\rightarrow2$, $J=4\rightarrow3$ in NGC\,4418 and Arp\,220 by ALMA
observation, indicating somewhat impact on the HCN rotational line. All of
our spectra do not show any detections of HCN v$_2$=1 \Jto\ including NGC\,4418
and Arp\,220, possibly due to limited S/N of APEX observation. Both NGC\,4418
and Arp\,220 show elevated $L'_{\rm HCN}/L'_{\rm HCO^+}$ ratio, especially
among SF-dominated galaxies (see Figure \ref{lineratio}), indicating possible
influence of IR pumping. NGC\,4418 and Arp\,220 are the most heavily influenced
galaxies by infrared pumping in our sample. However, both NGC\,4418 and
Arp\,220 do not show outliers in correlations of $L'_{\rm dense}$-$L_{\rm IR}$
and $\Sigma_{ L'_{\rm dense}}$-$\Sigma_{ L_{\rm IR}}$, indicating limited
enhancement towards \HCNto. Therefore, we do not expect a significant influence
by IR pumping in our sample.

\subsection{Comparison with other HCN and \HCOto\ work}
\label{section:Imanishi}

\cite{Imanishi22} reported ALMA observations of \HCNto\ and \HCOto\ towards ten
ULIRGs, most of which contain AGNs. They also found positive correlations
of $L_{\rm IR}$-$L^\prime_{\rm HCN}$ and $L_{\rm IR}$-$L^\prime_{\rm HCO^+}$. 

We obtained the data (Project code: 2017.1.00022.S and 2017.1.00023.S) from the
ALMA archive, and processed HCN and \HCOto\ datacubes with the standard
pipeline.  Overall we got roughly consistent results compared to those in
\cite{Imanishi22}.  We then adopted the size estimated from dust continuum
given by \cite{Imanishi22}. Details of the measured fluxes and sizes are shown
in Appendix \ref{app:Imanishmeasure} and Table \ref{table:Imanishflux}.
 
\begin{figure*}[ht]
\includegraphics[height=3.5in,width=7in]{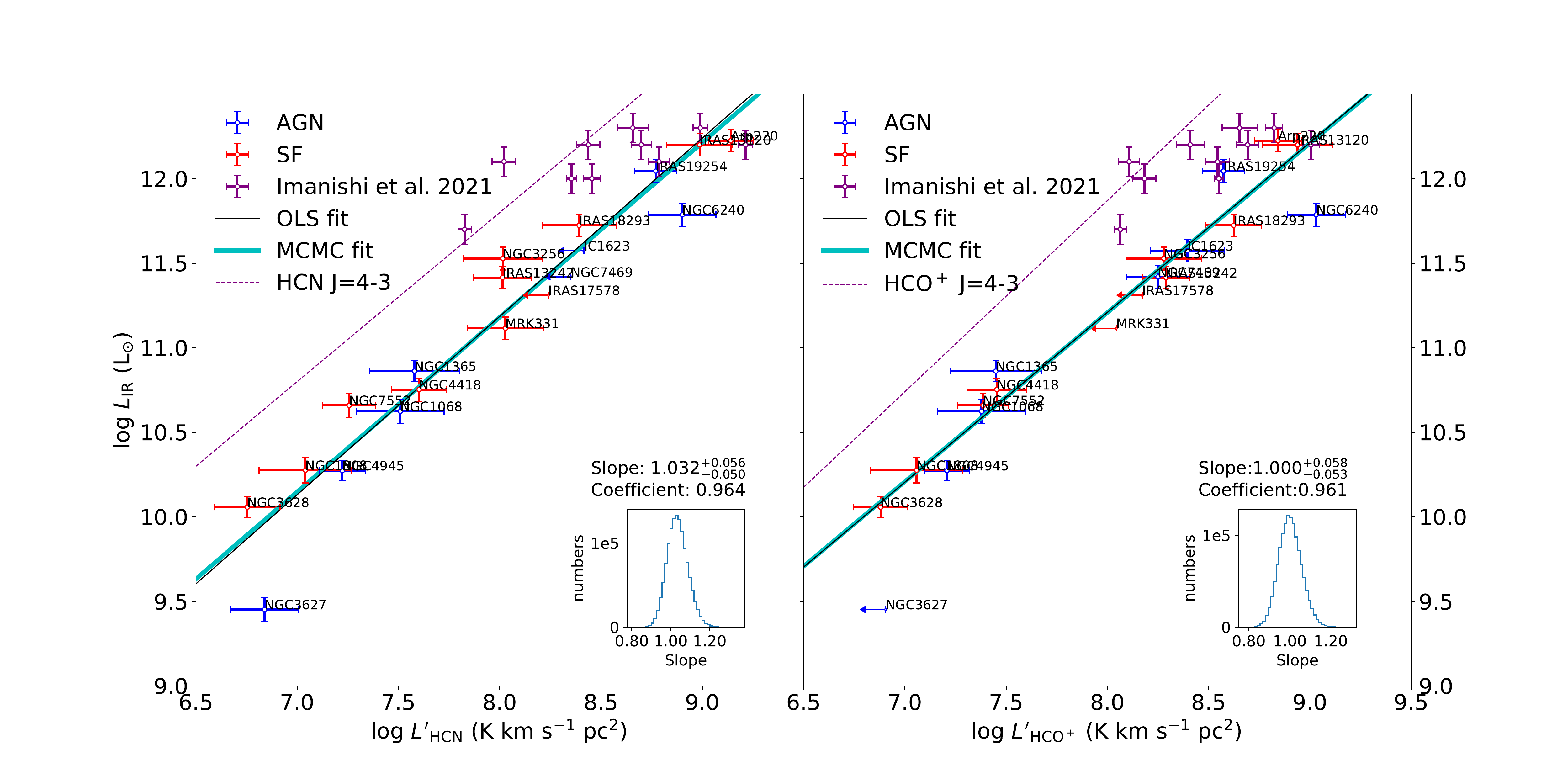}
\includegraphics[height=3.5in,width=7in]{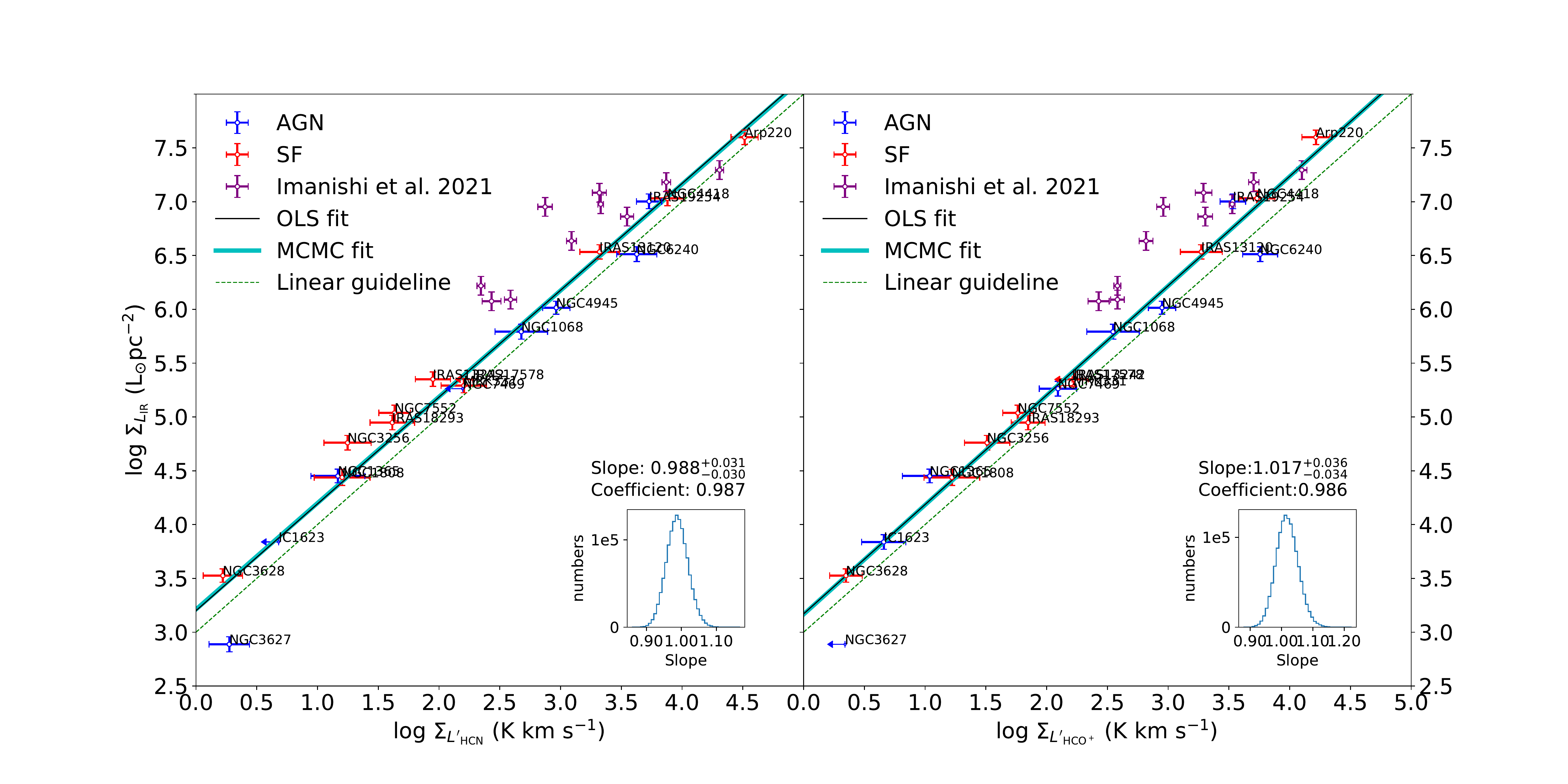}
\caption{ Correlations including ALMA-ULIRG sample from \citep{Imanishi22}.
{\it Top:} Correlations of $L'_{\rm HCN}-L_{\rm IR}$ ({\it top left})  and
$L'_{\rm HCO^+}-L_{\rm IR}$ ({\it top right}).  {\it Bottom:} Correlations of
$\Sigma_{L^\prime_{\rm HCN}}-\Sigma_{L_{\rm IR}}$ ({\it bottom left}) and
$\Sigma_{L^\prime_{\rm HCO^+}}-\Sigma_{L_{\rm IR}}$ ({\it bottom right}).
AGN-dominated and star-formation dominated galaxies are shown in blue and red
points, respectively.  The fitting results of Orthogonal Least Squares ({\sf
OLS}) and Markov chain Monte Carlo ({\sf MCMC})  are shown in black and cyan
lines, respectively. The green-dashed line shows a linear relation for
reference.  The insets present the probability density distributions of the
fitted slopes. Data points from \cite{Imanishi22} are shown in purple and not
included into fitting. The purple dashed lines are the fitting results of $\rm
HCN$ $J=4\rightarrow3$ and $\rm HCO^+$ $J=4\rightarrow3$ obtained from
\cite{Tan2018}.} \label{comparison} 
\end{figure*}

In Figure \ref{comparison}, we overplot the ALMA-ULIRG data on the \HCNto\ and
\HCOto\ correlations.  The ALMA-ULIRG data seem to be systemically above both
correlations of $L'_{\rm denseJ=2\rightarrow1}/L_{\rm IR}$ and $\Sigma_{L'_{\rm
dense}}-\Sigma_{L_{\rm IR}}$.  For our APEX data, the orthogonal scatters on
the correlation lines are $0.15$ and $0.11$ dex for $\rm HCN$ and $\rm HCO^+$,
respectively.  And the mean orthogonal offset of the ALMA-ULIRG galaxies are
$0.24\pm0.22$ and $0.27\pm0.16$ dex for $\rm HCN$ and $\rm HCO^+$,
respectively.  Only two and one ALMA-ULIRG galaxies lie in the 1-$\sigma$ range
of our $L_{\rm IR}-L^\prime_{\rm HCN}$ and $L_{\rm IR}-L^\prime_{\rm HCO^+}$
correlation, respectively. Two and three galaxies of ALMA-ULIRGs are off by
$>3\,\sigma$ from the correlation lines of $\rm HCN$ and $\rm HCO^+$,
respectively.

This offset is consistent with that found by \cite{Imanishi22}, who found that
the ALMA-ULIRG \Jto\ data follow $J=4\rightarrow3$ relation
given in \cite{Tan2018}.  \cite{Imanishi22} interpret that both HCN and \HCOp
are thermalised, which would bring the $\rm HCN$ and $\rm HCO^+$
$J=4\rightarrow3$ line luminosities comparable to those of
\Jto\ transitions. From our APEX data, the fitted linear lines
are systematically lower than those obtained from the $J=4\rightarrow3$ data
from \cite{Tan2018}, making the ALMA-ULIRGs above the \Jto\
relations.

Considering most of ALMA-ULIRGs are AGN-important ULIRGs, their $L_{\rm IR}$
might be dominated by AGNs, similar to the hot dust-obscured galaxies
(hot-DOGs) found with WISE mid-IR surveys \citep{Wu2012,Tsai2015}.  The IR
emission of hot-DOGs is mainly from the hot dust heated up by AGNs, which on
average contribute $> 75\%$ of the bolometric luminosity \citep{Fan2016}.  On
the other hand, the hot-DOGs also have high dust temperature ($\langle T_{\rm
dust} \rangle \sim 72 \,\rm K$), which seems to be the same for the
ALMA-ULIRGs.  The IRAS $f_{60\rm \mu m}/f_{\rm 100\mu m}$ ratio of the
ALMA-ULIRGs are all higher than those of the APEX galaxies, except for
NGC~4418, with average $f_{60\rm \mu m}/f_{\rm 100\mu m}$ ratios of
1.09$\,\pm\,$0.28 and 0.77$\,\pm\,$0.19 for the ALMA-ULIRGs and our APEX
sample, respectively. Therefore, we suspect that the ALMA-ULIRGs are
AGN-dominated, IR-overluminous galaxies, and they should behave similarly for
the \Joz\ lines of both HCN and \HCOp. Missing flux of ALMA observation may
also help to explain in a few targets (details see Appendix
\ref{app:Imanishmeasure}).

\section{Summary}
We present APEX observation towards \HCNto\ and \HCOto\ in 17 nearby
infrared-bright star-forming galaxies. Combining \HCNto\ and \HCOto\ data in
the literature, and with the total IR luminosity fitted from dust SED, we
correlations slopes of $1.03\pm 0.05$  and  1.00$\pm 0.05$ for $L'_{\rm
HCO^+}-L_{\rm IR}$ and $L'_{\rm HCN}-L_{\rm IR}$, respectively.    

To obtain correlations of surface densities, which could eliminate the biases
from uncertain distances, we use the size of 1.4\,GHz radio continuum to
normalise the luminosities of both IR emission ($\Sigma_{ L_{\rm IR}}$) and
dense gas tracers ($\Sigma_{ L'_{\rm HCN}}$ and $\Sigma_{ L'_{\rm HCO^+}}$).
These surface density correlations also show linear slopes of $0.99 \pm 0.03$
and $1.02 \pm 0.03$, for HCN and HCO$^+ J=2\rightarrow1$ lines, respectively.
The slope errors and $p$-values of the surface-density correlations are also
smaller than those of the luminosity correlations. The Spearman correlation
coefficients of the $J=2-1$ lines (0.98 and 0.96 for \HCNto\ and \HCOto) are
higher than those obtained from the $J=1-0$ \citep[0.94 for HCN
$J=1\rightarrow0$;][]{Gao2004b} and the $J=4-3$ \citep[0.89 and 0.84 for HCN
$J=4\rightarrow3$ and HCO$^+$ $J=4\rightarrow3$;][]{Tan2018} transitions,
indicating the advantage of the \Jto\ transitions of HCN and HCO$^+$ in tracing
the star-forming gas.

Comparing with the \HCNto\ and \HCOto\ data from star-forming clouds of the
Magellanic Clouds \citep{Galametz20}, we find that the low-metallicity
environment not only slightly deviates the data from the overall surface-density
correlations, but also would bias the \HCNto/\HCOto\ line ratio to low ratios.
This is consistent with previous findings in low-metallicity galaxies
\citep[e.g.,][]{Braine17}. The systematically lower HCN/\HCOp\ ratios are
possibly owing to the variation of [N/O] elemental abundance variation in
low-metallicity environments.

On the other hand, when comparing with the ULIRGs sample from
\cite{Imanishi22}, those AGN-important galaxies lay systematically above both
correlations of $L'_{\rm dense}$-$L_{\rm IR}$ and $\Sigma_{ L'_{\rm
dense}}$-$\Sigma_{ L_{\rm IR}}$ found in our APEX sample. This is likely due to
the AGN contribution to the total IR luminosity for such compact objects, which
could heavily overestimate SFRs. Therefore, the contribution from AGNs may not
be negligible in such extreme conditions.


\begin{acknowledgments}
We thank Dr. Maud Galametz for providing HCN and HCO$^+$ data of Magellanic Clouds
in \cite{Galametz20}. Z.Y.Z. and J.Z. acknowledge the support of the National
Natural Science Foundation of China (NSFC) under grants No. 12041305, 12173016,
the Program for Innovative Talents, Entrepreneur in Jiangsu, the science
research grants from the China Manned Space Project with No. CMS-CSST-2021-A08
and CMS-CSST-2021-A07. Chentao Yang acknowledges support from ERC Advanced Grant 789410. This
publication is based on data acquired with the Atacama Pathfinder Experiment
(APEX). APEX is a collaboration between the Max-Planck-Institut fur
Radioastronomie, the European Southern Observatory, and the Onsala Space
Observatory. $Herschel$ was an ESA space observatory with science instruments
provided by European-led Principal Investigator consortia and with important
participation from NASA.  PACS has been developed by a consortium of institutes
led by MPE (Germany) and including UVIE (Austria); KU Leuven, CSL, IMEC
(Belgium); CEA, LAM (France); MPIA (Germany); INAFIFSI/OAA/OAP/OAT, LENS, SISSA
(Italy); IAC (Spain). This development has been supported by the funding
agencies BMVIT (Austria), ESA-PRODEX (Belgium), CEA/CNES (France), DLR
(Germany), ASI/INAF (Italy), and CICYT/MCYT (Spain). SPIRE has been developed
by a consortium of institutes led by Cardiff University (UK) and including
University of Lethbridge (Canada); NAOC (China); CEA, LAM (France); IFSI,
University of Padua (Italy); IAC (Spain); Stockholm Observatory (Sweden);
Imperial College London, RAL, UCL-MSSL, UKATC, Univ. Sussex (UK); and Caltech,
JPL, NHSC, University of Colorado (USA). This development has been supported by
national funding agencies: CSA (Canada); NAOC (China); CEA, CNES, CNRS
(France); ASI (Italy); MCINN (Spain); SNSB (Sweden); STFC, UKSA (UK); and NASA
(USA). This work is based [in part] on observations made with the Spitzer Space
Telescope, which was operated by the Jet Propulsion Laboratory, California
Institute of Technology under a contract with NASA. This publication makes use
of data products from the Wide-field Infrared Survey Explorer, which is a joint
project of the University of California, Los Angeles, and the Jet Propulsion
Laboratory/California Institute of Technology, funded by the National
Aeronautics and Space Administration. This paper makes use of the following
ALMA data: ADS/JAO.ALMA\#2015.1.00717.S. ALMA is a partnership of ESO
(representing its member states), NSF (USA), and NINS (Japan), together with
NRC (Canada) and NSC and ASIAA (Taiwan) and KASI (Republic of Korea), in
cooperation with the Republic of Chile. The Joint ALMA Observatory is
operated by ESO, AUI/NRAO and NAOJ.

\end{acknowledgments}
\software{GILDAS/CLASS \citep{GILDAS}, Numpy \citep{numpy1,numpy2}, emcee \citep{emcee2013}, Photutils \citep{photutils}}

\appendix
\section{Background subtraction}
\label{app:bkg}
\begin{figure}
\includegraphics[height=2.5in]{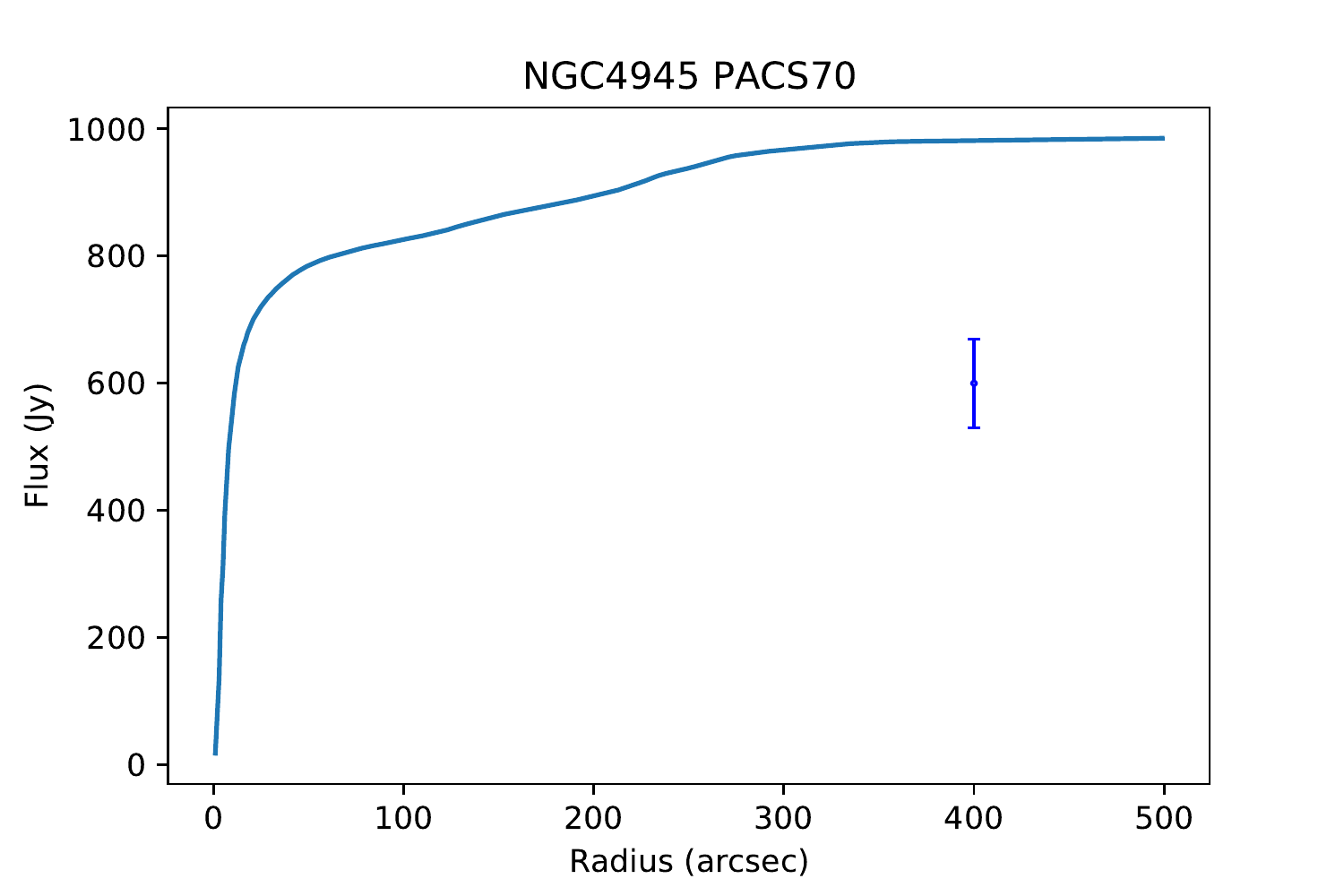}
\caption{Example: The curve of flux growth of PACS 70\,\um\ map of NGC\,4945. Y axis is flux within the circle of a certain radius. The error bar shows background rms plus 7\,\% absolute flux calibration error.}
\label{growthcurve}
\end{figure}

When measuring the infrared flux of each galaxy, we need to subtract the
background to avoid possible contamination from other nearby sources and
compute the error from root mean square error of infrared image. First, we
remove strong sources higher than $3\sigma$ using python package {\sc
photutils}\footnote{\url{https://pypi.org/project/photutils/0.4/}}. Then we
take the initial flat field subtraction using average flux of selected
background area. We compute the total flux in a small circle and increase the
radius of the circle until the flux in the circle do not increase. To show the
size which can cover total flux of the galaxy, we plot the flux vary with the
circle radius in Figure \ref{growthcurve}. Finally we choose the annulus within
$1.0\times$ and $1.5\times$ this radius as the background.

\section{Two-component dust SED fitting}
\label{app:SED_fitting}

\begin{figure}
\includegraphics[height=3.5in]{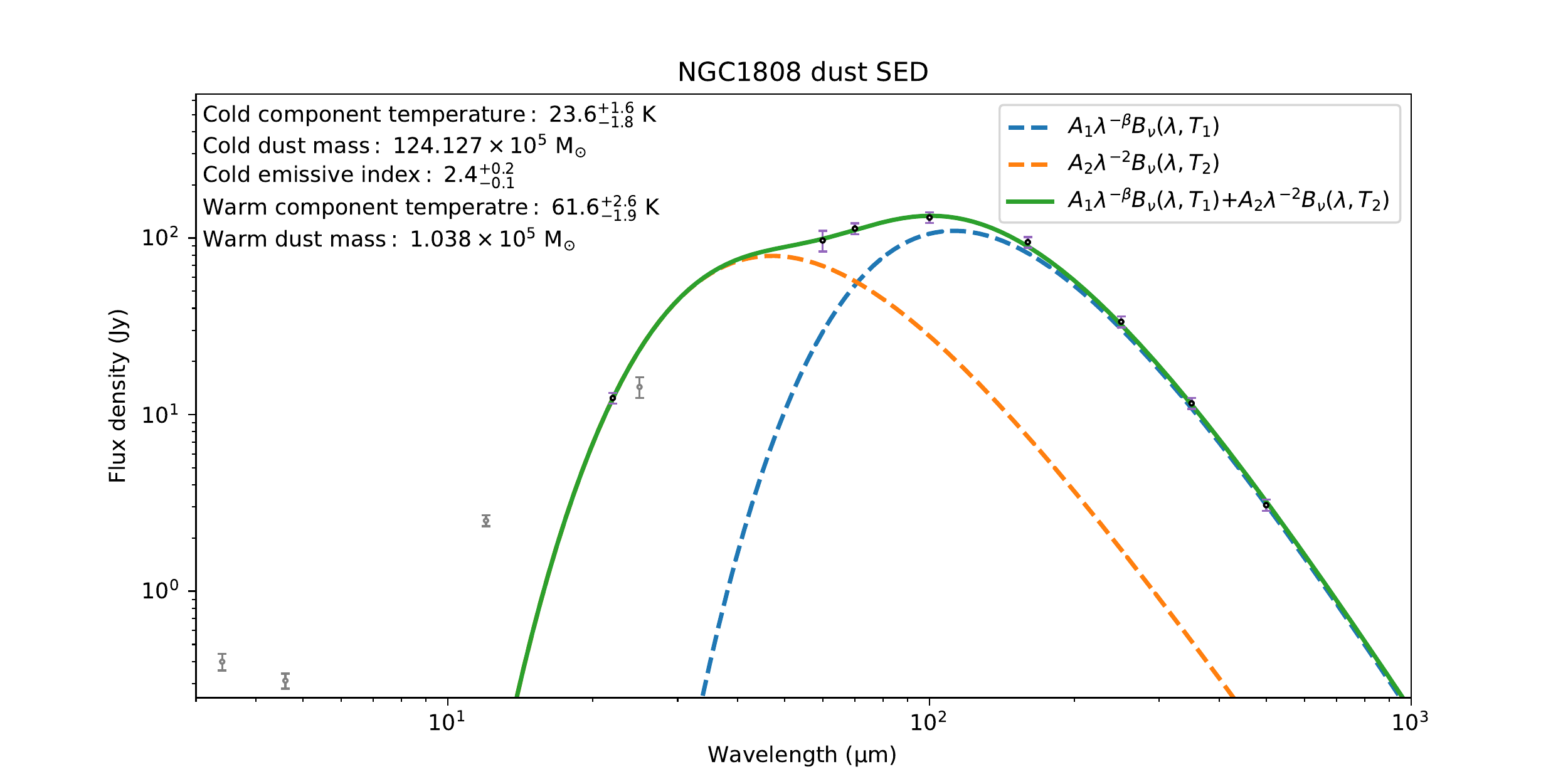}
\caption{An example of dust SED of NGC\,1808. The orange and blue-dashed lines
represent warm and cold components, respectively. The grey data points
represent bands not in the fitting, including WISE 3.4\,\um, 4.6\,\um, 12\,\um,
and IRAS 25\,\um due to short wavelength or low resolution.} 
\label{SED_plot}
\end{figure}

We download eleven bands infrared data observed by SPITZER and HERSCHEL from
3.4\,\um~to 500\,\um. We convolve them to same resolution and then do the
photometry. We try SED fitting with different bands data, and we find that
short wavelength bands (< 12\,\um) could suffer from contamination of stars or
PAH emission, and they are higher than that modified-blackbody expect. We also
put IRAS low resolution ($\sim4'$) flux for comparison, for most cases they
include total fluxes of the whole galaxies. We compute the scaling factor from
MIPS 24\,\um~and PACS 70\,\um~data. We include 60\,\um~and exclude 25\,\um~data
points . One example of dust SED fitting is shown in Figure \ref{SED_plot}.

\section{Different infrared bands as SFR tracer}
\label{app:compare_lum}

\begin{figure}
\includegraphics[height=3.5in]{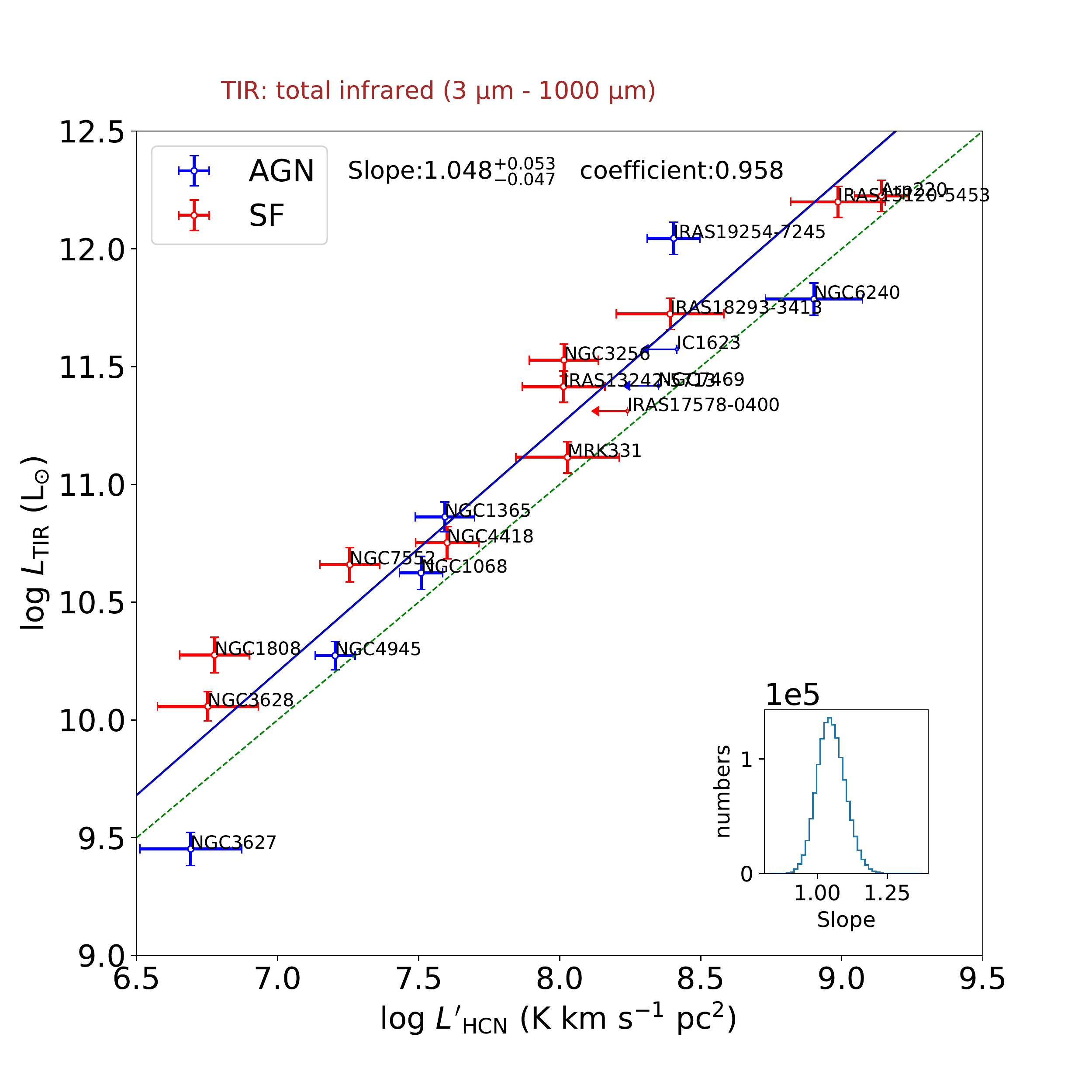}
\includegraphics[height=3.5in]{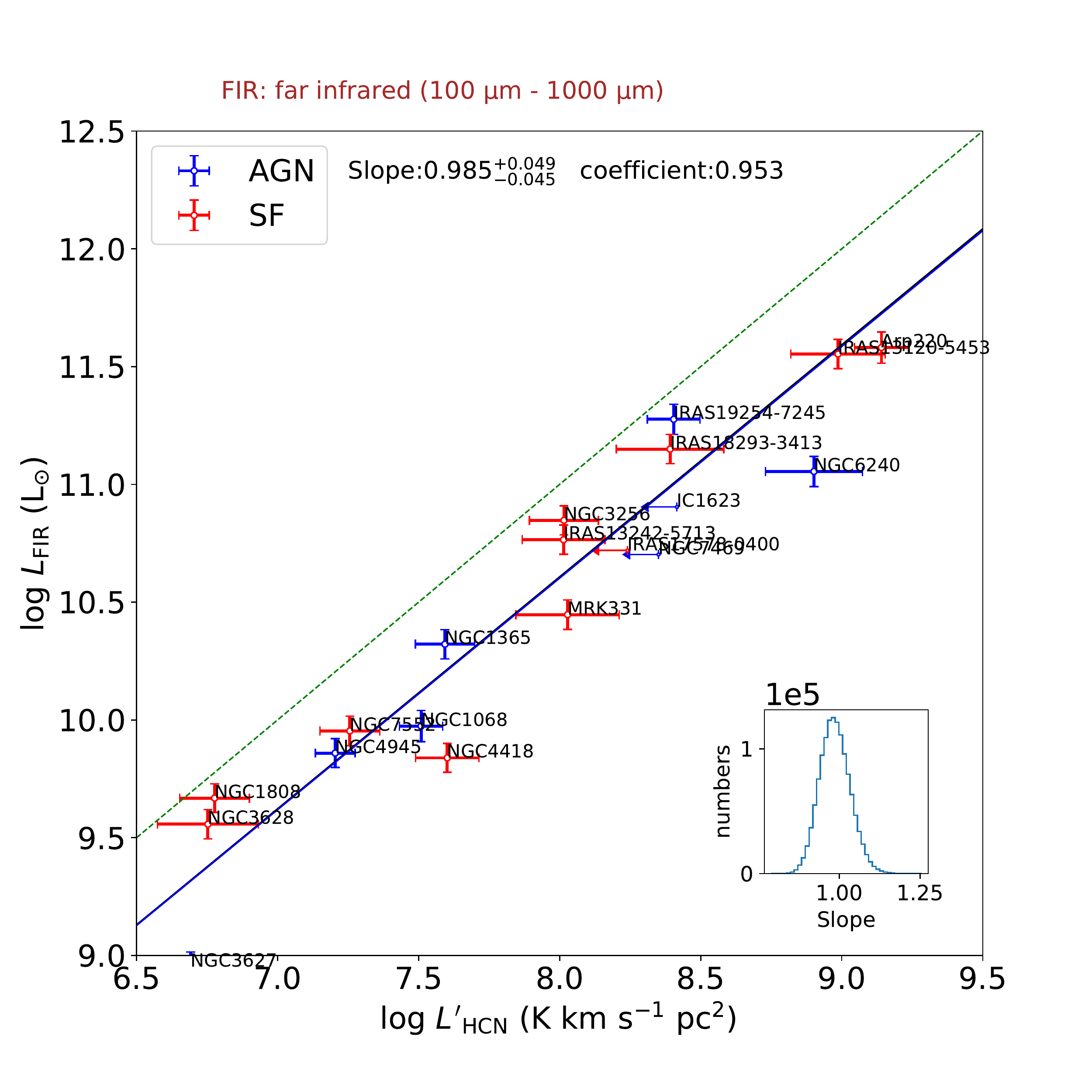}
\includegraphics[height=3.5in]{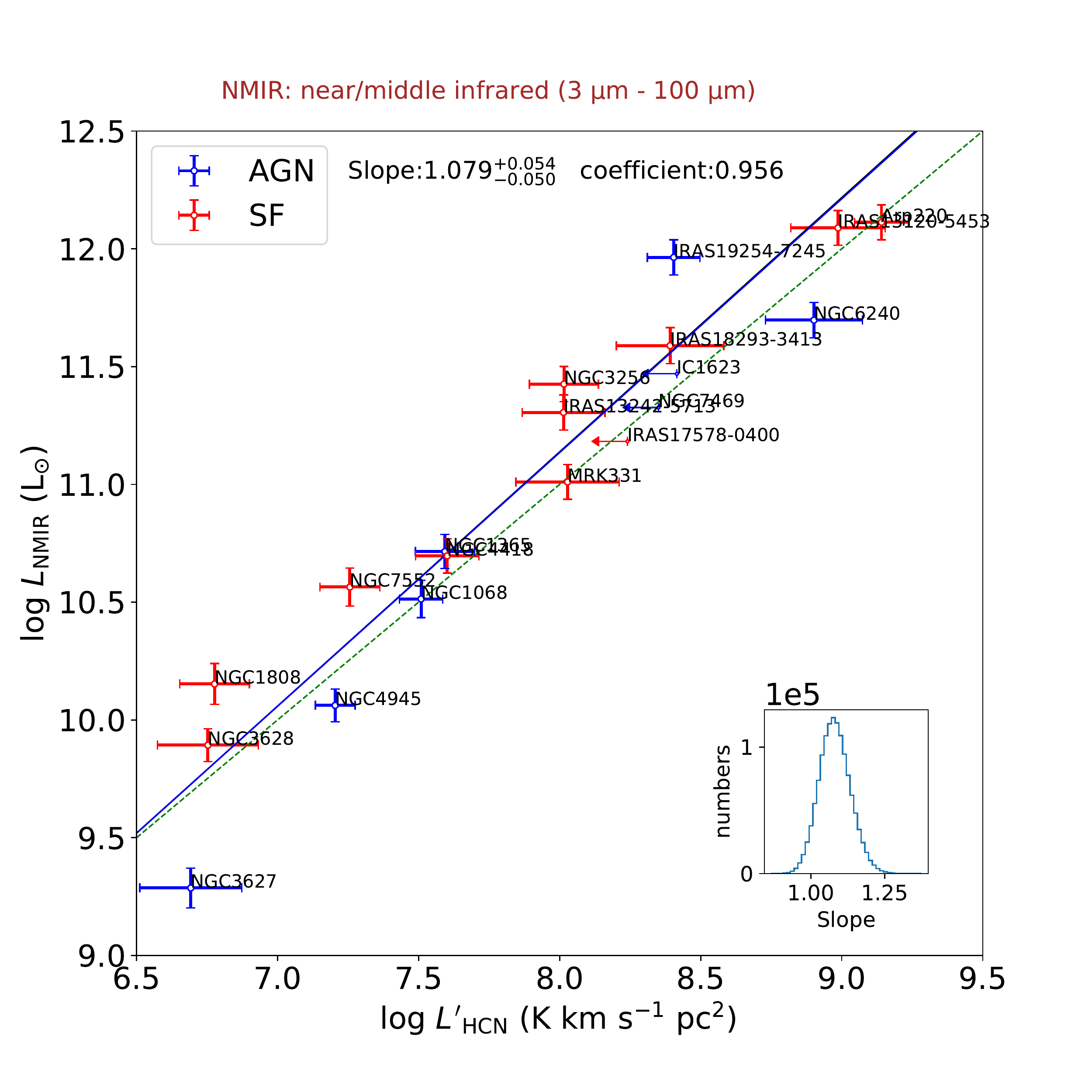}
\includegraphics[height=3.5in]{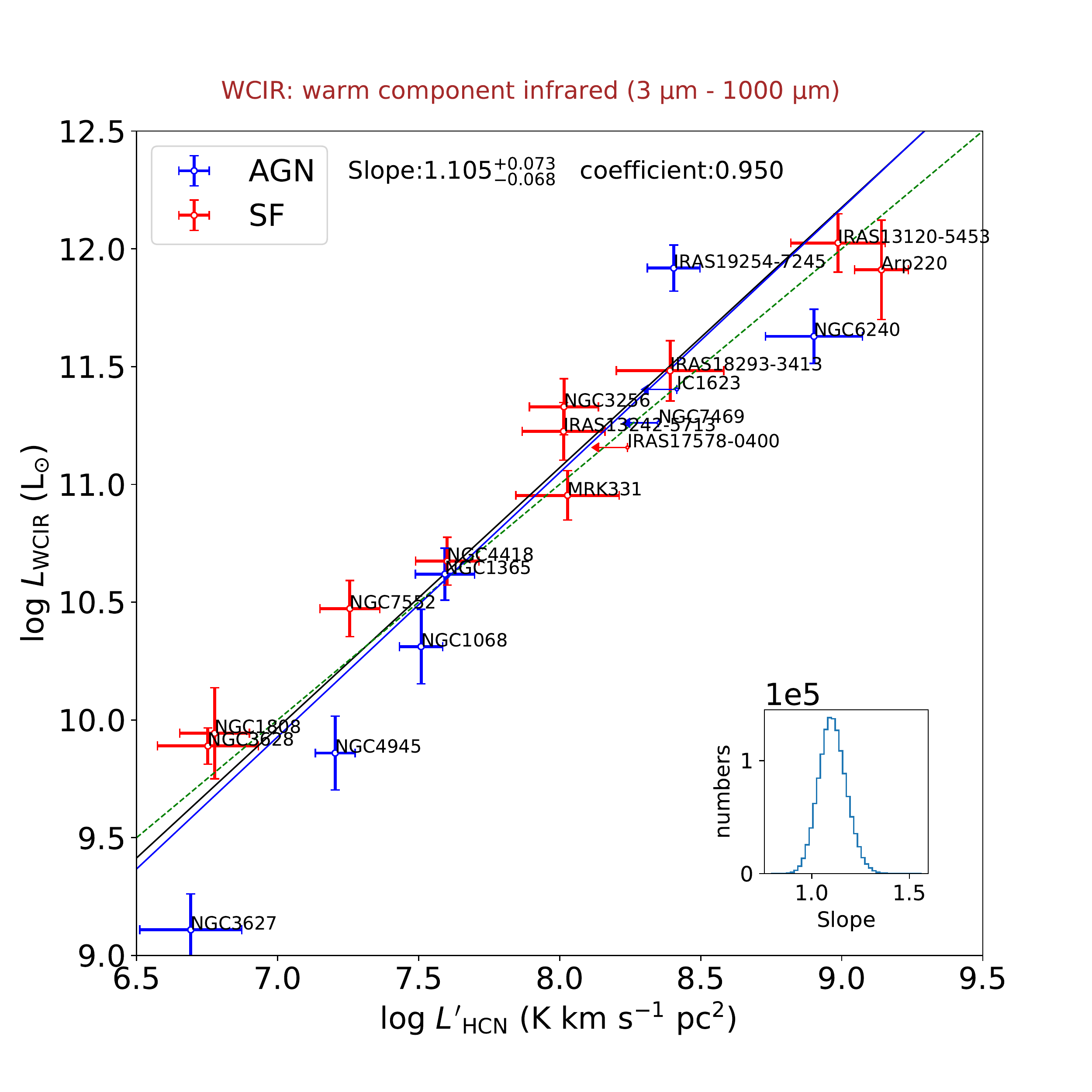}
\caption{Correlations between $L^\prime_{\rm HCN}$ and luminosities of different infrared bands. Top left: total infrared luminosity. Top right: far infrared luminosity. Bottom left: near/middle infrared luminosity. Bottom right: warm component infrared luminosity. The adopted wavelength ranges and dust components are labeled on the top of each panel.}
\label{compare_corr}
\end{figure}

To infer infrared band could better trace star-formation rate, we compare total
infrared (3-1000\,\um), far infrared (100-1000\,\um), near/middle infrared
(3-100\,\um) and warm component luminosity versus dense gas luminosity. We fit
the relationship with {\sf MCMC} and Least Square method to get the slope and compute
the coefficient with python package $\sc{numpy}$. We show the comparison plots
in Figure \ref{compare_corr}. We find that near/middle infrared contribute most
of total infrared luminosity and $\sim$ 3 times higher than far infrared
luminosity. However, far infrared luminosity is a little more stable than
near/middle infrared according to their slopes using different SED fitting
methods with different infrared bands. We finally choose total infrared
luminosity to trace star formation rate.

\section{Re-measure ALMA $\rm HCN$ and $\rm HCO^+$ \Jto\ flux}
\label{app:Imanishmeasure}

We reprocessed $\rm HCN$ and $\rm HCO^+$ \Jto\ data reported in \cite{Imanishi22}. For IRAS\,12127-1412, IRAS\,13509+0442, and Superantennae, there is no available $\rm HCN$ and $\rm HCO^+$ \Jto\ imaging products. So we downloaded their data and reprocessed with the pipeline and re-imaged with $''$robust=1.5$''$ with task {\sc tclean}. For others, we adopted the delivered data products. We re-measured line fluxes of $\rm HCN$ and $\rm HCO^+$ \Jto\ in all targets with the aperture size determined by curve of flux growth (described in Appendix \ref{app:bkg}) of moment 0 map of initial estimated $\rm HCN$ and $\rm HCO^+$. The re-measured line fluxes are shown in Table \ref{table:Imanishflux}. The fluxes of NGC\,1614, IRAS\,13509+0442 and IRAS\,22491-1808 are significantly higher than that reported in \cite{Imanishi22} (>80\%) and fluxes of three other galaxies are $\sim\,30\%$ higher. 

Missing flux of large scale structures during the interferometric observations
may also play a role in biasing the sample distributions, along with AGN bolometric contribution. For example, NGC\,1614
has been observed with CO \Joz\ and $^{13}$CO \Joz\ \citep{Konig2016}, which all show a gaseous disk
of $\sim 10-15''$ in diameter.  The ALMA \HCNto\ observations only revealed the
central emitting region of $\sim 2-3''$  in diameter, which may indicate
potentially extended fluxes filtered by ALMA.  While IRAS\,12112+0305 has been 
resolved into two nuclei, \cite{Evans2002} imaged its CO distribution and shown extended emission around the nuclei.

ALMA's shortest baseline is $\sim 15\,\rm m$, corresponding to largest scale of 30$''$. However, in the $uv$ data of IRAS\,12127-1412, IRAS\,13509+0442, and Superantennae, most shortest baselines are flagged possibly due to antenna shadowing. Thus, the largest recovered scale is 12$''$, similar to scales of NGC\,1614 and IRAS\,12112+0305. However it is difficult to evaluate how much would be the missing flux without single dish observations. 

\begin{deluxetable*}{ccccccc}
\tablenum{6}
\tablecaption{Line fluxes from ALMA observations}\label{table:Imanishflux}
\tablewidth{0pt}
\tablehead{
\colhead{Source name} & \colhead{Redshift} &
\colhead{Distance} &\colhead{$S_{\rm HCN~J=2\rightarrow1}$} & \colhead{$S_{\rm HCO^+~J=2\rightarrow1}$} &\colhead{HPBW$_{\rm dust}$} & \colhead{${\rm log}~L_{\rm IR}$} \\
\colhead{}& \colhead{}& \colhead{(Mpc)} & \colhead{($\rm Jy~km~s^{-1}$)}&\colhead{($\rm Jy~km~s^{-1}$)}&\colhead{(10$^{-3}$arcsec)}&\colhead{($\rm L_\odot$)}\\
\colhead{(1)}&\colhead{(2)}&\colhead{(3)}&\colhead{(4)}&\colhead{(5)}&\colhead{(6)}&\colhead{(7)}
}
\startdata
NGC1614        & 0.0160 & 68  & 11.56 $\pm$ 0.29 & 19.99 $\pm$ 0.31 & 1710    $\times$       1630 & 11.7 \\
IRAS06035-7102 & 0.0795 & 356 & 3.77  $\pm$ 0.25 & 3.72  $\pm$ 0.29 & 705     $\times$        613 & 12.2 \\
IRAS08572+3915 & 0.0580 & 256 & 1.44  $\pm$ 0.12 & 1.76  $\pm$ 0.13 & 318     $\times$        287 & 12.1 \\
IRAS12112+0305 & 0.0730 & 326 & 8.63  $\pm$ 0.27 & 5.89  $\pm$ 0.28 & 283     $\times$        186 & 12.3 \\
IRAS12127-1412 & 0.1332 & 620 & 0.79  $\pm$ 0.07 & 0.74  $\pm$ 0.08 & 151     $\times$        96  & 12.2 \\
IRAS13509+0442 & 0.1364 & 636 & 1.25  $\pm$ 0.16 & 1.24  $\pm$ 0.19 & 481     $\times$        367 & 12.3 \\
IRAS15250+3609 & 0.0552 & 243 & 4.31  $\pm$ 0.19 & 2.31  $\pm$ 0.19 & 445     $\times$        375 & 12   \\
Superantennae  & 0.0617 & 273 & 7.47  $\pm$ 0.54 & 4.27  $\pm$ 0.38 & 348     $\times$        284 & 12.1 \\
IRAS20551-4250 & 0.0430 & 188 & 5.51  $\pm$ 0.03 & 8.67  $\pm$ 0.03 & 367     $\times$        346 & 12   \\
IRAS22491-1808 & 0.0776 & 347 & 12.9  $\pm$ 0.36 & 8.00  $\pm$ 0.39 & 217     $\times$        131 & 12.2 \\
\enddata
\tablecomments{ Column 1: galaxy name. Column 2: Redshifts from \cite{Imanishi22}. Column
3: Luminosity distance from \cite{Imanishi22}. Column 4: \HCNto\ line flux we measured. Column 5: \HCOto\ line fluxed we measured. Column 6: galaxy sizes: Half Power Beam Width (HPBW) of 178\,$\rm GHz$ dust continuum from \cite{Imanishi22}.
Column 7: logarithmic value of infrared luminosity inferred from IRAS observation \citep{Imanishi22}}
\end{deluxetable*}

\section{Dust temperature in AGN-host and SF-dominated galaxies.}
\label{app:Tdust}
\begin{figure}
\includegraphics[width=0.48\textwidth]{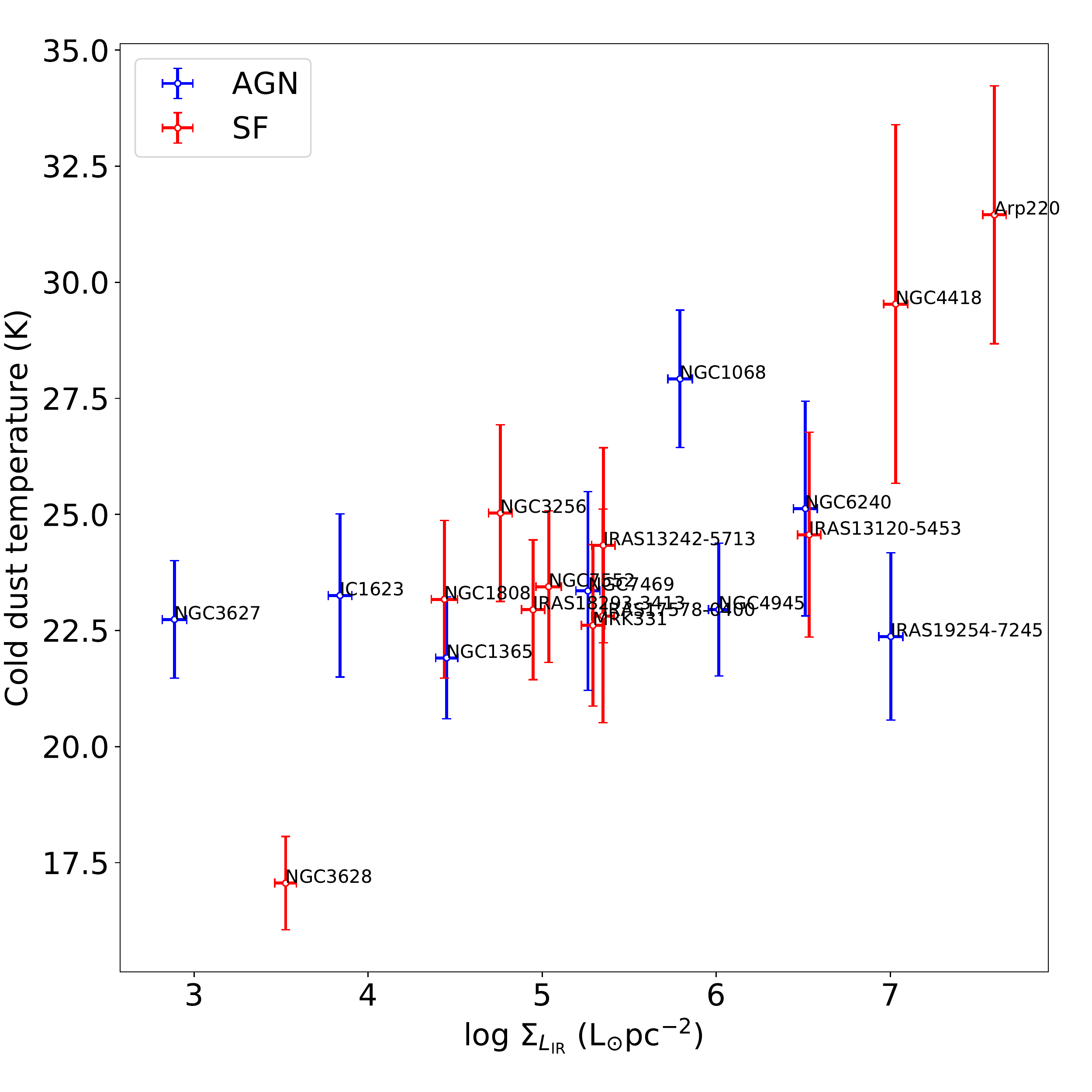}
\includegraphics[width=0.48\textwidth]{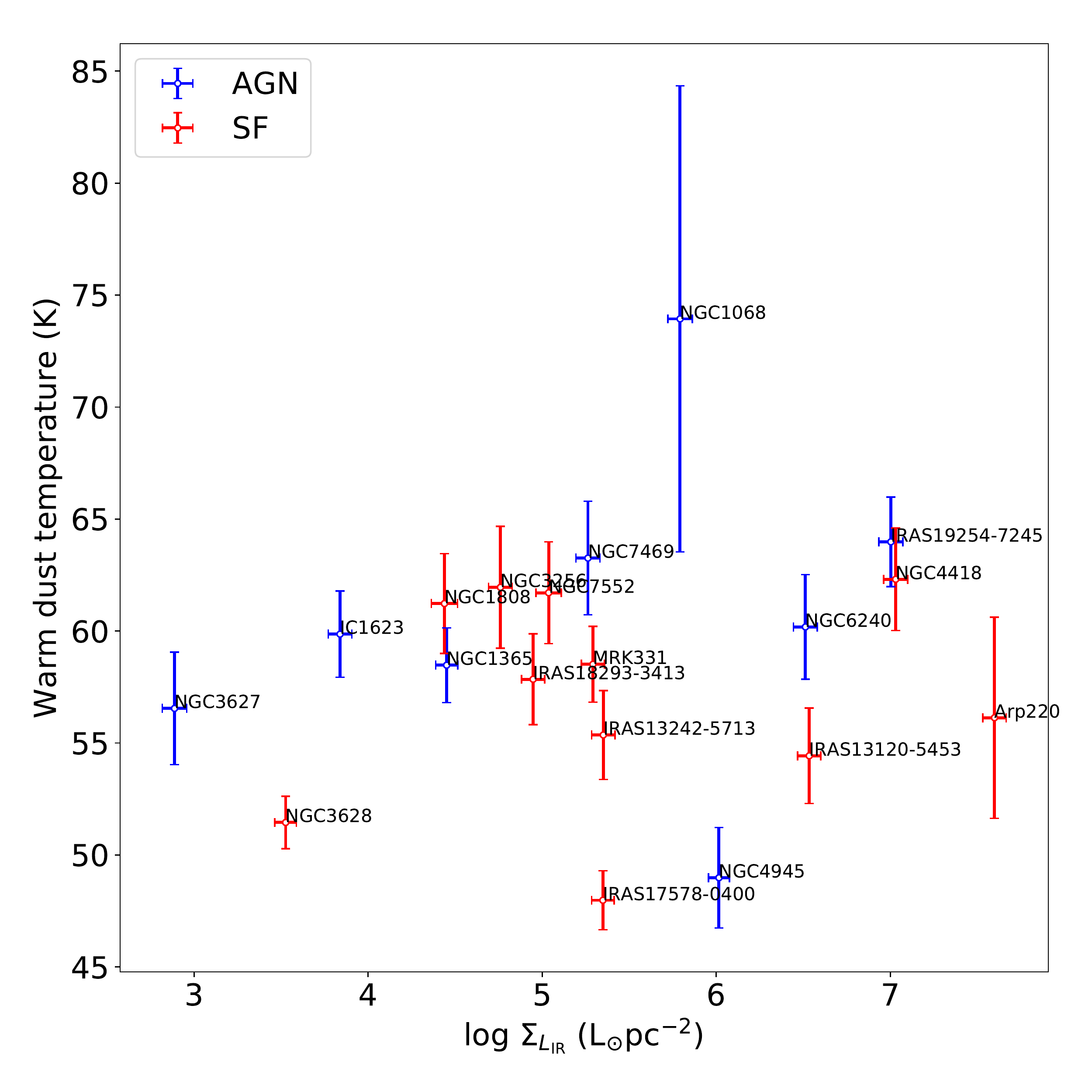}
\caption{(Left) Cold dust temperature ($T_{\rm c}$) as a function of infrared
luminosity surface density ($\Sigma_{\rm L_{\rm IR}}$). (Right) Warm dust temperature ($T_{\rm w}$) as
a function of $\Sigma_{\rm L_{IR}}$. AGN-dominated and star-forming dominated
galaxies are shown in blue and red points, respectively.} \label{fig:TdustIR}
\end{figure}

Infrared luminosity of ULIRGs in \cite{Imanishi22} may be enhanced by
AGN-heated hot dust (Section \ref{section:Imanishi}). Here we check if the dust
temperature of galaxies in our sample are influenced by AGN activities, by
plotting dust temperature versus infrared luminosity surface density in
different galaxy types ( see Figure \ref{fig:TdustIR}). Dust temperature are
obtained under assumption of a modified blackbody with a fixed warm component
emissivity index of 2. Figure \ref{fig:TdustIR} shows results of the
infrared luminosity surface density and the cold dust temperature fitted with
these assumptions in the left panel. The right panel of Figure
\ref{fig:TdustIR} shows  infrared luminosity surface density versus warm dust
temperature. The caption mentions $T_{\rm c}$ and $T_{\rm w}$ equals to cold
and warm dust temperature, respectively. There is no significant relation
between dust temperature and galaxy type or $\Sigma_{\rm L_{\rm IR}}$.

\bibliographystyle{aasjournal}
\bibliography{references}

\end{CJK*}

\end{document}